\documentclass[iop]{emulateapj}

\usepackage{multirow}

\usepackage{natbib}
\begin{document}

\title{Leveraging 3D-HST Grism Redshifts to Quantify Photometric Redshift Performance}
\author{Rachel Bezanson\altaffilmark{1}\footnotemark[$\dagger$], David A. Wake\altaffilmark{2,3}, Gabriel B. Brammer\altaffilmark{4}, Pieter G. van Dokkum\altaffilmark{5}, Marijn Franx\altaffilmark{6}, \\Ivo Labb\'{e}\altaffilmark{6},
Joel Leja\altaffilmark{5}, Ivelina G. Momcheva\altaffilmark{5}, Erica J. Nelson \altaffilmark{5}, Ryan F. Quadri \altaffilmark{7}, Rosalind E. Skelton\altaffilmark{8}, Benjamin J. Weiner\altaffilmark{1}, Katherine E. Whitaker\altaffilmark{9}\footnotemark[$\dagger$]}

\altaffiltext{1}{Steward Observatory, Department of Astronomy, University of Arizona, AZ 85721, USA}
\altaffiltext{2}{Department of Physical Sciences, The Open University, Milton Keynes, MK7 6AA, UK}
\altaffiltext{3}{Department of Astronomy, University of Wisconsin-Madison, Madison, WI 53706, USA}
\altaffiltext{4}{Space Telescope Science Institute, Baltimore, MD 21218, USA}
\altaffiltext{5}{Department of Astronomy, Yale University, 260 Whitney Avenue, New Haven, CT 06511, USA}
\altaffiltext{6}{Leiden Observatory, Leiden University, Leiden, The Netherlands}
\altaffiltext{7}{George P. and Cynthia W. Mitchell Institute for Fundamental Physics \& Astronomy, Department of Physics \& Astronomy, Texas A\&M University, College Station, TX 77843, USA}
\altaffiltext{8}{South African Astronomical Observatory, PO Box 9, Observatory, Cape Town, 7935, South Africa}
\altaffiltext{9}{Department of Astronomy, University of Massachusetts, Amherst, MA 01003, USA}
\footnotetext[$\dagger$]{Hubble Fellow}

\shorttitle{3D-HST Grism Redshifts and Photo-z Performance}
\shortauthors{Bezanson et al.}

\begin{abstract}
We present a study of photometric redshift accuracy in the 3D-HST photometric catalogs, using 3D-HST grism redshifts to quantify and dissect trends in redshift accuracy for galaxies brighter than $H_{F140W}<24$ with an unprecedented and representative high-redshift galaxy sample. We find an average scatter of $0.0197\pm0.0003(1+z)$ in the \citet{skelton3dhst} photometric redshifts. Photometric redshift accuracy decreases with magnitude and redshift, but does not vary monotonically with color or stellar mass. The 1-$\sigma$ scatter lies between $0.01-0.03$(1+z) for galaxies of all masses and colors below $z<2.5$ (for $H_{F140W}{<}24$), with the exception of a population of very red ($U-V > 2$), dusty star-forming galaxies for which the scatter increases to $\sim0.1(1+z)$. Although the overall photometric redshift accuracy for quiescent galaxies is better than for star-forming galaxies, scatter depends more strongly on magnitude and redshift than on galaxy type. We verify these trends using the redshift distributions of close pairs and extend the analysis to fainter objects, where photometric redshift errors further increase to $\sim0.046(1+z)$ at $H_{F160W}=26$. We demonstrate that photometric redshift accuracy is strongly filter-dependent and quantify the contribution of multiple filter combinations. We evaluate the widths of redshift probability distribution functions and find that error estimates are underestimated by a factor of $\sim1.1-1.6$, but that uniformly broadening the distribution does not adequately account for fitting outliers. Finally, we suggest possible applications of these data in planning for current and future surveys and simulate photometric redshift performance in the LSST, DES, and combined DES and VHS surveys.  
\end{abstract}

\section{Introduction}

Studies of the high-redshift Universe rely increasingly upon \emph{photometric redshifts} to identify and map the distribution of distant galaxies. These photometric redshifts are estimated from the overall spectral shapes as traced by catalogs of photometric data, as opposed to fitting one or more spectroscopic features. Photometric redshift surveys dramatically extend the possibilities of cosmological and galaxy evolutionary studies by vastly increasing the numbers and variety of galaxies beyond more observationally expensive spectroscopic galaxy surveys.

Because galaxy redshift is such a fundamental property, understanding the errors in photometric redshift estimates is crucial for interpreting empirical findings. For example, redshift uncertainties have been demonstrated to severely impact the measured evolution of the mass function \citep[e.g.][]{chen:03,marchesini:09,muzzin:13}. Photometric surveys can allow for studies of large scale structure and galaxy clustering that are inaccessible to spectroscopic surveys, but the modeling of results depends strongly on understanding the redshift uncertainties \citep[e.g.][]{chen:03,quadri:08, wake:11, mccracken:15, soltan:15}. In order to fully model the effects of photometric redshifts we must quantify their accuracy, which itself can depend on redshift and galaxy properties.

Traditionally, photometric redshift accuracy is tested by direct comparison between measured redshifts and \emph{true} redshifts for a subset of a catalog with followup spectroscopy \citep[e.g.][]{skelton3dhst, dahlen:13}. Alternatively, several groups have identified novel methods of testing photometric redshift accuracy using the clustering properties of galaxies \citep[e.g.][]{newman:08,benjamin:10, quadri:10}. Finally, a number of studies of photometric redshift accuracy have been conducted based on simulated mock galaxy catalogs \citep[e.g.][]{ascaso:15}. The first method is the most direct, but is typically biased towards very specific samples and the brightest galaxies for which spectroscopic redshifts are feasible: primarily at $z<1$ and for star-forming galaxies with bright emission lines. The second class of methods have different possible implementations, but in general these require large data sets, can lack sensitivity to certain types of systematic redshift errors or to catastrophic failures, and the results may be difficult to interpret. Although mock catalogs are an attractive alternative and require no additional data, they are fundamentally limited by their ability to match the empirical diversity of an evolving galaxy population.

Several methods of fitting photometric redshifts and many software packages and libraries and exist within the community. Given the same data, each method will produce subtly different results \citep[e.g.][]{hogg:98, hildebrandt:08, hildebrandt:10, abdalla:11}. Recently, \citet{dahlen:13} published an extensive study evaluating the accuracy of redshifts produced by various photometric codes, focusing on the direct comparison of objects with spectroscopic redshifts in the CANDELS (Cosmic Assembly Near-infrared Deep Extragalactic Legacy Survey) fields, including a sample with deeper Hubble Space Telescope (\emph{HST}) grism spectroscopic redshifts to extend the analysis to high redshift. Although the study investigated some trends in photometric redshift accuracy with galaxy properties, it is fundamentally limited to the availability of spectroscopic redshifts.

The 3D-HST survey \citep[PI: P. van Dokkum]{3dhst,skelton3dhst} provides a unique opportunity to directly test the photometric redshift accuracy in the CANDELS \citep{candels, candelsb} and 3D-HST fields. The data from this HST Legacy program combined with those from the AGHAST (A Grism H-Alpha SpecTroscopic) survey (PI: B. Weiner) include low-resolution grism spectroscopy across $\sim70\%$ of the CANDELS/3D-HST imaging footprint. This uniform spectroscopic coverage allows for unprecedented grism spectroscopic estimates of the true redshifts for thousands of galaxies beyond $z>1$. Using grism redshifts, we can quantify the redshift accuracy of photometric catalogs in these fields for a sufficiently large and unbiased sample of high-redshift ($z<3$) galaxies.  In this Paper, we evaluate the photometric redshift accuracy in the HST/WFC3(Wide Field Camera 3)-selected photometric catalogs produced by the 3D-HST collaboration \citep{skelton3dhst}. Although we focus our investigation on photometric redshifts derived by the \texttt{EAZY} code \citep{eazy}, we expect the conclusions to be similar for different algorithms given that \citet{dahlen:13} found no strong differences amongst different methodologies and codes for a similar dataset.  Additionally, although that study recommended median combining photometric redshifts using a multitude of fitting techniques, the \texttt{EAZY} code was run by three different groups and consistently produced relatively low scatter and outlier fractions amongst the suite of redshift tests. In this work, we aim to quantify trends in the scatter between photometric and true redshifts as a function of galaxy properties as well as the occurrence rates of catastrophic failures.

Given the ultimate goal of quantifying photometric redshift performance in the 3D-HST catalogs, this Paper is organized as follows. Section \ref{sect:data} briefly describes the 3D-HST dataset. Section \ref{sect:scatter} quantifies the accuracy of photometric redshifts of the full detected sample and as a function of galaxy properties by comparison with spectroscopic and grism redshifts in addition to an analysis of close pairs. Section \ref{sect:filters} discusses the relationship between photometric redshift accuracy and photometric bandpasses included in the redshift fitting. Section \ref{sect:pdfs} addresses the use of the full photometric probability distribution function of redshift as opposed to a single-valued photometric redshifts. Section \ref{sect:sim_surveys} extends the analysis of filter-dependence to simulate photometric redshift performance in the DES, DES plus VHS, and LSST surveys. Finally, we summarize the major results of the study in Section \ref{sect:summary}. 

Throughout this paper we assume a concordance cosmology ($H_0 = 70 \mathrm{km\,s^{-1}\,Mpc^{-1}}$, $\Omega_M = 0.3$, and $\Omega_{\Lambda}=0.7$) and quote all magnitudes in the AB system.

\section{Data}\label{sect:data}
\subsection{Sources of Data}
The primary data in this paper are collected from the HST/WFC3-selected v4.1 photometric \citep{skelton3dhst} and grism catalogs \citep{momcheva3dhst} produced by the 3D-HST collaboration over ${\sim}900$ square arcminutes in five extragalactic fields: AEGIS, COSMOS, GOODS-North, GOODS-South, and UDS. The photometric catalogs include PSF-matched aperture photometry from a multitude of multi-wavelength ($0.3\mu m$-$8.0\mu m$) ground and space-based images \citep{dickinson:03, steidel:03, capak:04, giavalisco:04, erben:05, hildebrandt:06, sanders:07, taniguchi:07, barmby:08, furusawa:08, wuyts:08, erben:09, hildebrandt:09, nonino:09, cardamone:10,  retzlaff:10, candels, candelsb, kajisawa:11, whitaker:11, 3dhst, bielby:12, hsieh:12, mccracken:12, ashby:13}.  Objects are detected from combined CANDELS/3D-HST HST/WFC3 images ($J_{F125W}$,$H_{F140W}$,and $H_{F160W}$). Photometric catalogs were produced using the MOPHONGO (Multiresolution Object PHotometry ON Galaxy Observations) code (I. Labb\'{e} et al., in preparation). 
\begin{figure*}
\centering
\includegraphics[width=\textwidth]{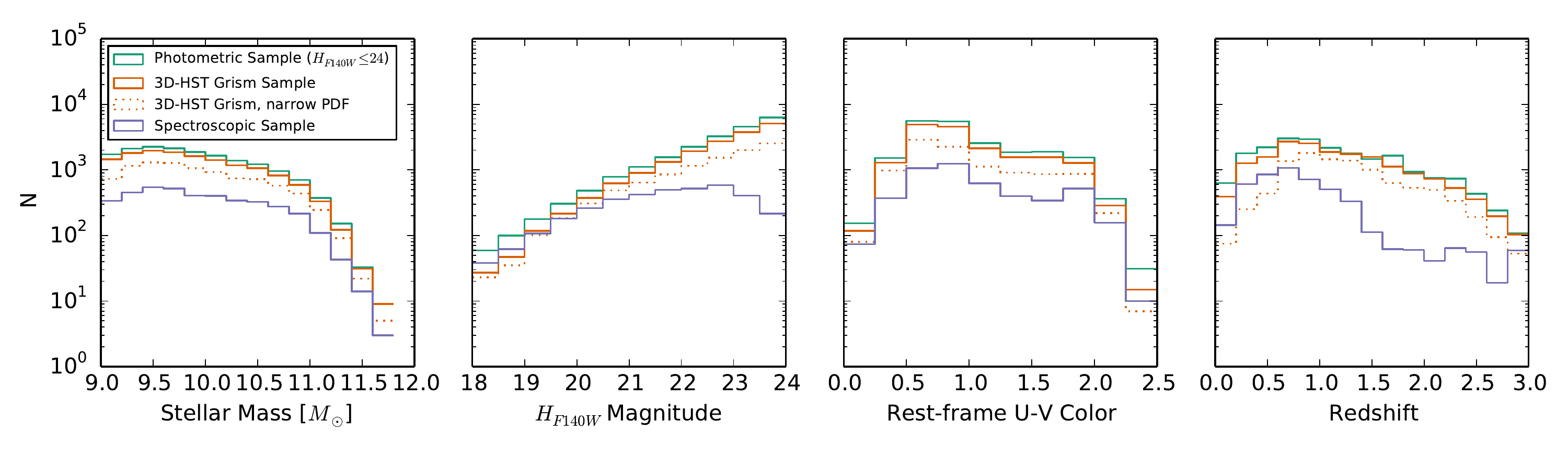}
\caption{Distribution of galaxies in the full photometric, grism, and spectroscopic redshift samples in stellar mass, apparent magnitude, rest-frame U-V color, and redshift. The grism redshift sample better reflects the distribution of properties of the photometric sample, particularly in faint and high redshift galaxies. Dotted orange histogram indicates the sample of grism redshifts that provide estimates of $z_{true}$ that are independent from photometric redshifts, as identified by decreased redshift uncertainty when the grism spectra are included in redshift fits.}
\label{fig:dist}
\end{figure*}

\begin{figure*}
\centering
\includegraphics[width=\textwidth]{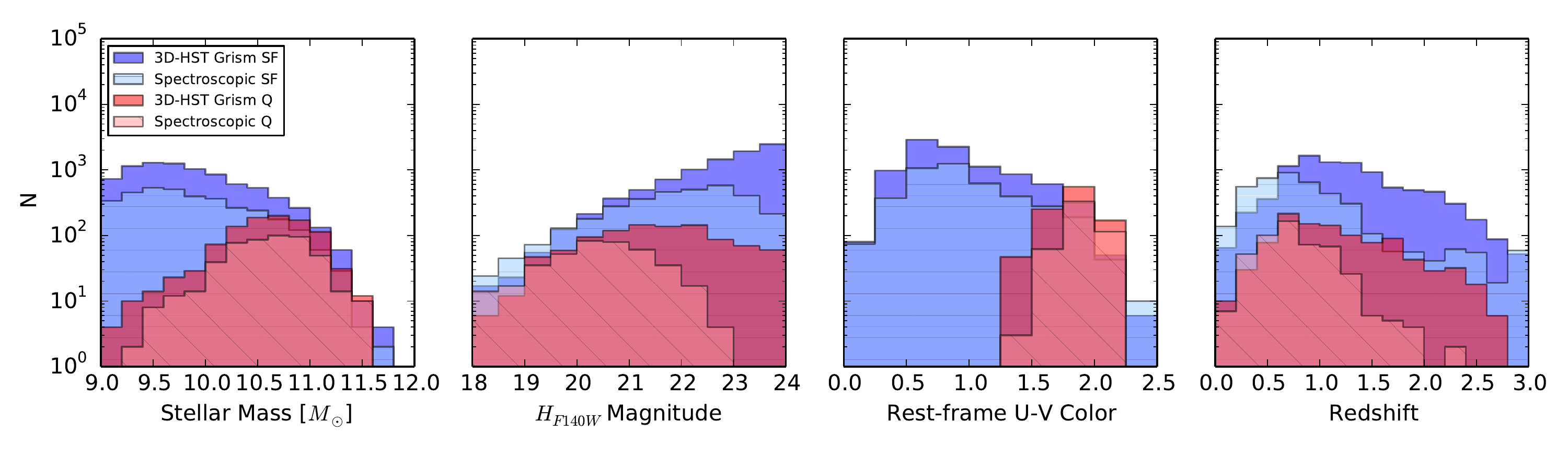}
\caption{Distribution of star forming (SF) and quiescent (Q) galaxies in the spectroscopic sample (hatched blue and red histograms) and the grism sample (solid blue and red histograms) with tightened redshift uncertainties in stellar mass, apparent magnitude, rest-frame U-V color, and redshift. In addition to its improved completeness at faint magnitudes ($H_{F140W} \gtrsim 21$) and high redshifts ($z\gtrsim1$), it is clear that the sampling of the galaxy populations in those regimes is dramatically improved for the 3D-HST grism redshifts.}
\label{fig:dist_sfq}
\end{figure*}

The 3D-HST Treasury Survey is primarily a 248 orbit grism spectroscopic survey, providing HST/WFC3 G141 near-infrared grism spectroscopy ($\lambda=1.1-1.7\mu m$) in four of the five CANDELS/3D-HST \citep{candels, candelsb} fields (AEGIS, COSMOS, GOODS-S, and UDS). Additional HST/WFC3 G141 grism spectroscopy in the GOODS-N field is included from the AGHAST survey (GO-11600, P.I.: B. Weiner). The combined dataset covers a total of $\sim600$ square arcminutes with an average two-orbit depth. Objects selected from the 3D-HST photometric catalogs are matched in the grism data, extracted, and analyzed uniformly by the 3D-HST collaboration \citep[][]{3dhst,momcheva3dhst}. All extracted spectra are jointly fit along with the photometric data to provide grism redshifts for all objects brighter than $JH_{IR} \leq 26$, where $JH_{IR}$ is based on flux in the combined $F124W$, $F140W$, and $F160W$ images. Grism spectra and redshift fits for all 23,564 galaxies brighter than $JH_{IR}<24$ are visually inspected to determine grism quality flags (use\_zgrism). Although redshift fits exist for fainter objects in the 3D-HST catalogs, we only include grism redshifts with these quality flags in this analysis. We adopt the term \emph{grism redshift} ($z_{grism}$) to describe these low resolution spectroscopic redshifts to distinguish from traditional high resolution \emph{spectroscopic redshifts} ($z_{spec}$). The uniform spectroscopic coverage of the survey is crucial to the current investigation. For a complete description of the 3D-HST survey see \citet{3dhst}, the photometric catalogs see \citet{skelton3dhst}, and the grism spectra see \citet{momcheva3dhst}.

The 3D-HST catalogs also include a vast collection of spectroscopic redshifts from ground-based spectroscopic surveys of these well-studied fields. In the AEGIS field, spectroscopic redshifts are matched with the DEEP2 DR4 survey \citep{cooper:12, newman:13}. In COSMOS, redshifts are collected from the zCOSMOS survey \citep{lilly:07}, and a collection of MMT/Hectospec redshifts (Kriek et al., in prep.). GOODS-N redshifts are included from \citet{kajisawa:10}, which includes data from a number of other surveys \citep{yoshikawa:10, barger:08, reddy:06, treu:05, tkrs, cowie:04, cohen:00, cohen:01, dawson:01}. In GOODS-S, redshifts are collected from the FIREWORKS catalog \citep{wuyts:08}. Finally, redshifts in the UDS are collected from the UDS Nottingham webpage, including data from \citep{yamada:05, simpson:06, breukelen:07, geach:07, ouchi:08, smail:08, ono:10, simpson:12, akiyama:15}, IMACS/Magellan redshifts \citep[][]{papovich:10}, an VLT X-shooter redshift from \citet{sande:13}, and Keck/DEIMOS redshifts \citep{bezanson:13b, bezanson:15}.

Photometric redshifts from \citet{skelton3dhst} catalogs are determined using the \texttt{EAZY} code \citep{eazy}, which fits the spectral energy distribution (SED) of each galaxy with a library of galaxy templates and outputs the full probability distribution function (PDF) with redshift; see \citet{skelton3dhst} for a complete description of this fitting. These fits utilize the default \texttt{EAZY} template set, which includes: five P\'EGASE \citep{fioc:97} stellar population synthesis models,  a young, dusty template, and an old, red galaxy template that is described in \citet{whitaker:11}. We adopt $z_{peak}$, or the peak redshift marginalized over the PDF as a galaxy's \emph{photometric redshift} ($z_{phot}$) in Sections \ref{sect:scatter} and \ref{sect:filters} of this paper. In \S \ref{sect:pdfs} we return to investigate the accuracy of the full photometric PDFs, assessing their overall widths. Grism redshifts that are obtained from joint fits to the photometry and HST - WFC3 slitless grism spectra from the 3D-HST survey. A full discussion of the redshift fitting can be found in \citet{momcheva3dhst}. In short, each two-dimensional grism spectrum is fit with a combination of EAZY continuum templates and a \citet{dobos:12} emission line template, with a prior imposed by the photometric redshift probability distribution function. 

Derived properties are included from the version 4.1.4 3DHST catalogs \citep{momcheva3dhst}. These fits assume either the spectroscopic or grism redshift of each galaxy when available \cite{momcheva3dhst} or photometric redshifts \citep{skelton3dhst} in the full CANDELS footprint, derived as follows. Rest-frame colors are estimated for all galaxies following \citet{brammer:11}, also using the \texttt{EAZY} code. Stellar population parameters are calculated using the \texttt{FAST} code \citep{kriek:09} using Single Stellar Population (SSP) models from \citet{bc:03} and assuming exponentially declining star formation histories, solar metallicity, and a \citet{chabrier:03} initial mass function. Galaxies with good photometry are identified by a use flag (use\_phot=1 flag in the 3D-HST catalogs), which indicates that an object is not a star, is not near a bright star, has at least two exposures in $F125W$ and $F160W$ images, is detected in $F160W$, and has non-catastrophic redshift and stellar population fits. 

We adopt the maximum probability redshift, z\_max\_gris, from the 3D-HST catalogs as the \emph{grism redshift} ($z_{gris}$) in this paper. A consequence of the inclusion of the photometric data in this fitting method is that the photometric and grism redshift estimates are not completely independent measurements. When investigating the scatter between the two correlated measurements, we only include galaxies for which the addition of the grism spectrum added significant information to the fit, as quantified by a tightened probability distribution, such that the 68\% confidence interval for the $z_{gris}$ is less than half of that for $z_{phot}$ (discussed in more detail in \S\ref{sect:corr}). 

\subsection{Properties of the Sample}
\begin{figure*}[t]
\centering
\includegraphics[width=0.95\textwidth]{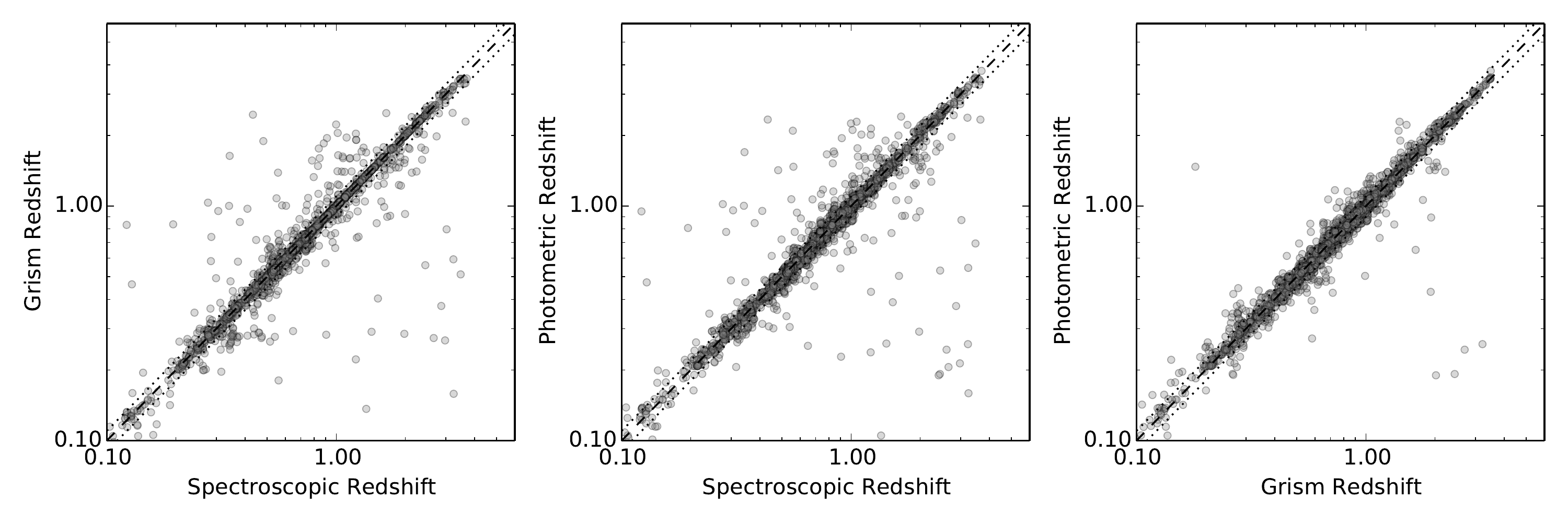}
\caption{Grism vs. Spectroscopic redshift, Photometric vs. Spectroscopic redshift, and Photometric vs. Grism redshift for all galaxies with spectroscopic redshifts in the 3D-HST catalogs. The scatter is lower between spectroscopic and grism redshifts than with photometric redshifts; however the outlier fraction is similar for grism and photometric redshifts.}
\label{fig:z_z}
\end{figure*}

\begin{deluxetable}{lccccc}
\tabletypesize{\scriptsize}
\tablecaption{Number of Galaxies in Each Sample}
\tablehead{\colhead{Total} & \colhead{AEGIS} & \colhead{COSMOS} & \colhead{GOODSN} & \colhead{GOODSS} & \colhead{UDS}}
\startdata
\sidehead{\textbf{Full Spectroscopic Sample}}
4805 & 1094 & 420 & 1836 & 1280 & 175 \\
\sidehead{\textbf{Full Grism Sample}}
17732 & 3139 & 3576 & 3338 & 4260 & 3419 \\
\sidehead{\textbf{Grism Sample with Narrowed PDFs\tablenotemark{a}}}
10190 & 2032 & 1719 & 2148 & 2234 & 2057 \\
\sidehead{\textbf{Quiescent Grism Sample with Narrowed PDFs\tablenotemark{a}}}
1026 & 180 & 175  & 204 & 257 & 210 \\
\sidehead{\textbf{Star-Forming Grism Sample with Narrowed PDFs\tablenotemark{a}}}
9164 & 1852 & 1544  & 1944 & 1977 & 1847 \\
\enddata
\tablecomments{Total number of galaxies in each redshift sample. The narrowed grism redshift sample is over twice the size of the spectroscopic redshift sample and is much more representative of high redshifts ($z\gtrsim1$), faint magnitudes ($H_{140W}\gtrsim21$), and for quiescent galaxies. The grism redshifts are evenly spread across the CANDELS/3D-HST fields.}
\label{tbl:numbers}
\tablenotetext{a}{Narrowed PDFs refer to galaxies for which the 68\% confidence interval for z\_gris is less than or equal to half that of z\_phot to minimize correlated measurement errors.}
\end{deluxetable}

Figure \ref{fig:dist} indicates the distribution of $H_{F140W}\leq24$ galaxies in the 3D-HST catalogs with photometric redshifts (green), grism redshifts (orange), and spectroscopic redshifts (purple) as a function of stellar mass, apparent $H_{F140W}$ magnitude, rest-frame U-V color, and redshift. The full grism sample is included as the orange histogram in Figure \ref{fig:dist} and the effect of excluding possibly correlated redshift fits is indicated by the dotted orange histogram. The number of galaxies in each sample, both overall and in each field, is included in Table \ref{tbl:numbers}. Although this cut is roughly uniform across galaxy properties, this has the effect of preferentially excluding low redshift ($z\lesssim0.7$) galaxies, where the wavelength coverage of the G141 grism provides little spectral information. While this diminishes the utility of grism redshifts at low redshift, we emphasize that at these redshifts spectroscopic samples are much more representative of the overall population of galaxies. We further investigate the extent and consequences of possible correlations between photometric and grism redshifts in \S \ref{sect:corr}. 

We highlight the bias of spectroscopic redshift surveys towards star-forming galaxies at the faint and high redshift ends of the distributions. To demonstrate this, we use rest-frame $U-V$ and $V-J$ color criteria to distinguish between star-forming and quiescent galaxies in the 3D-HST catalogs, using the thresholds defined by \citet{whitaker:12a}. Solid histograms in Figure \ref{fig:dist_sfq} show the number of star forming galaxies (blue) and quiescent galaxies (red) with 3D-HST grism redshifts and narrowed redshift PDFs. The distribution of galaxies with spectroscopic redshifts is indicated by dotted lines and lighter histograms. Although the distributions of spectroscopic and photometric redshifts are similar in stellar mass and $U-V$ color, the number of quiescent galaxies with spectroscopic redshifts dwindles dramatically fainter than $H_{F140W}\gtrsim21$ and at high redshift ($z\gtrsim1$). This is specifically the regime in which the grism redshifts are especially important.

It is clear from Figure \ref{fig:dist} that the number of galaxies in the grism sample is nearly an order of magnitude larger than for the spectroscopic sample, but more importantly it more closely follows the distribution of the photometric catalog. Primarily, these redshifts include many more faint objects and galaxies at high ($z>1$) redshifts.  Furthermore, the grism redshifts include vastly better sampling of the quiescent galaxy population improving by more than an order of magnitude on the number of quiescent galaxies at faint magnitudes and high redshifts, although these numbers are still small.

\section{Photometric Redshift Accuracy: Quantifying Scatter and Failure Rates}\label{sect:scatter}

The strongest test of photometric redshift performance given a fitting methodology can be obtained by comparing photometric redshifts to \textit{true} redshifts for a subset of detected objects that reflects the parameter space spanned by the photometric catalogs themselves. Spectroscopic surveys provide excellent datasets with which to perform these tests, but are often quite biased either due to selection criteria or measurement failures. Due to its untargeted nature, redshifts determined from the 3D-HST grism spectra are not susceptible to these selection biases. In fact, the distribution of galaxies in the grism sample very closely follows that of the full photometric sample down to $H_{F140W}\leq24$ with a slight offset due to the smaller footprints (see Figures \ref{fig:dist} and \ref{fig:dist_sfq}). Furthermore, we note that spectroscopic redshifts do not always represent the true redshift, either due to errors in spectroscopic analysis or misidentification of photometric counterparts.

In order to test the accuracy of the photometric redshifts in the 3D-HST/CANDELS fields, particularly for faint, high-redshift, and/or quiescent galaxies we benefit significantly by using grism redshifts, instead of those from higher resolution spectroscopy, as a proxy for the true redshifts of galaxies in the catalogs. In this Section we demonstrate the feasibility of using the grism redshifts in this way and test the photometric redshift performance in the 3D-HST catalogs. 

\subsection{Spectroscopic Sample}\label{sect:spec_scatter}

\begin{figure*}[!t]
\centering
\includegraphics[width=\textwidth]{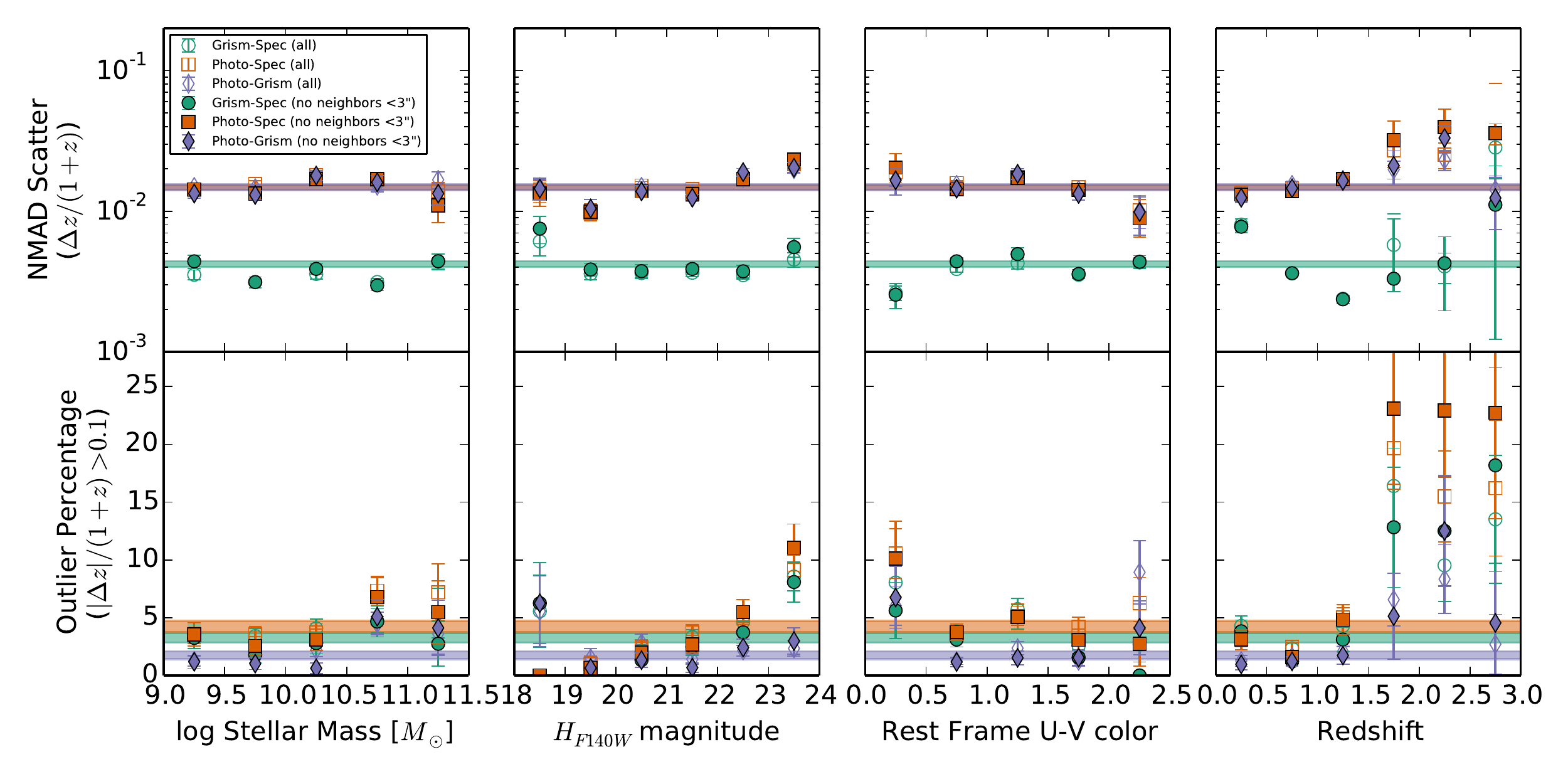}
\caption{Redshift Accuracy for 3D-HST galaxies with spectroscopic, photometric, \emph{and} grism redshifts. Each column includes NMAD scatter (top row) and outlier fraction (bottom row) for this sample as a function of stellar mass (left column), $H_{F140W}$ magnitude (second column), rest-frame U-V color (third column), and redshift (fourth column). Comparison between spectroscopic and grism redshifts is included with green symbols, spectroscopic and photometric redshifts in orange, and grism and photometric redshifts in purple. Filled symbols include only galaxies without neighbors within 3" to eliminate possible spec-z misidentifications; this does not significantly decrease outlier fractions. Scatter between grism and spectroscopic redshifts is much lower than for photometric redshifts, but the outlier fraction is similar. Scatter between photometric redshifts and grism redshifts is extremely similar to scatter with spectroscopic redshifts, suggesting grism redshifts can also be used as a proxy for true redshift.}
\label{fig:scatter_all}
\end{figure*}

We begin by identifying a subset of 2993 galaxies in the 3D-HST catalogs with photometric, grism, and spectroscopic redshifts. Taking the spectroscopic redshift to be the true value, the scatter between redshift estimates is indicative of the errors in the photometric and grism redshifts.  For the following tests, we compare all three redshifts for the full spectroscopic sample. Comparisons with the spectroscopic redshifts may yield the best estimate of redshift measurement errors, since these are more precise measurements of $z_{true}$ and the grism and photometric redshift measurements may be correlated. However the spectroscopic sample will always be smaller and more biased than the grism redshifts.

In Figure \ref{fig:z_z} we show the photometric, grism, spectroscopic redshift comparisons. Outlier thresholds of $|\Delta\,z|/(1+z) > 0.1$ are indicated by dotted lines in each panel. Qualitatively, the left panel (grism vs. spectroscopic redshift) exhibits less scatter than the center (photometric vs. spectroscopic redshift) panel. 

We quantify the scatter in $\Delta\,z/(1+z)$ using the normalized median absolute deviation (NMAD) as:

\begin{equation}
 \sigma_{NMAD} = 1.48 \times \mathrm{median}(|\Delta\,z|/(1+z)). 
 \end{equation}
 
 \noindent This measure of scatter is sensitive to the median deviations but less sensitive to catastrophic redshift failures than an RMS scatter. The outlier fraction is defined as the fraction of galaxies with $|\Delta z|/(1+z) >0.1$, although we find similar results with different definitions of this quantity. We emphasize that this definition of outliers does not include formal errors; we return to evaluating the redshift accuracy with respect to photometric redshift error estimates in \S \ref{sect:pdfs}. In this and subsequent sections, we only calculate scatter and outlier fraction for subsamples with more than ten galaxies.  Figure \ref{fig:scatter_all} shows the scatter and outlier fraction as a function of mass, $H_{F140W}$ magnitude, rest-frame U-V color, and redshift for the spectroscopic sample. All comparisons are made for the same sample of galaxies: photometric versus spectroscopic redshifts in orange, grism versus spectroscopic in green, and photometric versus grism in purple.  Errors in each measurement are estimated via bootstrap resampling of the full sample. The average value for each sample is indicated by the colored horizontal band in each panel and average scatter and outlier fractions are reported in Table \ref{tbl:scatter}.

One concern in interpreting accuracies derived from comparisons with spectroscopic redshifts is the possibility that published spectroscopic redshifts can also be erroneous. The spectroscopic redshift catalog contains only high-quality redshifts, as assessed by each independent study, however there is still the possibility that the spectroscopic measurement was not assigned to the correct object in the 3D-HST catalogs. Spectroscopic counterparts in the 3D-HST catalogs were matched within a radius of $0''.5$ \citep{skelton3dhst}. Although this is a conservative matching aperture, misidentification of photometric counterparts due to faulty astrometry or close neighbors, could falsely boost the measured rate of catastrophic failures in photometric redshift estimations. We can minimize this possibility by only including spectroscopic redshifts for galaxies with a unique counterpart in the 3D-HST photometric catalogs, removing galaxies from the sample for which there was at least one neighboring galaxy within $3''$ for which the spectroscopic redshift falls inside of the 95\% confidence interval of the photometric P(z). The scatter and outlier fractions for this sample are included as filled symbols in Figure  \ref{fig:scatter_all}. This aggressive cut decreases the sample to 1654 galaxies. However, the effect on scatter and outlier fractions is extremely subtle. Therefore, the catastrophic redshift failures cannot be explained simply by incorrect comparisons, but note that there additional errors in spectroscopic redshift identification could also contribute these outliers.

A number of overall trends appear in each column. Scatter between grism and spectroscopic redshifts is much lower than for photometric redshifts, but the outlier fraction is comparable. The outlier fractions are lower between grism and photometric redshifts, suggesting that the two measurements are correlated. The NMAD scatter between photometric redshifts and grism or spectroscopic redshifts is strikingly similar, both on average and as a function of galaxy properties.  In most cases the measurements completely overlap. This suggests that if grism redshifts are used to evaluate the accuracy of photometric redshifts, $\sigma_{NMAD}$ will be a robust indication of the scatter about $z_{true}$. 

The outlier fraction is $\sim2$ times lower for the grism redshifts compared to the spectroscopic redshifts, which suggests the existence of correlated errors if the grism catastrophic redshift failures are a subset of photometric failures. We investigate how much of this is driven by cases where the spectroscopic redshifts are not accurately identifying $z_{true}$. We visually inspect the spectral energy distributions and images of the 54 outliers ($|z_{spec}-z_{gris}|/(1+z_{spec}) > 0.1$), 7 of which are \emph{not} $|z_{phot} - z_{gris}| > 0.1$ outliers. First, we find that $\sim40\%$ (21) of the galaxies are below $z_{gris}=0.7$, where the G141 grism provides little additional information due to a lack of spectral features. Additionally, many of these outliers (33\%, 18) are in the GOODS-N MODS compilation \citep{kajisawa:10}, which does not have quality flags. Furthermore, the grism spectra caught emission lines for 10 (19\%) of these galaxies. Finally, the grism spectra are extracted using the photometric positions and therefore, in the absence of blending in the HST imaging, they are not susceptible to misidentification. We conclude that for a significant fraction of catastrophic redshift outliers, the grism provides estimates of true redshifts of the photometric objects in the catalog that are as good or better than the spectroscopic redshifts and this difference could account for the difference in outlier fractions.

\begin{figure*}[!t]
\centering
\includegraphics[width=0.9\textwidth]{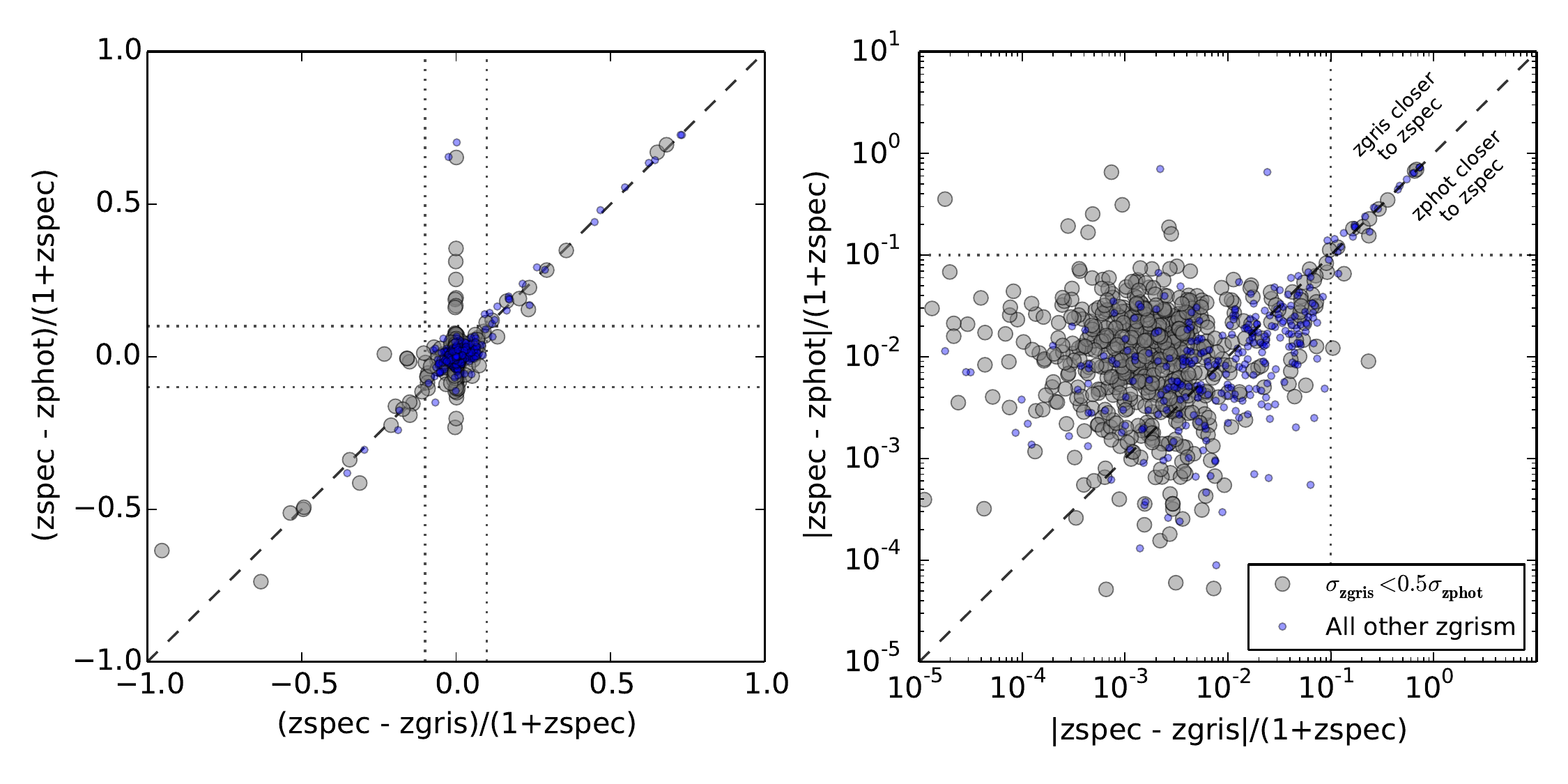}
\caption{Correlations between redshift errors in linear scale (left panel) and logarithmic scale (right panel). Large grey symbols indicate galaxies for which the 68\% confidence interval for $z_{gris}$ is $\leq 0.5$ that of the photometric redshift, small blue points mark galaxies with untightened PDFs. The vertical axes show residuals in photometric redshifts and horizontal axes show the residuals in grism redshifts about the known spectroscopic redshifts. Dotted lines indicate the $|\Delta z|/(1+z)>0.1$ outlier threshold. Diagonal trends (dashed lines) indicate correlated errors between photometric and grism redshifts for a small subset of the complete sample. Vertical trend in the left panel highlights a subsample of galaxies for which the grism redshifts catch the true (spectroscopic) redshift, while the photometric redshifts exhibit a fair amount of scatter. Minimal correlated residuals between photometric and grism redshifts suggest that grism redshifts provide an independent measurement of an object's true redshift.}
\label{fig:zall_zall_diff}
\end{figure*}

Photometric redshift errors increase with both $H_{F140W}$ magnitude and redshift in a comparison with either spectroscopic or grism redshifts for the spectroscopic sample, as found by \citet{dahlen:13}.  We find very little correlation between redshift accuracy and stellar mass, and a non-monotonic but clear trend of decreasing scatter for the reddest colors.  Although it is tempting to interpret trends in redshift accuracy shown in Figure \ref{fig:scatter_all}, we caution that these are based on a heterogeneous (and biased) spectroscopic sample. In particular, the outlier fraction increases dramatically with redshift at $z>1.5$. However, this is also where the size of the spectroscopic sample dwindles. One takeaway is that for this sample of galaxies for which spectroscopic redshifts are obtainable (and perhaps easy), the grism redshifts are excellent (NMAD scatter is low), but the outlier fraction is similar to that of the photometric redshifts. The spectroscopic subsample is too small to disentangle trends in both redshift and mass; for this we must utilize grism redshifts for a larger sample.

\subsection{How Correlated are Grism and Photometric Redshifts?}\label{sect:corr}

The 3D-HST grism redshift fits are made using a joint fit to the photometric catalogs and grism spectra; the resulting redshift estimates may be correlated with purely photometric redshifts. In this Section we assess the magnitude of this correlation and therefore the utility of grism redshifts as an independent estimate of \emph{true redshift}. For this test, we include the full sample with spectroscopic and grism redshifts, investigating the residuals between the spectroscopic, or true, redshift of a galaxy and its photometric and grism redshift. 

Figure \ref{fig:zall_zall_diff} shows the residuals with respect to spectroscopic redshift in photometric versus grism redshift.  In each panel, the large symbols indicate galaxies for which the grism redshift P(z) is tightened with respect to that of the photometric redshift (68\% confidence interval of $z_{gris}$ is narrower than that of $z_{phot}$ by a factor of 0.5), the small symbols show the remainder of the sample. This criterion does not severely impact the demographics of galaxies with grism redshifts (dashed orange lines in Figure \ref{fig:dist}), but does minimize the effect of correlated residuals. In this Section we aim to quantify the effects of correlated errors on this sample; in subsequent sections we will use only grism redshifts to test photometric redshifts. Dotted lines indicate our adopted outlier threshold ($|\Delta z|/(1+z) = 0.1$).

Figure \ref{fig:zall_zall_diff}(a) shows the residuals with linear scaling. An interesting feature of the left panel is the vertical trend, indicating a subsample of galaxies for which the grism identifies the spectroscopic redshift, but the photometric redshifts exhibit higher residuals. Only a small fraction of galaxies lie along the diagonal trend, on which photometric and grism redshifts exhibit strongly correlated residuals, particularly for the subset of galaxies with ``tightened'' PDFs. Only $\sim$3\% of galaxies are outliers in both photometric and grism redshifts for the total sample, however $\sim70\%$ of photometric outliers are also outliers in grism redshifts.  For galaxies with tightened PDFs, this correlated outlier rate is lower at $\sim2\%$. For this sample, $\sim50\%$ photometric outliers are also grism outliers. The true rate of correlated redshift failure could be even lower. From visual inspections of images, SEDs, and grism spectra, we find that 33\% of the galaxies with tightened PDFs and correlated residuals have possible neighbors that could contribute to $z_{spec}$ misidentification and 42\% of the grism spectra include an identified emission line, suggesting that the grism redshift is the true redshift.

The Figure \ref{fig:zall_zall_diff}(b) shows the absolute value of the photometric and grism redshift residuals. In logarithmic scaling, galaxies preferentially lie above the diagonal line, indicating that grism redshifts have smaller residuals than photometric redshifts. The scatter is higher for the photometric residuals ($\sim0.037$) than grism redshifts ( $\sim0.0145$), when correlated residuals ($>0.05$ in both) are excluded. The cut in grism redshift uncertainty eliminates a large fraction of high residual and correlated objects in this projection. Only a small fraction (2\% of tightened sample) of all galaxies lie on the diagonal trend of correlated errors. Therefore, scatter and outlier fractions between photometric and grism redshifts will be dominated by the independent accuracy of each redshift estimate, but will not be artificially reduced by correlated errors. 

\subsection{Beyond Spec-zs: Trends in Photometric Redshift Accuracy with Mass, Magnitude, Color, \& Redshift}
\begin{figure*}[!t]
\centering
\includegraphics[width=\textwidth]{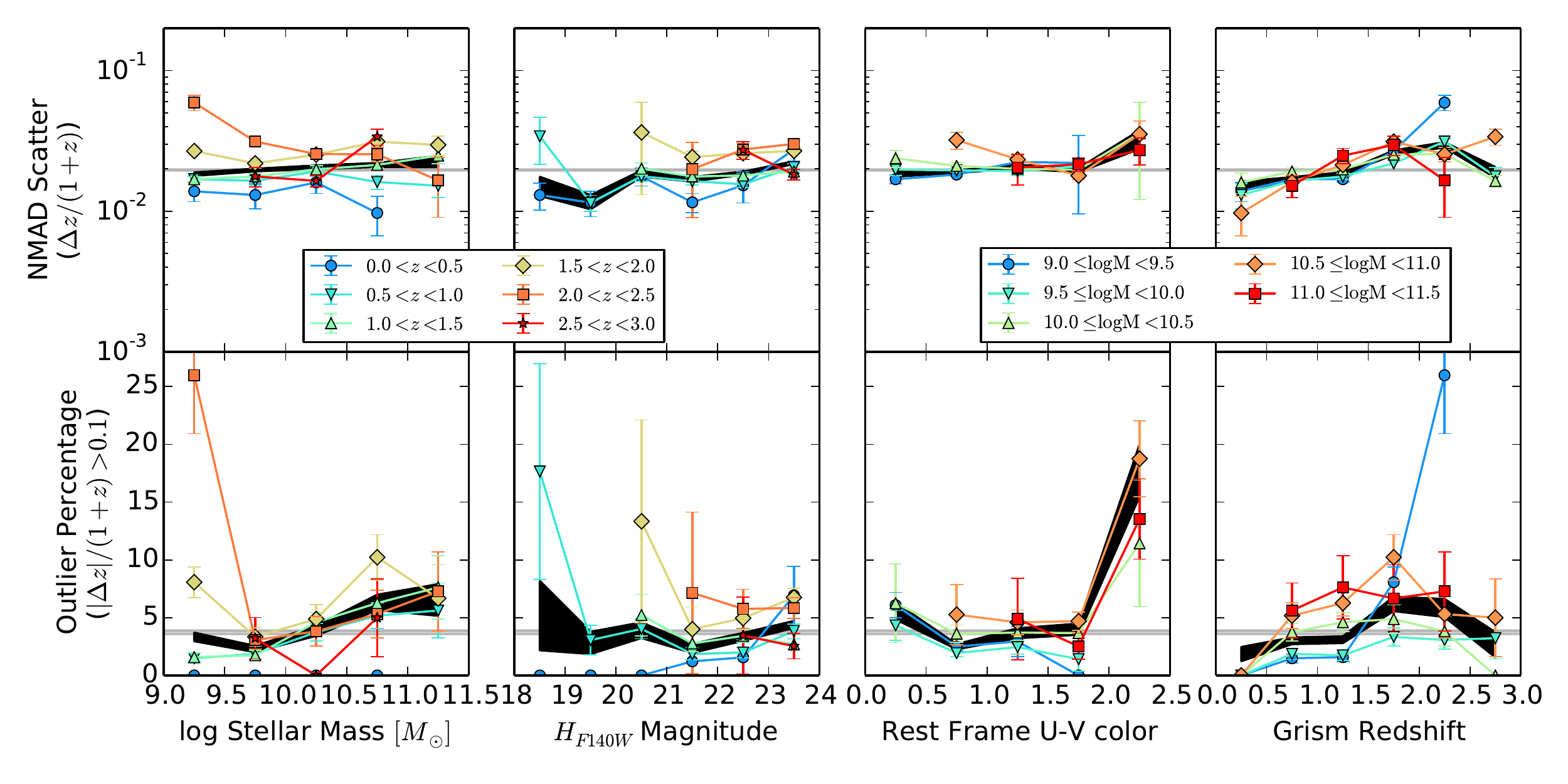}
\caption{NMAD scatter (top row) and outlier percentage (bottom row) for all grism redshifts with narrower P(z)s than for the photo-z (by a factor of $\leq0.5$) split by stellar mass, $H_{F140W}$ magnitude, U-V color, and redshift are indicated by filled black bands. Mean values are indicated by gray bands. Samples are further split by either redshift (left two panels) or stellar mass (right two panels). Photometric redshift accuracy depends primarily on magnitude and redshift, with non-monotonic variations as a function of galaxy mass or color.}
\label{fig:zphot_zgris_scatter_tight}
\end{figure*}

\begin{figure*}[!t]
\centering
\includegraphics[width=\textwidth]{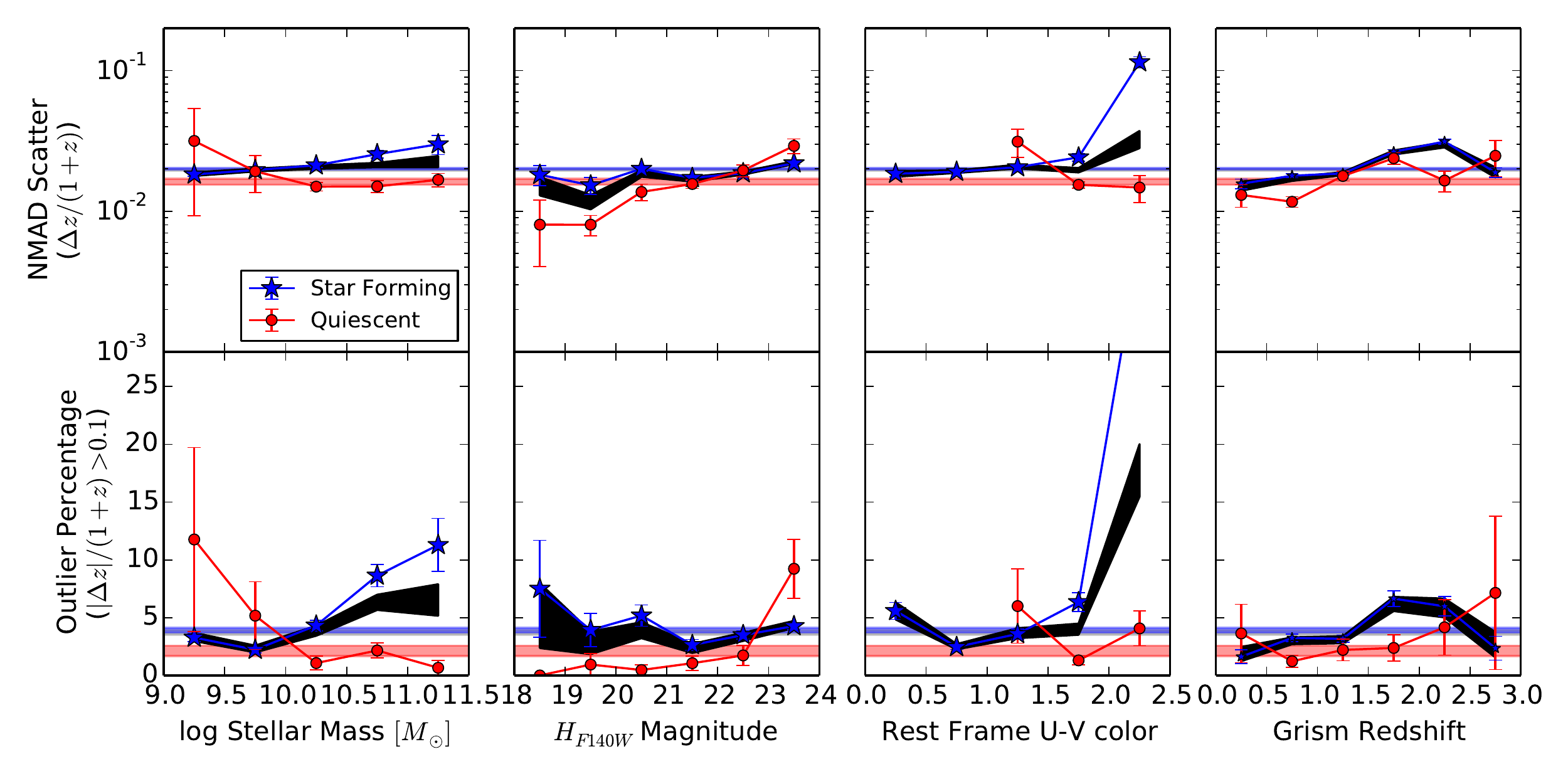}
\caption{NMAD scatter (top row) and outlier percentage (bottom row) compared to grism redshifts (with narrower P(z)s than for the photo-z by a factor of $\leq0.5$) split by stellar mass, $H_{F140W}$ magnitude, U-V color, and redshift are indicated by filled black band. The sample is split into star-forming (blue stars) and quiescent (red circles) galaxies based on their UV and VJ colors. Mean values are indicated by blue and red bands. Quiescent galaxies have more accurate photometric redshifts than star-forming galaxies, however this accuracy is strongly dependent on magnitude and redshift.}
\label{fig:zphot_zgris_scatter_tight_sfq}
\end{figure*}

\subsubsection{Testing Photometric Redshifts with Grism Redshifts}

\begin{deluxetable}{llccc}
\tabletypesize{\scriptsize}
\tablecaption{Scatter and Outlier Fraction in Spectroscopic Sample}
\tablehead{\colhead{S1} & \colhead{S2} &\colhead{$\sigma_{NMAD}$} & \colhead{Outlier \%}}
\startdata
\sidehead{\textbf{Full Spectroscopic Sample\tablenotemark{a}}}

Phot & Spec & $0.0158\pm0.0005$  & $4.6\%\pm0.4$ \\
Gris & Spec & $0.0038\pm0.0001$ & $4.2\%\pm0.4$ \\
Phot & Gris & $0.0156\pm0.0005$ & $2.1\%\pm0.3$ \\

\sidehead{\textbf{Spectroscopic Sample without possible mis-IDs\tablenotemark{b}}}
Phot & Spec & $0.0148\pm0.0006$  & $4.2\%\pm0.5$ \\
Gris & Spec & $0.0042\pm0.0002$  & $3.3\%\pm0.4$ \\
Phot & Gris & $0.0148\pm0.0007$ & $1.8\%\pm0.3$ \\

\sidehead{\textbf{Grism Sample with Narrowed PDFs\tablenotemark{c}}}
Phot & Gris & $0.0197\pm0.0003$  & $3.7\%\pm0.2$ \\

\sidehead{\textbf{Grism Sample with Narrowed PDFs (Star-Forming)\tablenotemark{c}}}
Phot & Gris & $0.0201\pm0.0003$ &  $3.9\%\pm0.2$ \\
\sidehead{\textbf{Grism Sample with Narrowed PDFs (Quiescent)\tablenotemark{c}}}
Phot & Gris & $0.0162\pm0.0008$  &$2.1\%\pm0.5$ \\

\enddata
\tablecomments{Average scatter and outlier fraction between photometric, grism, and spectroscopic redshifts in the 3D-HST survey.}

\tablenotetext{a}{Sample selection: z\_spec$>0$, use\_phot$=1$, use\_zgrism$=1$}
\tablenotetext{b}{Sample selection: z\_spec$>0$, use\_phot$=1$, use\_zgrism$=1$, no neighbors within $3''$ for which z\_spec falls within 95\% confidence interval for z\_phot.}
\tablenotetext{c}{Sample selection: use\_phot$=1$, use\_zgrism$=1$, 68\% confidence interval for z\_gris less than or equal to half that of z\_phot.}
\label{tbl:scatter}
\end{deluxetable}

We have demonstrated that 3D-HST grism redshifts can be used to provide a measurement of $z_{true}$ and assess photometric redshift quality, improving upon the severe biases inherent with using spectroscopic redshifts. In this Section we utilize the full sample of grism redshifts to investigate the variation in photometric redshift performance. For this test, we include all galaxies with good photometry and grism redshifts (use\_phot = 1, use\_zgrism =1) and narrowed PDFs (as defined in the previous Section).  The uniformity and size of the sample of galaxies with grism redshifts, as opposed to a spectroscopic sample (see Figures \ref{fig:dist} and \ref{fig:dist_sfq}), allows us to dissect trends photometric redshift accuracy in mass, apparent magnitude, galaxy color, and redshift. 

Figure \ref{fig:zphot_zgris_scatter_tight} demonstrates trends in NMAD scatter (top row) and outlier fraction (bottom row) as a function of stellar mass and magnitude in the $H_{F140W}$ imaging (first and second columns: split into redshift ranges) and redshift and U-V color (third and fourth columns: split by stellar mass). The average scatter and fraction are indicated by the gray band in each panel. On average, the scatter between $z_{phot}$ and $z_{gris}$ is slightly higher than that of the spectroscopic sample ($\sigma_{NMAD}=0.0197\pm0.0003$ versus $\sigma_{NMAD}=0.0148\pm0.0006$ for the $z_{spec}$ comparison). There are certain mass and redshift ranges for which the outlier fraction increases dramatically, but part of this seems to be driven by uncertain grism redshifts or small subsample size. 

The NMAD scatter does not depend strongly on stellar mass or UV color, with the minor exception of galaxy populations such as extremely red high redshift galaxies that are likely to be ill-fit (lower left panel). However, by using this unique dataset, it is apparent that the fraction of photometric redshifts that will catastrophically fail in estimating the true redshift of a galaxy depends strongly on the properties of and distance to the galaxy. For example, the outlier fraction for low mass galaxies is extremely low ($\lesssim5\%$) at low redshift ($z<1.5$) and for those with blue colors, but increases by a factor of $\sim2-3$ at higher redshifts. The outlier fraction of massive galaxies ($\log (M_{\star}/M_{\odot}) > 10.5$) is a factor of $\sim2$ higher than average at all but the highest and lowest redshift bins. 

We note that increased scatter or outlier fractions in this sample could indicate regimes in which either photometric or grism redshifts, or both, are less accurate. For example low-mass galaxies at $z\sim2$ exhibit large outlier fractions, even though the $\sigma_{NMAD}$ is less dependent on these properties. 

Another key question is how photometric redshift performance depends on galaxy type. Directly testing this is uniquely possible with the 3D-HST dataset. Figure \ref{fig:zphot_zgris_scatter_tight_sfq} includes the same trends in scatter and outlier fraction, but now indicates the trends for $U-V$ and $V-J$ identified star-forming (blue stars) and quiescent (red circles) galaxies. Overall, photometric redshifts are more accurate for quiescent galaxies than star-forming galaxies in scatter and outlier fractions, as indicated by the red and blue bands. This can be readily understood because quiescent galaxies have stronger Balmer/4000$\AA$ breaks which are easily identified in broad or medium band photometry. Above $z\sim2.5$, the Lyman break for star-forming galaxies begins to fall into the optical photometric bands and improves photometric redshift accuracies \citep[see also e.g.][]{whitaker:11}.

Additionally, trends in these panels are extremely helpful in interpreting Figure \ref{fig:zphot_zgris_scatter_tight}. For example, although the scatter of the full sample does not depend on stellar mass, scatter increases to $\sim0.03(1+z)$ for star-forming galaxies above $M_*>10^{11}M_{\odot}$. Similarly, the increasing outlier fraction (up to $\sim10\%$) is due to star-forming galaxies alone; photometric redshift accuracy does not appear to depend on stellar mass for quiescent galaxies.  On the other hand, photometric redshift accuracy decreases more dramatically with magnitude for quiescent galaxies (rising from $\sigma_{NMAD}\sim0.008$ at $H_{F140W}\sim18$ to $\sim 0.03$ at the faint end versus star-forming galaxies, which exhibit $\sigma_{NMAD}\sim0.02$ at all magnitudes.  
\begin{figure*}[t]
\includegraphics[width=\textwidth]{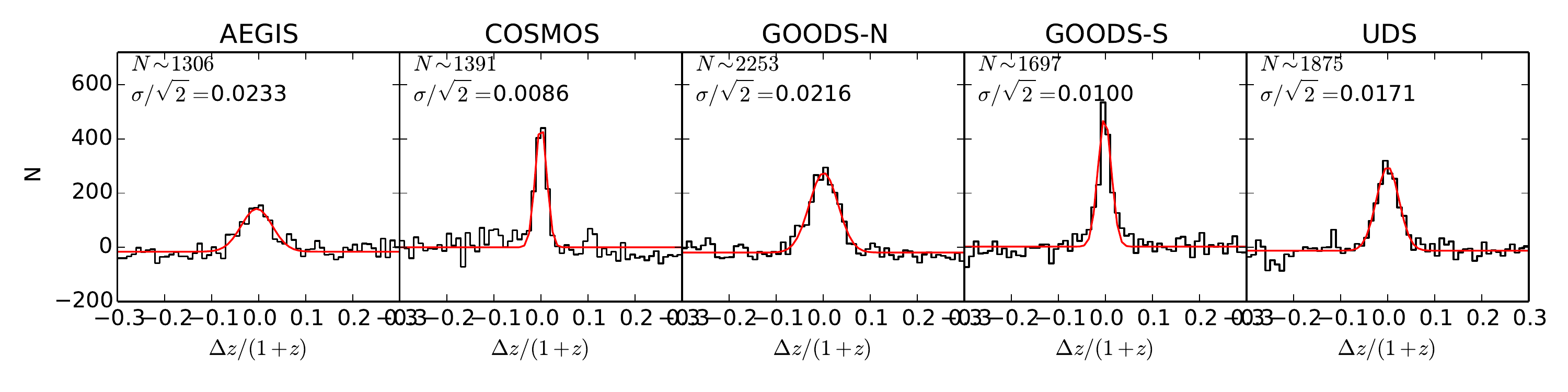}
\caption{Distribution of photometric redshift differences ($(z_1{-}z_2)/(1{+}z_{avg})$) for close pairs in each 3D-HST fields The characteristic 1$\sigma$ error in photometric redshift is approximated by $\sigma/\sqrt{2}$ of the distribution \citep{quadri:10}. The approximate number of pairs, calculated by integrating the gaussian fits, and measured photometric redshift errors are indicated in the upper left corner of each panel. Photometric redshift accuracy varies significantly amongst fields, at least in part by differing photometric coverage.}
\label{fig:pairs}
\end{figure*}

\begin{figure*}[]
\includegraphics[width=\textwidth]{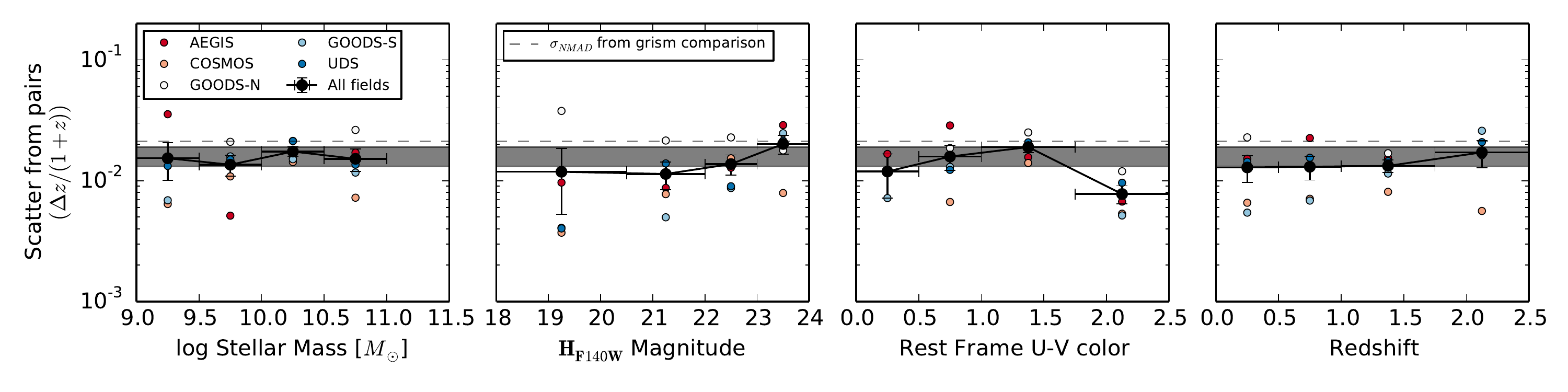}
\caption{Photometric redshift accuracy from close pairs analysis as a function of stellar mass (left panel), $H_{F140W}$ magnitude (second panel), rest-frame U-V color (third panel), and redshift (right panel). Average error derived from pairs analysis is indicated by gray shaded band and average NMAD scatter between photometric and grism redshifts by the gray dashed horizontal line. Redshift errors derived from close pairs are slightly lower than those derived from direct grism redshift comparisons, however trends of increasing errors with magnitude and redshift persist.}
\label{fig:pairs_trends}
\end{figure*}

Perhaps the most striking trend is with rest-frame color, where photometric redshift error and outlier fractions dramatically increase to $\sigma_{NMAD}=0.11$ and $\sim37\%$ outliers at the reddest $U-V$ colors. This trend was also apparent in Figure \ref{fig:zphot_zgris_scatter_tight}, but it is now apparent that only star-forming galaxies are contributing to the increase in scatter and outlier fraction. These galaxies must be extremely dusty to explain their red colors and they appear to have highly degenerate redshifts with the current template set (Brammer et al., in prep), despite the inclusion of the old, dusty template. It is noteworthy that this trend does not exist in the spectroscopic sample, highlighting the importance of the 3D-HST grism redshifts in fully characterizing photometric redshift performance. 

These red, dusty star-forming galaxies are an increasingly prevalent population at high redshift \citep[e.g.][]{marchesini:10, muzzin:13, marchesini:14}. Estimating the photometric redshifts for galaxies that are both red in U-V and V-J colors is helped by including an appropriate dusty starburst template \citep{marchesini:10}, but in general estimating their photometric redshifts becomes more difficult as the dust degrades the prominence of the break. Not accounting for this growing population of galaxies can systematically place them at the wrong photometric redshifts \citep{marchesini:10}, significantly influence the observed evolution of the stellar mass function for star-forming galaxies \citep{muzzin:13}, and underestimate star formation rates \citep[e.g.][]{fumagalli:14}. However, the \citet{skelton3dhst} photometric redshift fits already include a dusty and old template in the \texttt{EAZY} template set. In this case, the scatter and outlier fraction of the reddest galaxies points to a subset of extremely red star-forming galaxies for which photometric redshifts still consistently fail.

\subsubsection{Photometric Redshift Accuracy from Close Pairs}

The analysis in the previous subsection depended on the use of 3D-HST grism redshifts to estimate $z_{true}$ for galaxies in the photometric catalogs. We perform an independent test of the photometric redshift accuracy by following the close pairs analysis described by \citet{quadri:10}. Due to the clustered nature of galaxies throughout space, galaxies which appear very near to one another projected on the sky are likely to be physically associated. If both galaxies are at the same true redshift, then differences in measured redshift will be due to the measurement and template/fitting errors. Although true physical pairs cannot be individually identified, \citet{quadri:10} defined a statistical method to utilize the distribution of these redshift differences, subtract out the contribution of chance alignment to the sample of close pairs in a survey, and calculate the photometric redshift errors. Figure \ref{fig:pairs} shows the redshift distributions of close pairs, after subtracting the contribution of random superpositions, for the entire photometric sample (use\_phot=1) in each of the 3D-HST fields (in the $F140W$ footprint) down to the approximate magnitude limit of the grism redshift sample ($H_{F140W}\leq24$). Pairs are selected within 2-30 arcseconds of separation, with the lower limit to avoid erroneous correlations due to blending in the IRAC images. Errors on the photometric redshifts of the pairs in the sample are related to a gaussian fit to the distribution as:

\begin{equation}
\sigma (phot) = \sigma \left(\frac{z1-z2}{1+z_{avg}}\right)\frac{1}{\sqrt{2}}
\end{equation}

As with the grism redshift comparisons, measured photometric redshift uncertainty exhibits clear field-to-field variation (see Section \ref{sect:filters} for a more detailed discussion of inhomogeneous photometric data). We calculate the overall scatter as the average of all five fields and utilize jackknife resampling to estimate errors given that the formal errors on each individual gaussian fit are negligible with respect to field-to-field variation. We calculate the photometric redshift errors following this method for galaxy pairs in bins of mass, magnitude, color and redshift, only including pairs for which both galaxies are included in the selection. Figure \ref{fig:pairs_trends} shows the measured photometric redshift errors as a function of stellar mass (left panel), $H_{F140W}$ magnitude (second panel), rest frame U-V color (third panel), and redshift (right panel). There is significant variation amongst fields (see \S \ref{sect:filters}), however we find reasonable agreement between the average error estimated by this methodology (gray bands) and the $\sigma_{NMAD}$ from direct comparison between grism and photometric redshifts (shown as gray dashed lines in each panel for comparison). For an evaluation of grism redshift accuracy via close pairs analysis, we refer the reader to \citet{momcheva3dhst}.

Again these estimates of photometric redshift errors do not exhibit strong trends with stellar mass or U-V color (as in Figure \ref{fig:pairs_trends} from grism redshift tests), although there is a clear decrease in scatter at the reddest colors. We note that galaxy pairs may sample the galaxy mass distribution differently than that of the full grism sample, which could lead to subtle differences in the estimates of redshift accuracies. On the other hand, the pairs-derived photometric errors depend slightly more strongly on $H_{F140W}$ magnitude. Part of this is due to a small number of bright ($<20.5$) pairs of galaxies, however the trend extends to faint magnitudes. This may in part be due to the fact that the grism redshift sample is less complete below $z\sim0.7$ (see Figure \ref{fig:dist}). Although we only use grism redshifts with tightened PDFs to test photometric redshift accuracy, we note that if residual correlations between photometric and grism redshifts depend on galaxy properties, this could alter trends in redshift performance. In this case, trends evaluated by the analysis of close pairs could be stronger than for the grism comparisons. We expect this correlation to be higher for galaxies without emission lines, particularly those with fainter continuum. This effect could contribute to the differences in scatter with magnitude. On the other hand, the photometric redshift errors estimated using close pairs could be artificially diminished by redshift ``attractors'' in photometric redshift space that artificially place galaxies at the same redshifts \citep{quadri:10}. 

\begin{figure}[!t]
\centering
\includegraphics[width=0.45\textwidth]{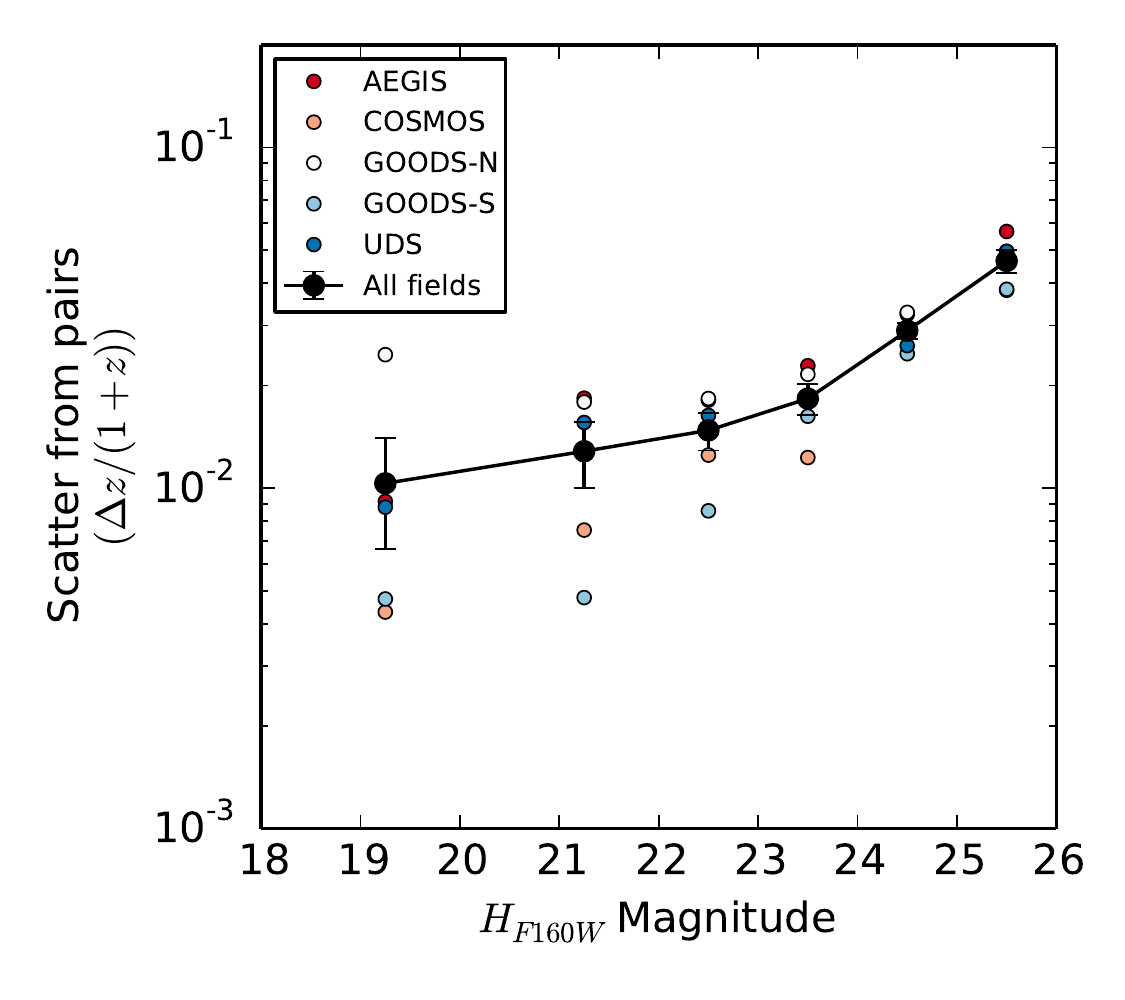}
\caption{Scatter as a function of magnitude as determined by pairs analysis of galaxies below the magnitude limits of the grism redshifts. Photometric redshift accuracy continues to diminish with redshift, with increased scatter of $0.046\pm0.005(1+z)$ for the faintest objects with $25\leq H_{F160W} <26$.}
\label{fig:pairs_mag}
\end{figure}

Finally, one benefit of photometric redshift accuracy based on statistical pairs analysis is that it is limited to the photometric depths, not those of the grism redshifts. In the case of the 3D-HST grism catalogs, redshift fits can be made to an arbitrary limit, however we expect that these will only be valuable for galaxies with emission lines at fainter magnitudes. Therefore, we have limited our analysis to galaxies with visually inspected redshift fits at $H_{F140W}<24$. We now extend the study of close pairs down to a fainter limiting magnitude ($H_{F160W}<26$), beyond the limits of the grism redshift estimates. Figure \ref{fig:pairs_mag} shows the measured scatter as a function of $H_{F160W}$ magnitude, which is slightly deeper than $H_{F140W}$. Indeed photometric redshift accuracy diminishes significantly beyond the limits of the grism redshifts, with scatter increasing by over a factor of two below $H_{F160W}=24$ and a factor of four across the whole magnitude range.

\section{The dependence of redshift accuracy on filters}\label{sect:filters}

All of the five extragalactic fields have extraordinary photometric coverage; the excellent photometric redshift accuracy ($\langle \sigma_{NMAD}\rangle \lesssim 0.02$) results from analysis of the fully sampled galaxy SEDs. In this Section, we investigate the importance of various categories of photometric data in determining photometric redshifts. For this analysis, we rerun \texttt{EAZY} for the 3D-HST photometric catalogs, including only specific subsets of filters. The \texttt{EAZY} code includes a redshift prior in the fitting, for which we use the K band magnitude in the default fitting. In cases where a $K$ or $Ks$ filter is not included in the subset, we use an $R$ band magnitude prior. Filter combinations for each field are included in Table \ref{tbl:filters}, including appropriate references. 

\begin{figure*}[]
\centering
\includegraphics[width=0.8\textwidth]{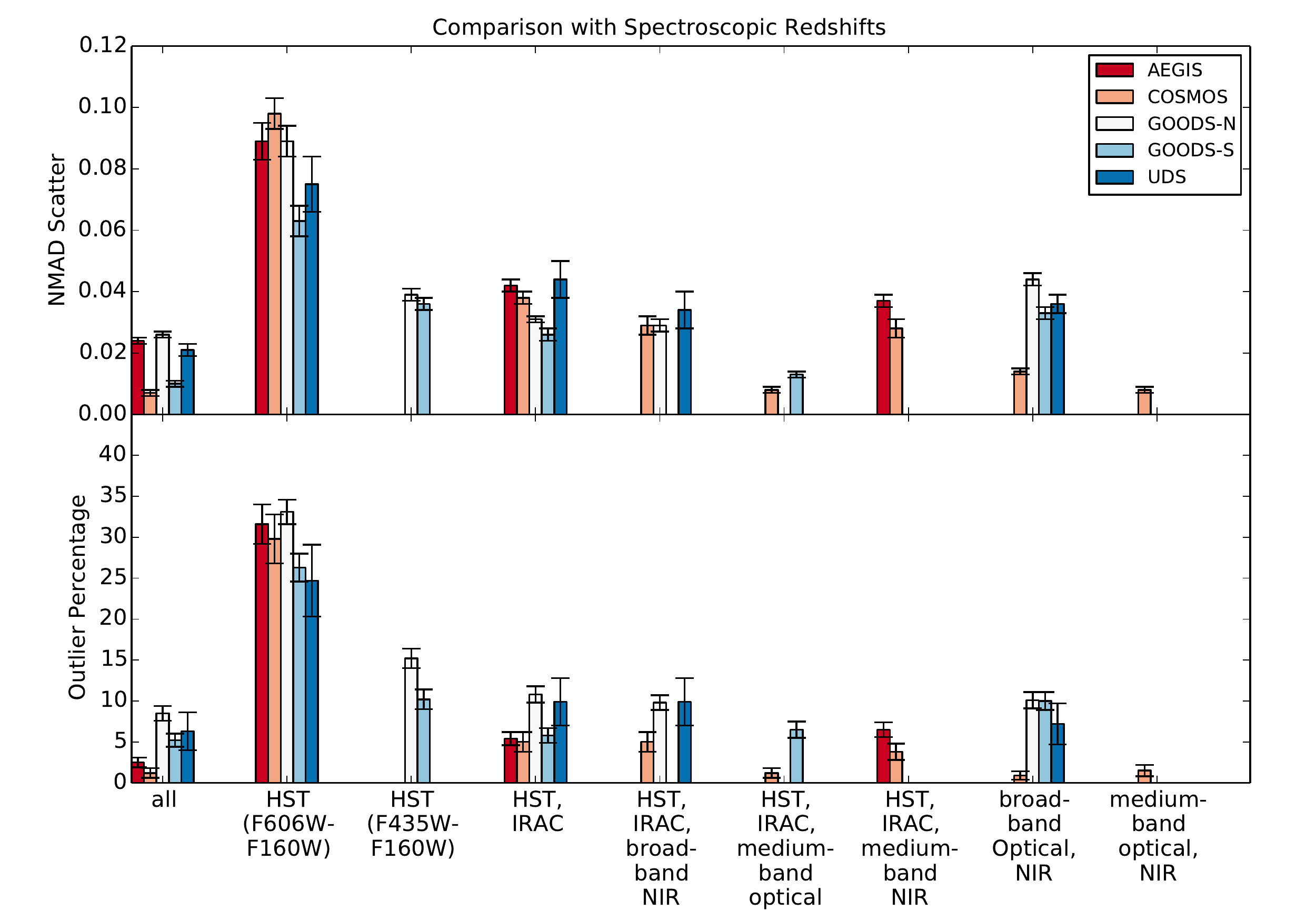}
\caption{Photometric redshift accuracy with different filter combinations compared to spectroscopic redshifts in each 3D-HST field. Photometric redshift accuracy depends strongly on the photometric bandpasses included in redshift fitting; at these magnitudes and redshifts blue optical imaging is crucial. Redshift accuracy varies strongly amongst fields, even when similar datasets are included. Some of this variation may be due to heterogeneous spectroscopic redshift samples across the fields.}
\label{fig:barplot_spec}
\end{figure*}

\begin{figure*}[]
\centering
\includegraphics[width=0.8\textwidth]{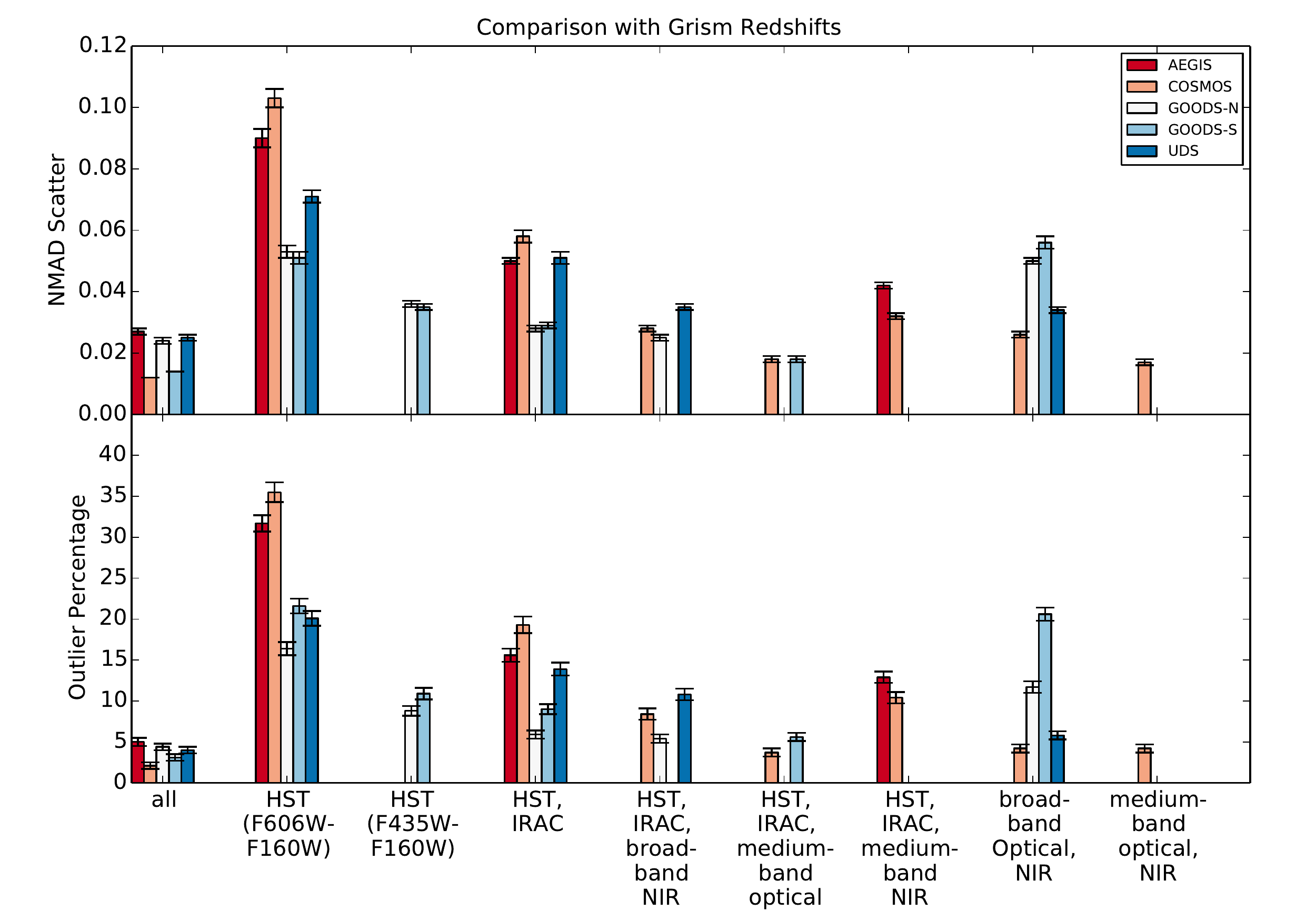}
\caption{Photometric redshift accuracy with different filter combinations compared to tightened grism redshifts. Again this figure demonstrates the strong bandpass dependence of photometric redshift accuracy, with uniformly distributed grism redshifts. When all photometry is included in the fit, there is less variation in redshift quality between fields than in spectroscopic comparison (Figure \ref{fig:barplot_spec}), but the comparison sample is insufficient to explain all field-to-field variation.}
\label{fig:barplot_gris}
\end{figure*}

\begin{deluxetable*}{cccll}[h] \label{tbl:filters}
\tabletypesize{\scriptsize}
\tablecaption{3D-HST/CANDELS Field Filter Subsets}
\tablehead{
\colhead{Field} & \colhead{Subset} & \colhead{Descriptive Label} & \colhead{Filters} & References}
\startdata
AEGIS & HST & HST(F606W-F160W) & F606W, F814W, F125W, F140W, F160W  &  1, 2, 3 \\
& IRAC & IRAC & 3.6, 4.5, 5.8, 8.0 $\mu\,m$ & 4, 5 \\
& NMBS & medium-band NIR & J1, J2, J3, H1, H2, K & 6 \\
& CFHTLS & broad-band optical & u, g, r, i, z & 7, 8 \\
\\
COSMOS & HST & HST(F606W-F160W) & F606W, F814W, F125W, F140W, F160W &  1, 2, 3 \\
& IRAC & IRAC & 3.6, 4.5, 5.8, 8.0 $\mu\,m$ & 4, 9 \\
& UltraVISTA & broad-band NIR & Y, J, H, K & 10 \\
& NMBS & medium-band NIR & J1, J2, J3, H1, H2, K & 6 \\
& CFHTLS & broad-band optical & u, g, r, i, z& 7, 8 \\
& Subaru & broad and medium-band optical & \multirow{2}{2.6in}{B, V, r', i', z', IA427, IA464, IA484, IA505, IA527, IA624, IA679, IA709, IA738, IA767, IA827} & 11  \\
\\
\\
GOODS-N & HST & HST(F435W-F160W) & \multirow{2}{2.5in}{F435W, F606W, F775W, F850LP, F125W, F140W, F160W}  & 1, 2, 3, 12 \\
\\
& IRAC & IRAC & 3.6, 4.5, 5.8, 8.0 $\mu\,m$ & 4, 13 \\
& HDFN & broad-band optical & U, B, V,  $\mathrm{r_c}$, $\mathrm{i_c}$, z' & 14 \\
& MODS & broad-band NIR & J, H, Ks & 15 \\ 
\\
GOODS-S & HST & HST(F435W-F160W) & \multirow{2}{2.6in}{F435W, F606W, F775W, F814W, F850W, F125W, F140W, F160W} & 1, 2, 3, 12 \\
\\
& IRAC & IRAC & 3.6, 4.5, 5.8, 8.0 $\mu\,m$ & 4, 13  \\
& GaBoDs & broad-band optical & U38, B, V, $\mathrm{R_c}$, I & 16, 17 \\
& MUSYC & medium-band optical & \multirow{2}{2.6 in}{IA427, IA445, IA505, IA527, IA550, IA574, IA598, IA624, IA651, IA679, IA738, IA767, IA797, IA856} & 18 \\
\\
& FIREWORKS & broad-band NIR & J, H, Ks & 19, 20 \\ 
\\
UDS & HST & HST(F606W-F160W) & F606W, F814W, F125W, F140W, F160W & 1, 2, 3 \\
& IRAC & IRAC & 3.6, 4.5, 5.8, 8.0 $\mu\,m$ & 4, 21\\
& SXDS & broad-band optical & B, V, $\mathrm{R_c}$, i', z' & 22 \\
& UKIDSS & broad-band NIR & J, H, Ks & 23 \\ 
\label{tbl:filters}
\tablerefs{
(1) \citet{candels}, (2) \citet{candelsb}, (3) \citet{3dhst}, (4) \citet{ashby:13}, (5) \citet{barmby:08},  (6) \citet{whitaker:11}, (7) \citet{erben:09}, (8) \citet{hildebrandt:09},
(9) \citet{sanders:07}, (10) \citet{mccracken:12}, (11) \citet{taniguchi:07}, (12) \citet{giavalisco:04}, (13) \citet{dickinson:03}, (14) \citet{capak:04}, (15) \citet{kajisawa:11},
(16) \citet{hildebrandt:06}, (17) \citet{erben:05}, (18) \citet{cardamone:10}, (19) \citet{wuyts:08}, (20) \citet{retzlaff:10}, (21) J. Dunlop et al. in prep., (22) \citet{furusawa:08},
(23) O. Almaini et al. in prep.
}
\end{deluxetable*}

For this test, we compare derived photometric redshifts with the spectroscopic and grism redshift measurements of $z_{true}$. Calculated scatter and outlier fractions between photometric and spectroscopic and photometric and grism redshifts are included in Table \ref{tbl:photoz_filters}. In the latter comparison, we again only include grism redshifts with tightened PDFs to minimize the effects of correlated errors on the measured scatter. Full comparisons are included in Appendix A. Although there are still variations in the details of photometry in each field, we classify subsets of photometry into the following categories: (1) all filters, (2) HST imaging (F606W-F160W), (3) HST imaging (F435W-F160W), (4) HST and IRAC, (5) HST, IRAC, and broad-band, ground-based near-IR imaging, (6) HST, IRAC, and medium-band optical imaging, (7) HST, IRAC, and medium-band near-IR imaging,  (8) broad-band, ground-based optical and near-IR imaging, (9) medium-band, ground-based optical and near-IR imaging. However, we note that the data included in each category will still vary in specific filter sets, photometric depths, and data quality. Scatter and outlier fractions between photometric and spectroscopic redshifts in each of these categories and fields is included in Figure \ref{fig:barplot_spec} and for photometric and grism redshifts in Figure \ref{fig:barplot_gris}.

The first thing to notice in Figure \ref{fig:barplot_spec} is that the scatter between photometric and spectroscopic redshifts varies significantly from field to field. This is due to a number of different factors, including heterogeneity in both the available photometry and spectroscopic followup in addition to cosmic variance. Overall scatter is lowest in the COSMOS field ($\sigma_{NMAD}=0.008$) and highest in GOODS-N ($\sigma_{NMAD}=0.027$).  This may in part be due to the optical and near-IR medium band photometry in the COSMOS field. Scatter is also low in GOODS-S, where photometry includes medium-band filters in the optical from the MUSYC survey, whereas the scatter in AEGIS is somewhat higher ($\sigma_{NMAD}=0.023$) even though the NMBS near-IR medium band filters are included. The relative importance of optical medium-band filters is partially due to the redshift distribution of the spectroscopic comparison sample, most of which are at low redshift. Using the grism redshifts, we can overcome this bias and assess the importance of filters in setting the photometric redshift accuracy for a more representative sample. 

Comparing to grism redshifts has the effect of normalizing variable spectroscopic redshift quality and quantity (Figure \ref{fig:barplot_gris}). In this case, the scatter and outlier fractions are somewhat more uniform across the five fields when all filters are included. In fields where HST ACS F435W imaging is not available (AEGIS, COSMOS, UDS), the scatter and outlier fraction of photometric redshifts are significantly higher when only HST imaging is fit. The inclusion of the blue filter in GOODS-N and GOODS-S introduces a decrease in scatter when only HST imaging is used to estimate photometric redshifts, although the effect is weaker than in the spectroscopic comparison. This emphasizes the importance of including blue wavelengths to identify spectral features such as the Lyman Break and the stellar bump.  Improvements in identifying these features can be made by including deep optical imaging, in particular medium-band imaging. Another striking improvement in the photometric redshift accuracy is gained with the addition of Spitzer IRAC imaging, in some cases decreasing the scatter by over a factor of two.

It is interesting to note that the photometric redshift accuracy is uniformly worse when compared to the grism redshift sample than relative to the spectroscopic redshift sample, especially in the outlier fractions. This is likely because the grism redshifts probe regions of parameter space where photometric redshifts are harder to measure: fainter galaxies, higher redshifts, and different galaxy types. Even with HST imaging and Spitzer IRAC photometry, the outlier fractions range from $\sim6\%$ to $\sim19\%$ without the bluer HST imaging. Systematics in these redshift measurements are examined in individual fields in the Appendix. This highlights the utility of the 3D-HST sample, but also emphasizes the importance of assessing the breadth of any spectroscopic sample used to evaluate photometric redshift performance.

We emphasize that collapsing the redshift accuracy into these three measures disguises systematics introduced by different filter combinations. For example, low redshift ($z\lesssim0.5-1$) galaxies are particularly driving the increased scatter in photometric redshift when only HST imaging ($F616W$-$F160W$) is included. We included detailed figures including the redshift scatter in each individual field in Appendix \ref{sect:fields}.
\begin{figure*}[!t]
\centering
\includegraphics[width=0.9\textwidth]{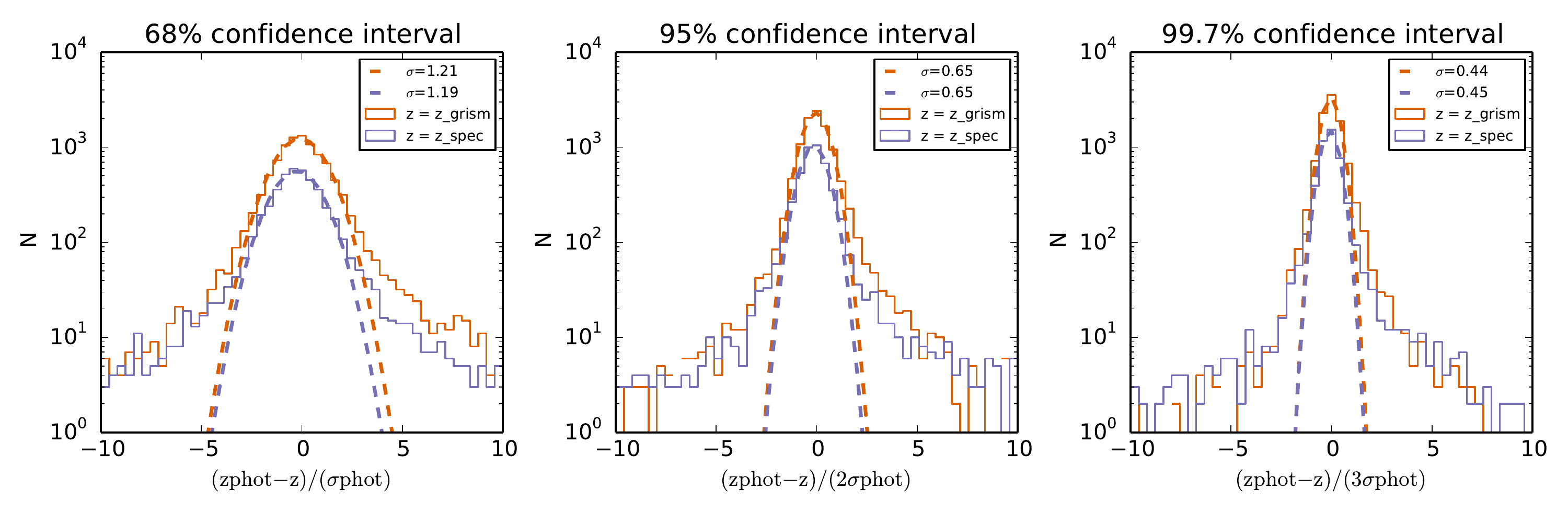}
\caption{Redshift deviations normalized by 68\%, 95\%, and 99,7\% confidence intervals for photometric redshifts ($(\mathrm{zphot}-\mathrm{ztrue})/(\mathrm{\sigma phot})$ in the left panel, $(\mathrm{zphot}-\mathrm{ztrue})/(\mathrm{2\sigma phot})$ in the center panel and $(\mathrm{zphot}-\mathrm{ztrue})/(\mathrm{3\sigma phot})$ in the right panel). Comparisons between photometric redshifts and spectroscopic redshifts (purple) or narrowed grism redshifts (orange) yield similar distribution shapes. Both exhibit roughly gaussian distributions (fits are indicated with dashed lines) but $\sigma\sim 1.2,0.6,0.4$ suggesting that the redshift PDFs are narrower than the observed scatter in redshift by a factor of $\sim1.2$.}
\label{fig:conf_int}
\end{figure*}

\begin{figure*}
\centering
\includegraphics[width=\textwidth]{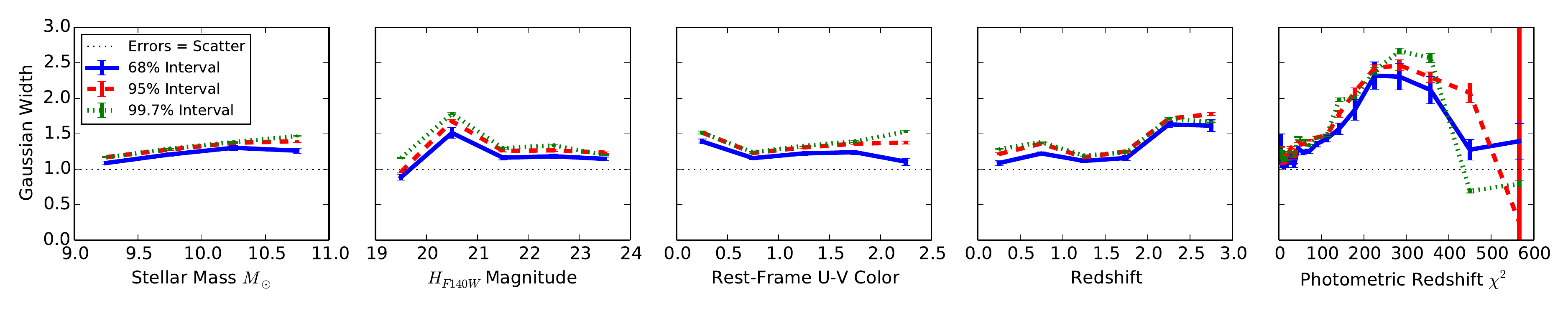}
\caption{Underestimate factor for photometric redshift errors as a function of galaxy properties: stellar mass, apparent magnitude, U-V color, redshift, and photo-z $\chi^2$ as measured by fitting gaussians to the scatter in redshift deviations from grism redshifts normalized by photometric error as in Figure \ref{fig:conf_int}. The dotted black line at unity indicates the value at which errors explain the observed redshift scatter. Solid lines indicate scatter normalized by 1$\sigma$ error, which is always slightly lower than the $2\sigma$($3\sigma$) width multiplied by a factor of two(three), indicating that the PDFs are too narrow to explain the photometric redshift scatter, particularly in the tails of the distribution. }
\label{fig:conf_int_trends}
\end{figure*}

\section{Photometric Redshift Probability Distribution Functions}\label{sect:pdfs}

In addition to fitting a single-valued photometric redshift estimate, the \texttt{EAZY} code produces individual probability distribution functions.  These PDFs provide an estimate of the likelihood that the galaxy lies at a given true redshift. Until this point, we have adopted the redshift with the maximum likelihood ($z_{peak}$) as the photometric redshift for each galaxy. For certain applications, we would like to incorporate the uncertainly on photometric redshift and ideally utilize the entire PDF function to describe the probability that the galaxy lies at a given $z_{true}$. This technique has been proven to significantly improve measurement uncertainties, for example it can increase the S/N of clustering measurements by a factor equivalent to an increase in survey size of $\sim2-3$ \citep{myers:09}. In this Section, we investigate the ability of the \texttt{EAZY}-generated PDFs to predict the $z_{true}$ values for the ensemble of 3D-HST galaxies.

\subsection{Photometric Redshift Confidence Intervals}

A key question regarding photometric redshift performance is whether the scatter between measured and true redshifts is primarily driven by uncertainties in the photometric redshift estimates. For this, we test the redshift deviations for individual galaxies relative to their estimated confidence intervals. Figure \ref{fig:conf_int} we show the distribution of deviations ($z_{phot}-z_{true}$) normalized by the 68\% (left panel), 95\% (center panel), and 99.7\% photometric redshift confidence intervals. Each panel includes two samples: the orange histogram indicates the distribution for a comparison with independent grism redshifts (with narrowed PDFs) and a purple histogram for the spectroscopic sample. Although the normalization between the two samples and confidence intervals differs, each is characterized by a gaussian central peak in addition to broader wings. Best-fit gaussians are fit to each distribution and standard deviations are indicated in the legend of each panel. These gaussian distributions contain roughly 90\% of galaxies in each panel.

For fully representative photometric redshift errors, we would expect a gaussian of width $\sigma=1.0$ for the 68\% confidence interval and $\sigma=1/2,1/3$ when redshift deviations are normalized by the $2,3\sigma$ errors. Like \citet{dahlen:13}, we find that the photometric redshift PDFs are too narrow in each confidence interval. However, the factor by which the PDFs would need to be broadened differs for each test. At the 68\% confidence level, the photometric PDFs are a factor of $\sim1.2$ too narrow, whereas the tails of the PDFs are further underestimated, requiring a factor of $\sim1.6$ to explain the observed scatter between photometric and true redshift.

To complicate the situation, this discrepancy is not uniform amongst galaxy types. Figure \ref{fig:conf_int_trends} further dissects the trends in uncertainty underestimation by galaxy stellar mass, apparent magnitude, rest-frame U-V color,  redshift, and $\chi^2$ from the photometric redshift fit using the sample of galaxies with narrowed grism PDFs.  As for the total sample, the underestimation of the 68\% confidence interval for photometric redshift errors is less than for the 95\% and 99.7\% confidence intervals. Furthermore, we see clear trends that this depends on galaxy properties. In contrast with the measured scatter in photometric redshifts with stellar mass, photometric redshift errors are decreasingly well calibrated with increasing mass. Aside from the brightest galaxies, which appear to have appropriate error estimates, the normalized scatter does not depend strongly on apparent magnitude or galaxy color. However, the uncertainties are underestimated by an increasing $\sim1.1$ at low redshifts to $\sim1.6$ at $z\sim2.5$. Finally, for very poorly fit photometric redshifts ($\chi^2\gtrsim100$), the scatter in redshifts is vastly under predicted by \texttt{EAZY} by up to a factor of $\sim2.5$. The right panel indicates strong correlation between photometric redshift scatter, as normalized by the redshift confidence intervals; at the largest $\chi^2$ values this normalization may include the entire allowed redshift range, driving gaussian widths back to 1.0.

\subsection{PDF width and quantifying catastrophic outliers}

In the previous Section, we demonstrated that the error estimates for the majority of galaxies are underestimated by approximately a factor of $1.2-1.3$ by looking at the distribution of scatter between photometric and grism redshifts. However, there are tails of the distribution of redshift scatter for which the errors cannot be described by a gaussian distribution. We now investigate the properties of these outliers. Following \citet{dahlen:13}, if the redshift error estimates are accurate for the entire population of galaxies, $\sim68\%$ of galaxies will have 68\% confidence intervals that include $z_{true}$ and likewise for the 95\% and 99.7\% confidence intervals. 

\begin{figure*}
\centering
\includegraphics[width=0.7\textwidth]{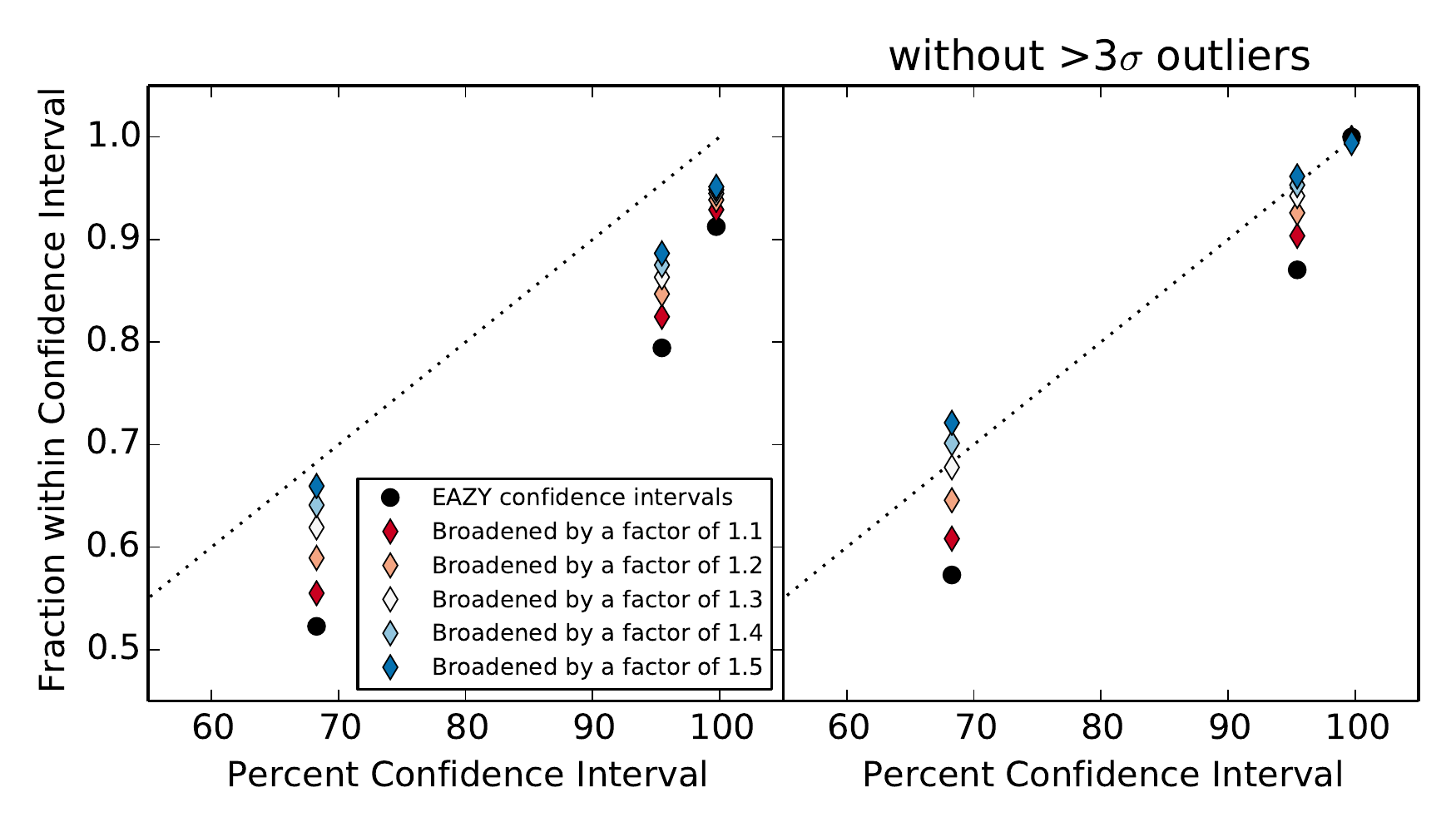}
\caption{Fraction of galaxies with $z_{grism}$ within photometric redshift confidence intervals. The full sample is included in the left panel, with only galaxies for which $z_{grism}$ falls within the $99.7\%$ confidence intervals. Overall $\sim10\%$ of galaxies will have grossly underestimated photometric redshift uncertainties, and confidence intervals are too narrow by a factor of $\sim1.2$.}
\label{fig:conf_fracs}
\end{figure*}

\begin{figure*}[!t]
\centering
\includegraphics[width=\textwidth]{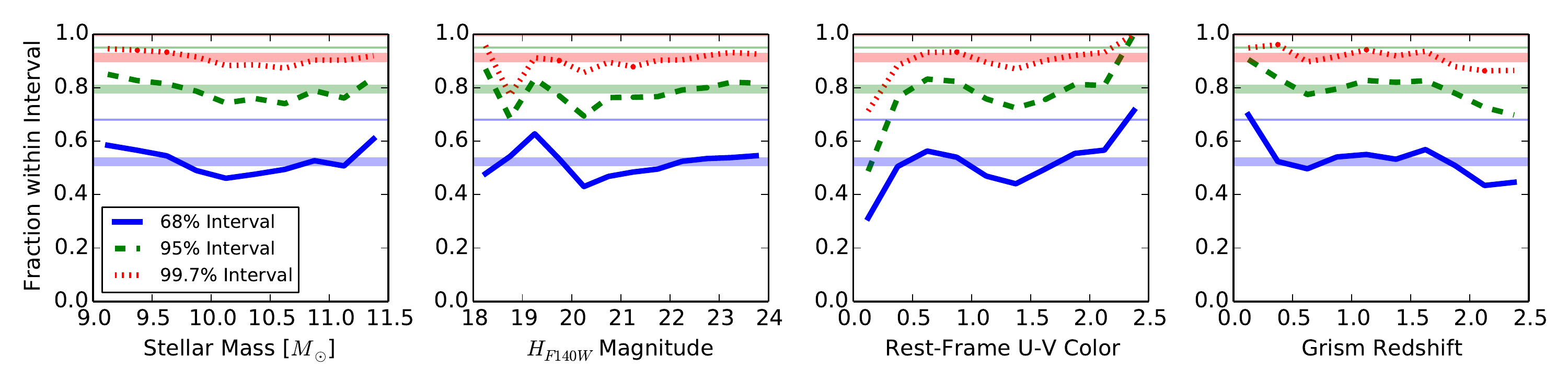}
\caption{Trends in the fraction of galaxies for which grism redshifts fall within photometric redshift confidence intervals (blue solid, green dashed, red dotted correspond to $1,2,3-\sigma$. Average values are indicated by thick horizontal lines and expected 68\%, 95\%, and 99.7\% values are indicated by thin colored lines.}
\label{fig:conf_fracs_trends}
\end{figure*}

Figure \ref{fig:conf_fracs} indicates the fraction of galaxies within the 68\%, 95\% and 99.7\% confidence intervals as black circles for the entire sample (left panel) and the sample for which $z_{grism}$ falls within the 99.7\% photometric redshift confidence interval (right panel). Colored diamonds demonstrate the fractions measured by artificially broadening the confidence intervals. One-to-one correspondence relations are included as black dotted lines. Clearly, the fraction of galaxies within a given confidence interval is well below the predicted value, partially due to underestimated errors. Even by extending the confidence intervals by a range of factors, there is always a fraction of galaxies for after cropping catastrophic outliers, defined such that $z_{grism}$ lies well outside the $3\sigma$ error estimates. These catastrophic fitting failures drive the overall fractions lower than can be explained by inflating error bars alone. In the right panel, we demonstrate that by excluding these outliers (approximately 10\% of galaxies) and then broadening the error estimates by a factor of $1.2-1.3$ found in the previous Section producing general agreement between the confidence intervals and redshift distributions of galaxies.

These fractions do not depend strongly on galaxy properties. In Figure \ref{fig:conf_fracs_trends} we show trends in these fractions as a function of stellar mass, $H_{F140W}$ magnitude, rest-frame U-V color, and grism redshift. Average values are indicated as solid, horizontal lines and trends with galaxy properties are shown in blue for the $1\sigma$ confidence interval, dashed green for $2\sigma$, and dotted red for $3\sigma$. In each case, roughly 10\% of galaxies lie outside of the $3\sigma$ confidence intervals. When using a purely photometric sample of galaxies, this will correspond to noise in galaxy counts that will not be accounted for by photometric redshift uncertainties.  This outlier rate is significantly higher than the outlier rates in $z_{phot}$ versus $z_{grism}$ comparisons but may also be important to include for studies that include photometric redshift error estimates. 

\subsection{How well do photometric PDFs predict true redshifts?}

\begin{figure*}
\centering
\includegraphics[width=\textwidth]{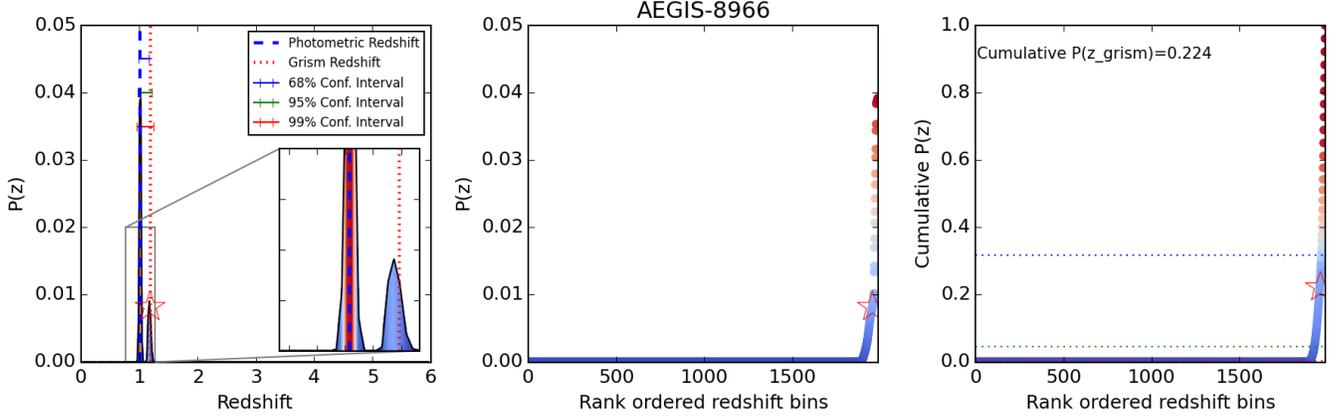}
\caption{Photometric PDF for sample galaxy AEGIS-8966. Photometric redshift is labeled with vertical, dashed blue line, grism redshift with red dotted line and red star and color-coding in each panel corresponds to the P(z) value. Note that the photometric redshift lies in the most likely peak of the PDF, but the true redshift lies on the second peak. Second and third panels include rank-order distribution of P(z) bins (and cumulative P(z)) with the location of $z_{grism}$ indicated again by the star.}
\label{fig:pdf}
\end{figure*}

In this Section we investigate the use of the full photometric redshift PDF as opposed to a single valued photometric redshift with errorbars. In particular, this could be important for galaxies that have multipeaked PDFs. We show an example galaxy from the catalog in Figure \ref{fig:pdf}. The $P(z)$ for the galaxy in redshift bins is included in the left panel, along labeled photometric redshift (blue) and grism redshift (red dotted line and star). Confidence intervals are indicated by blue (68\%), green (95\%), and red (99.7\%) errorbars. For this galaxy, the photometric redshift is assigned at the center of the most dominant peak of the PDF however the grism redshift reveals that the second peak is the location of the true redshift for this galaxy. In this specific case, the full PDF gives a clearer understanding of the uncertainties on the photometric redshift. Although the errorbars are somewhat broad, the actual redshift is well constrained between two narrower ranges.

\begin{figure*}
\includegraphics[width=\textwidth]{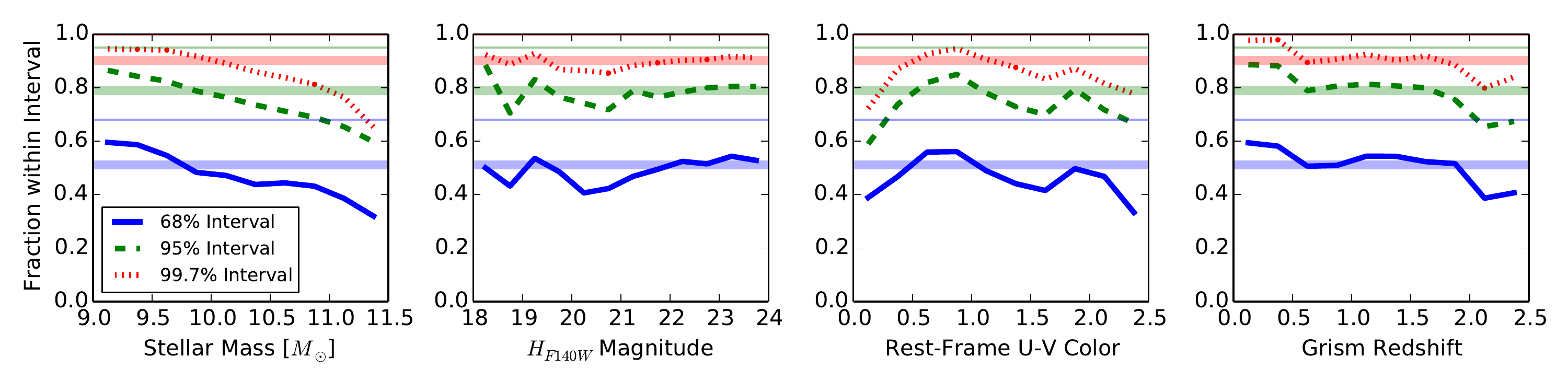}
\caption{Trends in the fraction of galaxies for which grism redshifts fall within photometric redshift confidence intervals as measured from rank-ordered photometric PDFs. Color coding and resulting average fractions are extremely similar to those measured from confidence intervals in Figure \ref{fig:conf_fracs_trends}, which suggests that confidence intervals estimated by \texttt{EAZY} from the CDFs are sufficient.}
\label{fig:conf_cum_fracs_trends}
\end{figure*}

To test the impact of such multipeaked PDFs, we rank redshift bins by P(z) in the PDF of every galaxy (as in the center panel of Figure \ref{fig:pdf}) and estimate the cumulative probability that corresponds to the redshift bin in which the grism redshift lies (right panel).  In this specific example, grism redshift lies on a second redshift peak, outside of the 68\% confidence interval; calculated in this way the $P(z\_grism)=0.224$. With this computed for each galaxy in the tightened grism sample, we repeat the test of enclosed fractions as a function of galaxy properties. Figure \ref{fig:conf_cum_fracs_trends} presents the fraction of galaxies for which $z_{grism}$ falls within the three defined confidence intervals. The average values for these fractions are remarkably similar to those derived via standard uncertainties.  One can imagine problematic PDF with widely-separated multiple peaks for which the confidence intervals defined by the \texttt{EAZY} code could overestimate the uncertainty, even though the errors on the whole are narrow with respect to the observed scatter. In practice, we don't find evidence that this is a dominant effect, likely because the subset of these significantly multipeaked PDFs is small and therefore the average fraction of true redshifts that fall within a given confidence interval is largely independent of the method used to determine confidence intervals. Although for the most part we find similar trends in fractions with galaxy properties as in Figure  \ref{fig:conf_fracs_trends}, we do find a fairly strong trend of decreasing fractions of correctly estimated errors within each interval with increasing stellar mass. This is consistent with the trend in uncertainty normalized scatter which increases with stellar mass, as shown in Figure \ref{fig:conf_int_trends}. 

\begin{figure}
\centering
\includegraphics[width=0.5\textwidth]{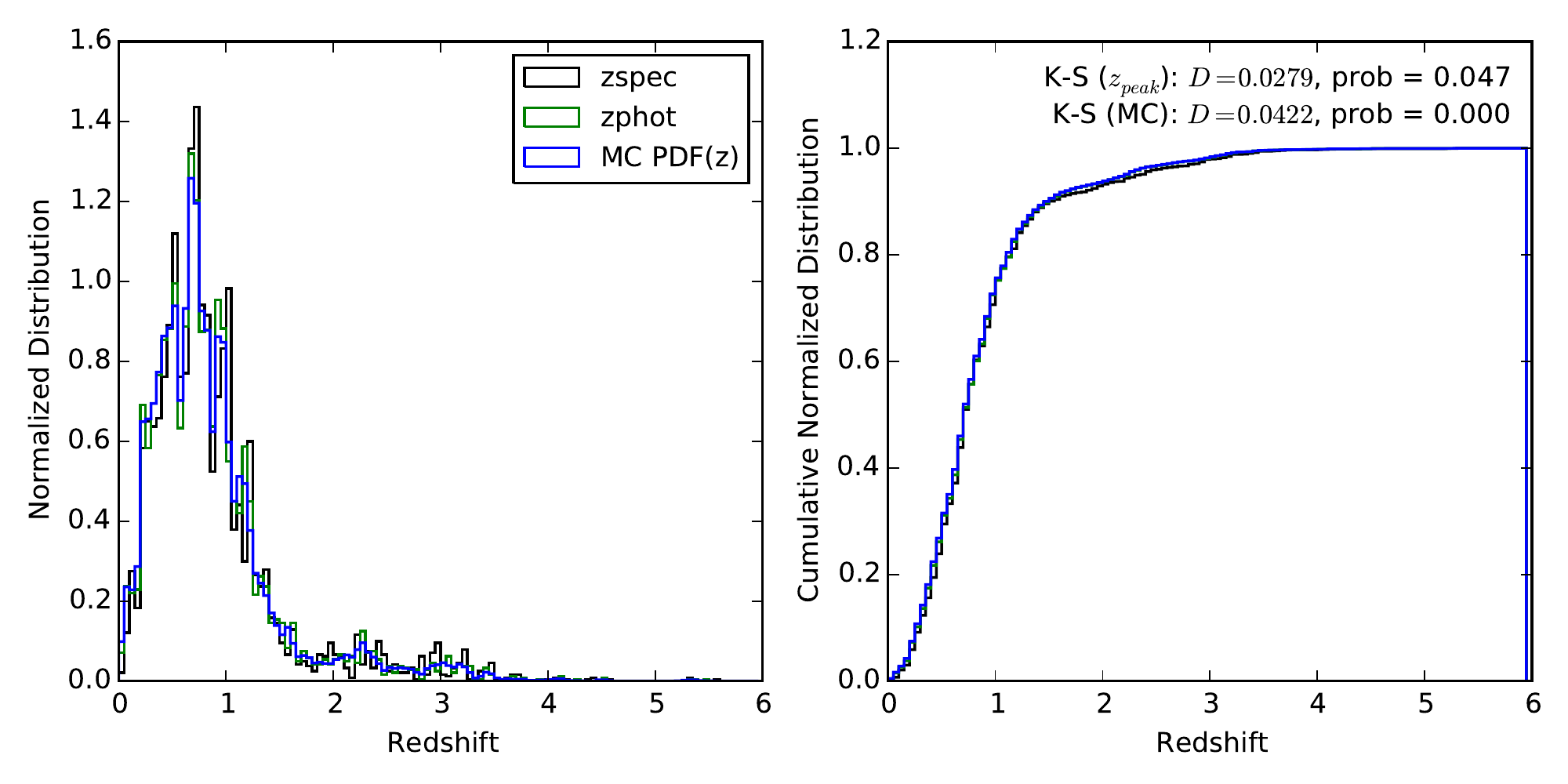}
\caption{Distribution and cumulative distribution of spectroscopic redshifts (black histograms), photometric redshifts (green), and Monte-Carlo sampling of photometric PDFs (blue). Full distributions agree extremely well and although a KS-test between the spectroscopic distribution and that of the photometric samples do not suggest that they are drawn from the same distribution, the distributions deviate by less than 3\% for the photometric redshifts and $\sim4\%$ for the MC sampled PDFs.}
\label{fig:KS_spec}
\end{figure}
\begin{figure}
\centering
\includegraphics[width=0.5\textwidth]{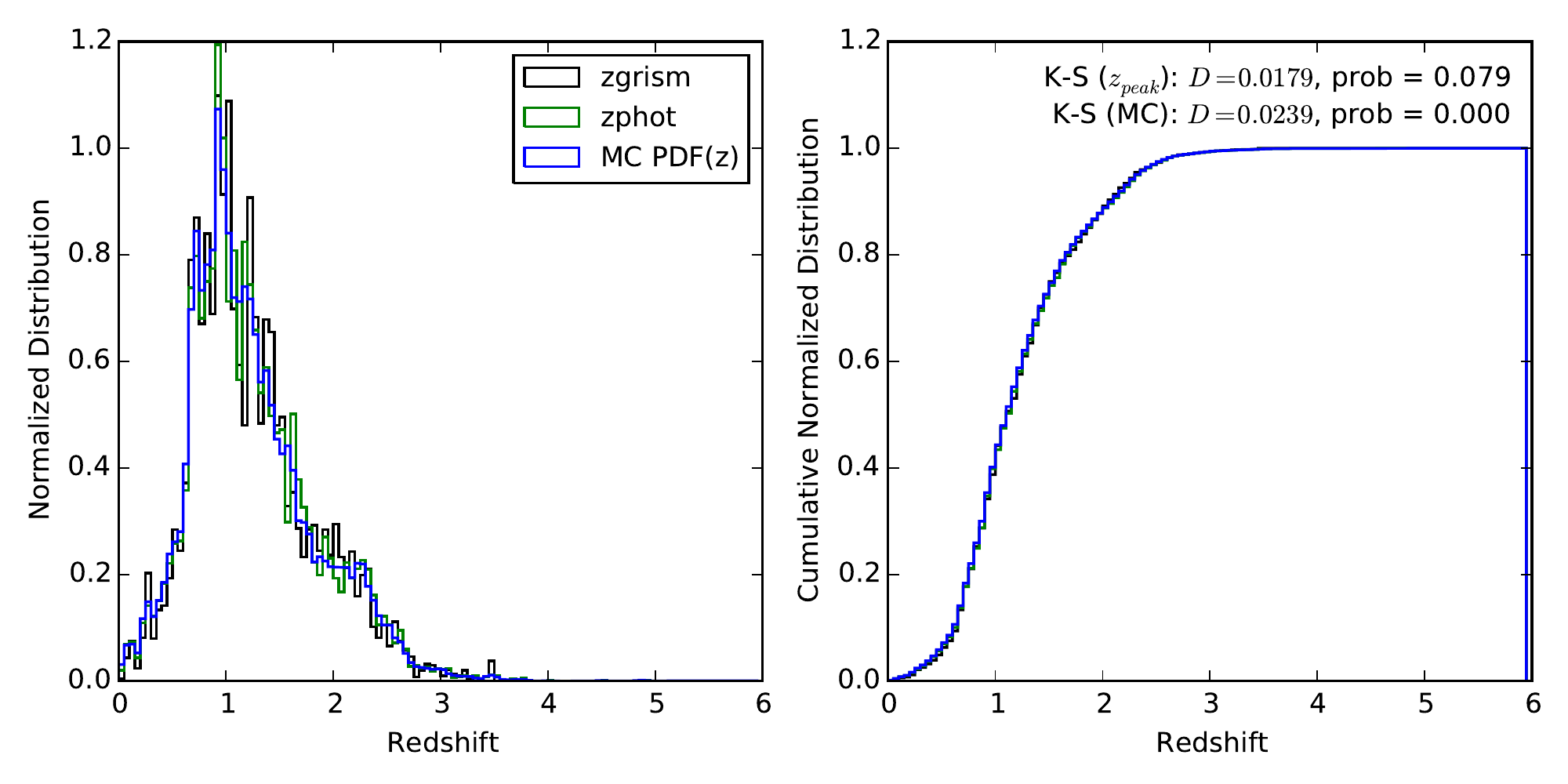}
\caption{Distribution and cumulative distribution of redshifts, as in Figure \ref{fig:KS_spec}, for comparison with grism redshifts. Again the cumulative distributions of grism and photometric redshifts agree to within $\sim2\%$ and MC sampled PDFs to within $\sim3\%$}
\label{fig:KS_gris}
\end{figure}

We now shift to a related issue, and investigate how well the photometric redshifts can recover redshift distributions of a sample of galaxies. We compare the overall redshift distribution (both of the spectroscopic sample and of the tightened grism sample) to the distribution of single-valued photometric redshifts and those derived by bootstrap resampling of the individual photometric PDFs. The redshift histogram and cumulative distributions for the spectroscopic sample is presented in Figure \ref{fig:KS_spec} and for the grism sample in Figure \ref{fig:KS_gris}. We emphasize that while we expect the photometric and true redshift distributions to be similar, measurement errors and catastrophic outliers will broaden the photometric redshift distributions. Therefore, these distributions should fail a simple Kolmogorov-Smirnov (K-S) test from a statistical standpoint, however the K-S statistic $D$, or the maximum distance between each cumulative distribution functions, is still an informative metric of the relative similarity of the two distributions.

\begin{figure*}
\centering
\includegraphics[width=\textwidth]{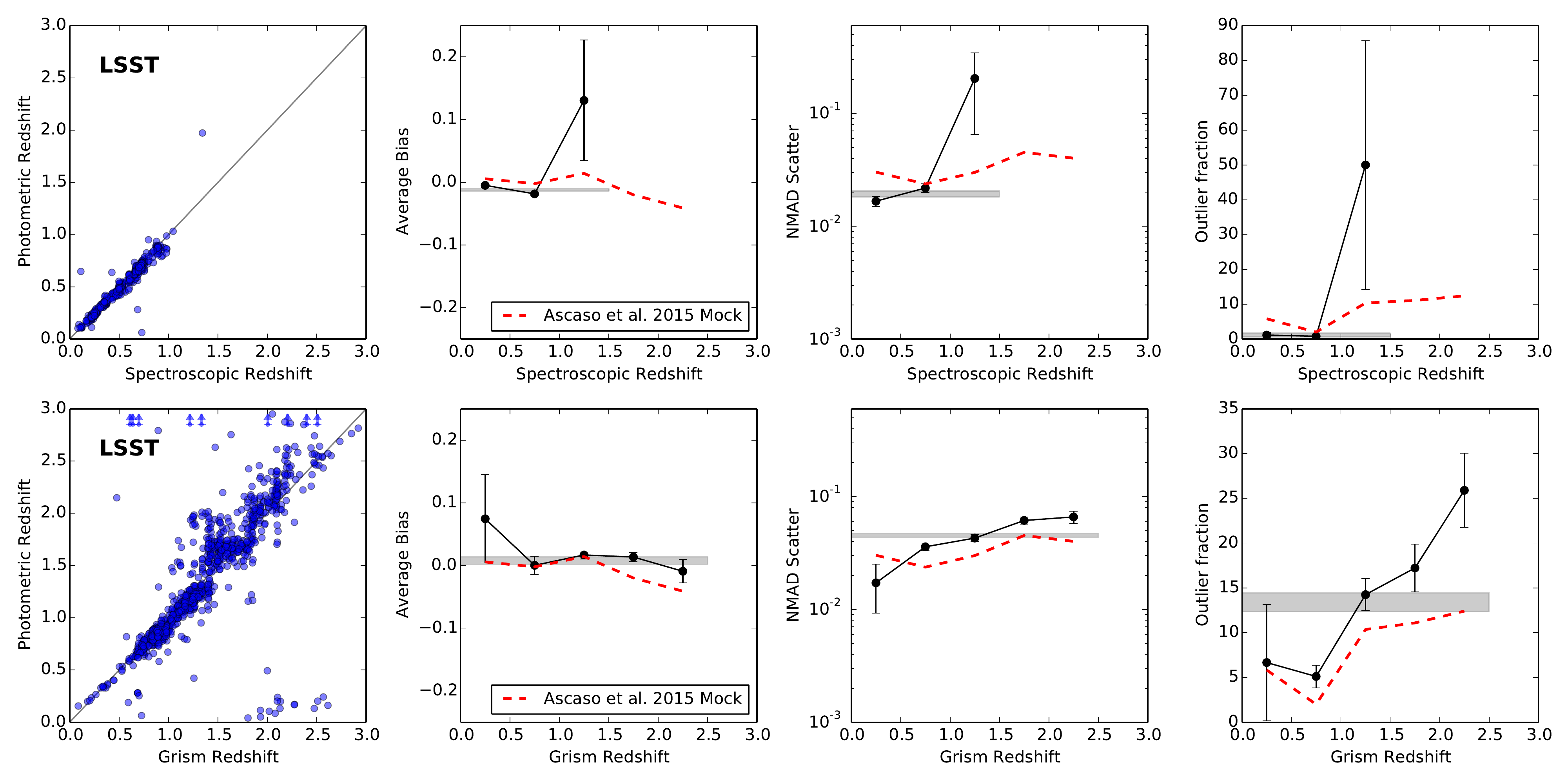}
\caption{Simulated photometric redshift performance for the LSST survey with $u,g,r,i,z,y$ filters created by adding noise to the 3D-HST photometric catalog in COSMOS. Scatter predicted by mock galaxy catalogs is indicated by a dotted red line \citep{ascaso:15}. Below $z\sim1$, the accuracy is quite good when compared to the spectroscopic redshift sample $\Delta z/(1+z)\sim0.02$. However, when the fainter and more representative grism redshift comparison sample is included, the scatter clearly depends strongly with redshift, increasing to $\sim0.04$ by $z\sim1.25$ and $0\sim0.07$ by $z\sim2.5$. Additionally, while the outlier fraction is excellent below $z\sim1$ for brighter spectroscopic redshifts, the percentage of $| \Delta z | > 0.1$ outliers increases from $\sim5\%$ to $\sim25\%$ at $z\sim2.5$.}
\label{fig:lsst}
\end{figure*}

In Figure \ref{fig:KS_spec}, the two-sided KS test between the spectroscopic and z\_peak redshift distributions indeed suggests a very low probability that the two samples are drawn from the same distribution. However we find excellent agreement between the two histograms, with maximum deviations of less than $3\%$. The distribution of Monte-Carlo sampled redshifts exhibits a larger deviation, but still agrees to within 4\% with the spectroscopic redshifts. We interpret this as due to the additional scatter to each galaxy due to the PDF resampling. The comparison with grism redshifts (Figure \ref{fig:KS_gris}) yields similar results: 2\% deviation for the comparison with z\_peak and 3\% deviations for the resampled redshifts.

The full redshift distributions are not statistically consistent in either case, however given their similarity there are benefits to using the full redshift PDF.  We note that although the distributions are more similar for the comparison with the single-valued $z_{peak}$ photometric redshift estimates, there are instances in which the full redshift PDF carries additional useful information. For example, in the case of a multi-peaked PDF the $z_{peak}$ carries only information about the most likely redshift, even when the secondary peak may be nearly as significant. Sampling the entire PDF will scatter the redshifts for all galaxies, hence the increased deviations, but for multi-peaked PDFs will capture all possible solutions, given the fitting methodologies and templates. We conclude that photometric redshifts can reproduce the true redshift distribution of galaxies in a sample to a few percent accuracy, but emphasize that because these deviations may depend on the galaxy properties the effects of using photometric redshifts should be carefully modeled.

\section{Photometric redshift accuracy in simulated surveys}\label{sect:sim_surveys}

In addition to characterizing the performance of photometric redshifts in the 3D-HST survey, this vast dataset can be used to predict or estimate the redshift accuracy in other surveys with similar photometric data. In this Section, we extend our analysis to predict the photometric redshift accuracy in three major planned datasets using the 3D-HST catalogs: the Large Synoptic Survey Telescope (LSST), the Dark Energy Survey (DES), and DES combined with the Vista Hemisphere Survey (VHS). Because both of these surveys are planned to include a Y-band filter, we limit this exercise to the COSMOS field where ground-based Y band imaging is included from the UltraVista Survey \citep{mccracken:12}. In each case we add noise to the 3D-HST catalogs to match the cited target depths of these surveys. Although the catalogs are based on real data, we refer to them as data simulations. We use the \texttt{EAZY} code to fit photometric redshifts to the resulting catalogs and analyze the photometric redshift accuracy for each simulated survey.

\subsection{LSST Survey}
The LSST survey is planned to image $>$10,000 $\mathrm{deg^2}$ in $u,g,r,i,z,y$ filters down to $26.1,27.4,27.5,26.8,26.1,24.9$ 5-$\sigma$ limiting magnitudes in the final coadded images \citep{ivezic:08}. The 3D-HST catalog in the COSMOS field includes imaging in each of these bands, however the $g, r$ and $z$ band imaging from the CFHTLS survey \citep{erben:05,hildebrandt:09} is slightly shallower than the proposed LSST depths. The differences are extremely small, less than $\sim0.3$ magnitudes in each case. We add noise to the other catalog fluxes ($u,i,Y$) to the planned depths prior to fitting photometric redshifts.
\begin{figure*}
\centering
\includegraphics[width=\textwidth]{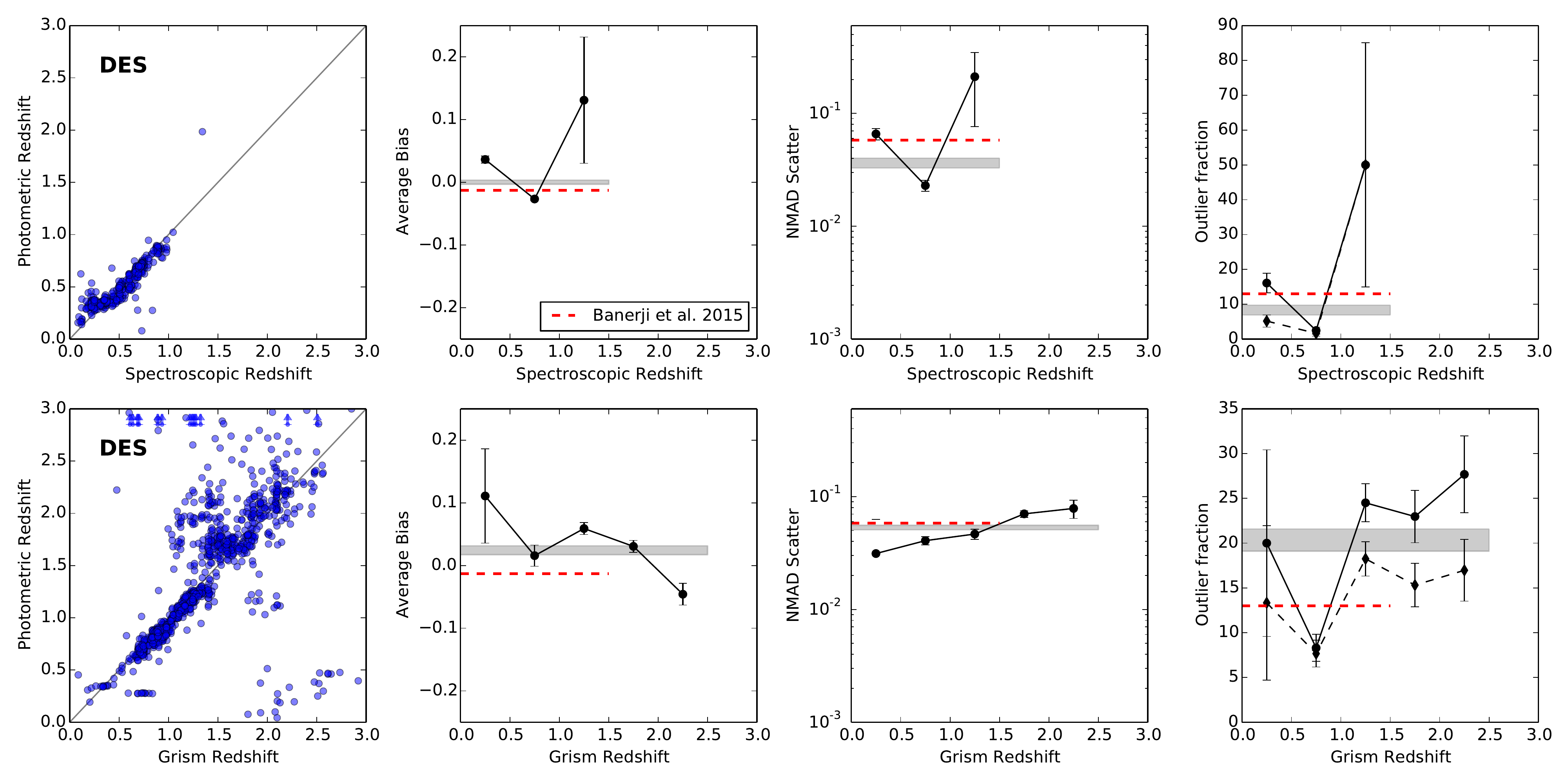}
\caption{Photometric redshift performance in the simulated DES survey ($g,r,i,z,y$ filters). Photometric redshift accuracy will depend strongly on redshift, with a significant outlier fraction ($\gtrsim20\%$) except at $0.5<z<1.0$. These systematics will be improved significantly by including near-IR photometry, e.g. from the VHS survey, as shown in Figure \ref{fig:des_vhs}. Red dashed lines indicate measured redshift performance with DES science verification data as estimated for a sample of bright objects with spectroscopic redshifts \citep{banerji:15}.}
\label{fig:des}
\end{figure*}

Results of the photometric redshift performance for the simulated LSST survey are shown in Figure \ref{fig:lsst}.  The top row includes a comparison of photometric and spectroscopic redshifts, the bottom row of photometric and grism redshifts.  Each row includes a $z_{phot}$ versus $z_{true}$ scatter plot, average bias $\langle \Delta z \rangle = (zphot-ztrue)/(1+ztrue)$, scatter, and outlier fraction as a function of redshift.  \citet{ascaso:15} also conducted a simulation of the LSST survey using mock redshift catalogs based on dark matter halos from the Millennium Simulation \citep{springel:05} and GALFORM semi-analytic models \citep{cole:00,bower:06}.  Average values from the current study are indicated by gray bands and predictions from \citet{ascaso:15} study are indicated by red dotted lines.

We find that a comparison with only spectroscopic redshifts yields a fairly optimistic view of $z<1$ photometric redshifts in the LSST survey, predicting $\sigma_{NMAD}\sim0.02$ and very few outliers. However, when the full sample of grism redshifts is included in the test, the measured scatter increases significantly, spanning from $\sim2\%$ at $z\sim0.25$ to $\sim7\%$ at $z\sim2$.  Additionally, the number of catastrophic outliers increases dramatically with redshift from 5\% to 25\%.  \citet{ascaso:15} performed similar tests using mock catalogs and found slightly more optimistic results, with NMAD scatter spanning $\sim0.03 - 0.04$ (red dashed lines in NMAD and outlier panels). Aside from the lowest redshift bin, where the COSMOS field is small and the grism adds little to the redshift determination, photometric redshift accuracy predicted from the mock catalogs is optimistic. Estimates from mock catalogs are lower by up to a factor of two compared to simulations leveraging real data.

\subsection{DES and VHS Surveys}
The DES survey is a photometric survey of 5000 $\mathrm{deg^2}$ of the southern sky that includes $grizY$ imaging using the Dark Energy Camera \citep{flaugher:05, diehl:12}. According to the survey description document \footnote{\texttt{http://www.darkenergysurvey.org/survey/des-description.pdf}}, the target 5-$\sigma$ limiting magnitudes for point sources in the DES survey will be $26.5,26.0,25.4,24.7,$ and $23.0$. As each of these limits is shallower than the imaging in COSMOS, we are able to accurately create a simulated catalog.  Additionally, the DES footprint overlaps with the VISTA Hemisphere Survey (VHS) which can complement the DES optical data with near-IR JHK imaging over $\sim20,000$ $\mathrm{deg^2}$ down to limiting magnitudes of $21.5,21.16,20.3$ \citep{mcmahon:13}.

\begin{figure*}
\centering
\includegraphics[width=\textwidth]{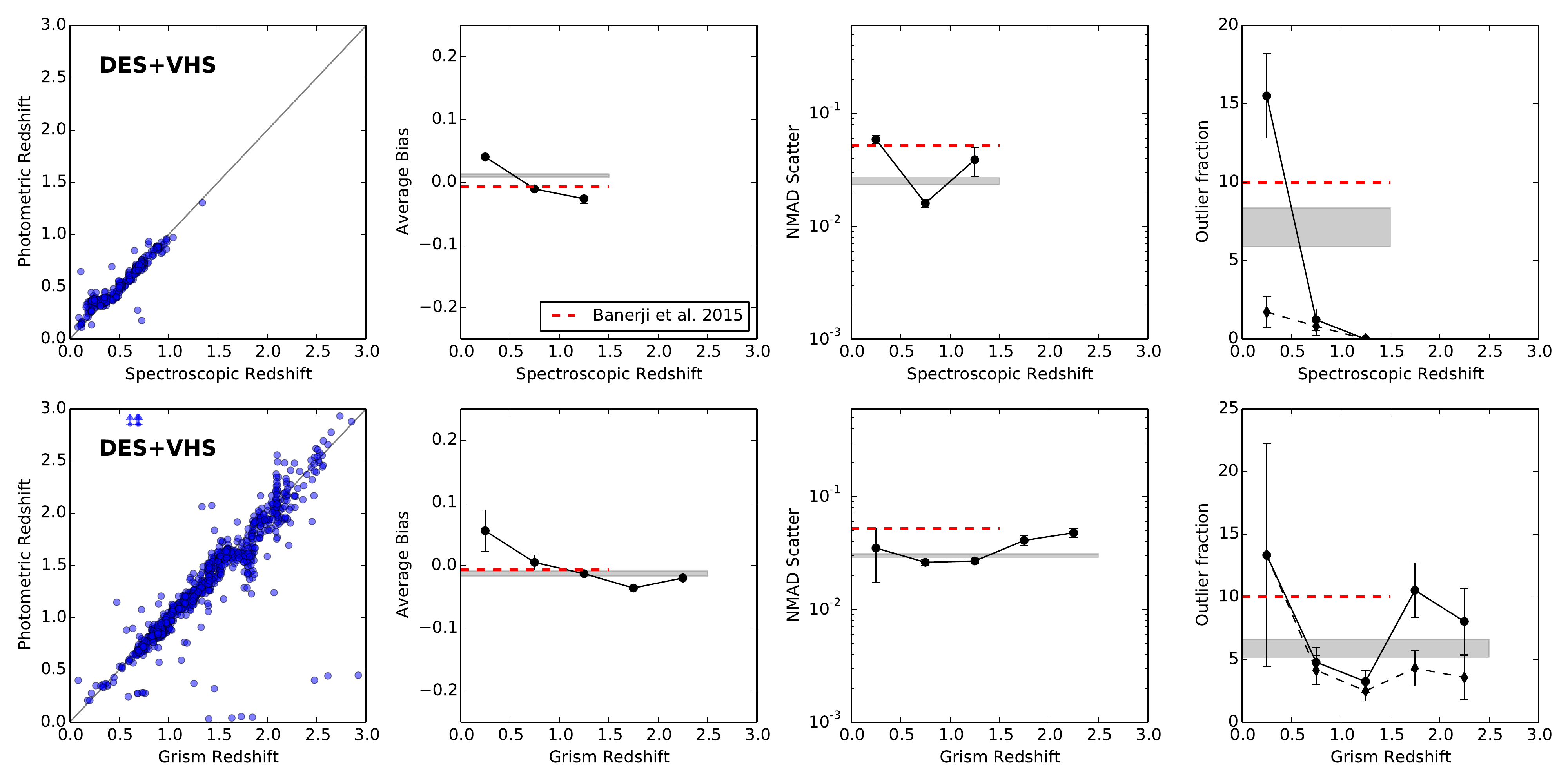}
\caption{Photometric redshift performance in the simulated combined DES and VHS surveys ($ugrizyJHKs$ filters) from spectroscopic and grism samples. Errors in photometric redshifts will be $\sim3\%$ on average, ranging up to $\sim5\%$ at $z\sim2.5$ with $\sim3-13\%$ outliers. Photometric redshift scatter and outlier fraction is lower in this dataset than in \citet{banerji:15}, suggesting the importance of including near-IR photometry from the VHS survey for photometric redshift performance.}
\label{fig:des_vhs}
\end{figure*}

Figure \ref{fig:des} demonstrates the photometric redshift performance for the spectroscopic and grism samples in the simulated DES survey and Figure \ref{fig:des_vhs} includes both DES and VHS filters.  For the DES filters alone, we find that the photometric redshift scatter will be higher than for the LSST ($\sim5\%$) and will increase to $\sim8\%$ by a redshift of 2.5.  This is partially due to the omission of the $u$ filter and shallower depths of the DES survey. Additionally, from the grism sample, we predict that the outlier fraction will be quite high with only DES imaging ($\sim20\%$).  These estimates are somewhat different from those found in a study by \citet{banerji:15} (red dashed line on Figures \ref{fig:des} and \ref{fig:des_vhs}), which found very similar scatter, but lower outlier fractions. We note that \citet{banerji:15} adopted a slightly different definition ($\Delta z/(1+z) > 0.15$). We include outliers defined in this way as a black dashed line in Figures \ref{fig:des} and \ref{fig:des_vhs} and find closer agreement between the two studies.

We find that the addition of near-IR photometry with the VHS specifications improves the photometric redshift performance dramatically, in contrast with the findings of \citet{banerji:15}. When these data are included, the mean NMAD scatter decreases to $\sim3\%$, with an increase to $\sim5\%$ above $z\gtrsim1.5$. In a comparison with the grism redshifts, which will be less biased than the spectroscopic catalogs used by \citet{banerji:15}, we find a lower outlier fraction of $\sim6\%$ versus 10\%. This average value decreases even further with the less strict outlier threshold ($\Delta z/(1+z) > 0.15$)

We emphasize that simulations like those presented in this Section are overly simplistic. The filters used in planned or on-going surveys may not exactly match those used in the 3D-HST catalogs. Furthermore, although we attempt to match quoted catalog depths, there will naturally be differences in image quality (e.g. seeing) and redshift fitting methodology that will influence photometric redshift performance. In particular, the COSMOS catalogs are HST-detected, therefore ground-based photometry will suffer more dramatically from blending. Furthermore, the redshift accuracy in the COSMOS field is excellent when only the broad-band optical and near-IR imaging is included (see e.g. Figure \ref{fig:barplot_gris}), therefore these estimates could further be a generous estimate of photometric redshift performance in the planned surveys. We emphasize the discrepancies between the mock and empirical predictions presented in this Section and suggest the importance of including empirical tests with representative spectroscopic samples in addition to mock simulations in order to robustly predict photometric redshift accuracy.

\section{Summary}\label{sect:summary}

Studies of the high redshift Universe are increasingly reliant on photometric redshifts to probe fainter targets in scope and variety than are inaccessible to even the most ambitious spectroscopic campaigns. The goal of this Paper is to assess and quantify the photometric redshift accuracy in the 3D-HST photometric catalogs. We summarize the major findings below:

\begin{itemize}
\item The 3D-HST photometric catalogs consist of PSF-matched aperture photometry across $\sim900$ square arcminutes in the CANDELS extragalactic field, including ground and spaced-based imaging from $0.3-8.0\mu m$. Overall, photometric redshift quality in the catalogs, calculated using \texttt{EAZY} \citep{eazy}, is excellent, with an overall characteristic scatter of $\Delta z/(1+z)\sim0.02$ down to $H=24$. This result is fairly robust to measurement technique, e.g. comparison with spectroscopic or grism redshifts versus galaxy pair counts, although it does vary amongst the five fields by $\pm0.006$.

\item The characteristic, or NMAD, scatter does not depend strongly on galaxy stellar mass or U-V rest-frame color, however we do find significant variations in the fraction of catastrophic outliers ($\Delta z/(1+z) > 0.1$). Photometric redshift scatter increases by $\sim1-2\%(1+z)$ with apparent magnitude (down to the limiting magnitude of $H_{F140W}=24$) and redshift (out to $z\sim2.5-3$). Analysis of close pairs suggests that redshift accuracy further degrades for fainter objects, reaching $\sim0.046(1+z)$ at $H_{F160W}=26$.

\item We confirm that the error estimates and PDFs produced by the \texttt{EAZY} code are narrow with respect to the photometric redshift scatter, but this underestimation cannot be improved by uniformly broadening the PDFs. Furthermore, errors in photometric redshift estimates do not capture the outlier behavior. However, the effect on the derived overall properties of a sample may be subtle; the overall spectroscopic/grism and photometric redshift distributions as probed by single valued estimates and full PDFs agree to within ${\sim}3-4\%$. In many specific cases, such as deriving luminosity or mass functions, scatter and outliers can tend to bias the derived properties. Although the size of this bias is not immediately calculable, it must be simulated for any given survey, magnitude limit, and redshift range.

\item Finally, a fraction of the field-to-field variation in photometric redshift quality can be attributed to the heterogenous nature of available imaging bands. We investigate the contribution of various filter combinations on the derived redshift accuracy, highlighting the dramatic impact driven by the inclusion of Spitzer-IRAC photometry, blue (F435W) HST photometry, and medium-band filters particularly in the optical. 

\end{itemize}

The conclusions from this paper extend far beyond the use of the 3D-HST catalogs and can be applied in the interpretation of current surveys for which grism spectroscopy is not available. Furthermore, the systematics in redshift accuracy can be used in the planning of future surveys. To illustrate this possibility, we included simulations of photometric redshift performance in the LSST, DES, and DES plus VHS datasets in \S \ref{sect:sim_surveys}. This type of empirical simulation could more realistically reflect the input galaxy population than a spectroscopic or mock catalog, which could yield overly optimistic estimates for redshift accuracy.

Additionally, the demonstrated filter-dependence can influence survey design choices. For example, the inclusion of blue (F435W) imaging in the GOODS fields significantly improved both the scatter and outlier fractions. One key question in planning photometric surveys is the balance between depth in broad filters and shallower imaging in narrower filters. The significant improvement in photometric redshift accuracy, especially in the outlier fraction, due to the inclusion of medium band imaging for the current sample, centered at $1<z<2$, can inform future studies of the earlier Universe. Similar medium band imaging in the Near-IR, as used by the Newfirm Medium Band Survey (NMBS) \citep{whitaker:11} and FourStar Galaxy Evolution Survey (ZFOURGE) survey (I. Labb\'{e} et al., in preparation) could be crucial for future studies at higher redshift to maximize confidence in redshift estimates for individual galaxies as opposed to their statistical properties.

\acknowledgments
{RB and KEW gratefully acknowledge support by NASA through Hubble Fellowship grants \#HF-51318 and \#HF2-51368 awarded by the Space Telescope Science Institute, which is operated by the Association of Universities for Research in Astronomy, Inc., for NASA, under contract NAS 5-26555. This research made use of Astropy, a community-developed core Python package for Astronomy \citep{astropy}. This work is based on observations taken by the 3D-HST Treasury Program (GO 12177 and 12328) with the NASA/ESA HST, which is operated by the Association of Universities for Research in Astronomy, Inc., under NASA contract NAS5-26555. Finally, this work is also based on observations taken by the CANDELS Multi-Cycle Treasury Program with the NASA/ESA HST, which is operated by the Association of Universities for Research in Astronomy, Inc., under NASA contract NAS5-26555.}

\appendix

\section{Field to field Variation in Photometric Redshift Accuracy}\label{sect:fields}

For the most part, we have treated the 3D-HST catalogs as a uniform photometric sample. However, aside from the availability of optical and near-IR HST imaging and Spitzer IRAC photometry, each field includes a heterogenous collection of photometry and spectroscopic redshifts. In this Appendix, we show the scatter between photometric redshifts and true redshifts (grism and spectroscopic) for each field and subset of photometry (Figures \ref{fig:scatt_fields_spec} and \ref{fig:scatt_fields_gris}). Panels in each figure are divided into fields, with the redshifts from the full photometric catalog included in the top left panel (gray points) and redshifts measured from subsets of filters in the additional panels (blue points). Figure \ref{fig:scatt_fields_spec} includes comparisons with spectroscopic redshifts and Figure \ref{fig:scatt_fields_gris} with grism redshifts. Measured scatter and outlier fractions are included in Table \ref{tbl:photoz_filters}.

These figures illuminate some of the reasons for the strong field-to-field variance in photometric redshift scatter and outlier fractions shown in Figures \ref{fig:barplot_spec} and \ref{fig:barplot_gris}. In Figure \ref{fig:scatt_fields_spec} it is apparent that the spectroscopic redshift follow-up varies wildly from field-to-field.  For example, GOODS-N exhibits significantly more scatter than the other fields, however it also includes much better sampling of $z>1$ galaxies e.g. than COSMOS, which has a amazingly tight relationship between photometric and spectroscopic redshifts. On the other hand, GOOD-S also includes a large number of high redshift galaxies, but much lower NMAD scatter.

Differences in spectroscopic datasets do not explain the full field-to-field variation as the variations persist in comparisons with grism redshifts (Figure \ref{fig:scatt_fields_gris}), where the redshift coverage should be approximately uniform. Still COSMOS and GOODS-S, which both have medium band optical imaging, have the most accurate photometric redshifts. These optical filters appear to have a greater effect than medium band filters in the near-IR, which are included in the AEGIS and COSMOS fields from the NMBS Survey \citep{whitaker:11}. We expect this is due to the redshift distribution probed by the grism; at higher redshifts, such deep medium band imaging should be increasingly important.

We find many systematics are introduced by including various subsets of photometric bands in the redshift fitting. Although we have discussed some of these trends in Section \ref{sect:filters}, we include all tests in separate panels in these figures to illustrate some of the systematic redshift failures that are behind the increased scatter and catastrophic failure rates. For example, it is clear that IRAC photometry breaks an important degeneracy that systematically narrows the distribution of photometric redshifts towards $z\sim0.7$, improves the accuracy at $z\sim2$, and discriminates between low redshift ($z<1$) and very high redshift galaxies ($z\sim4{-}5$).

\begin{figure*}
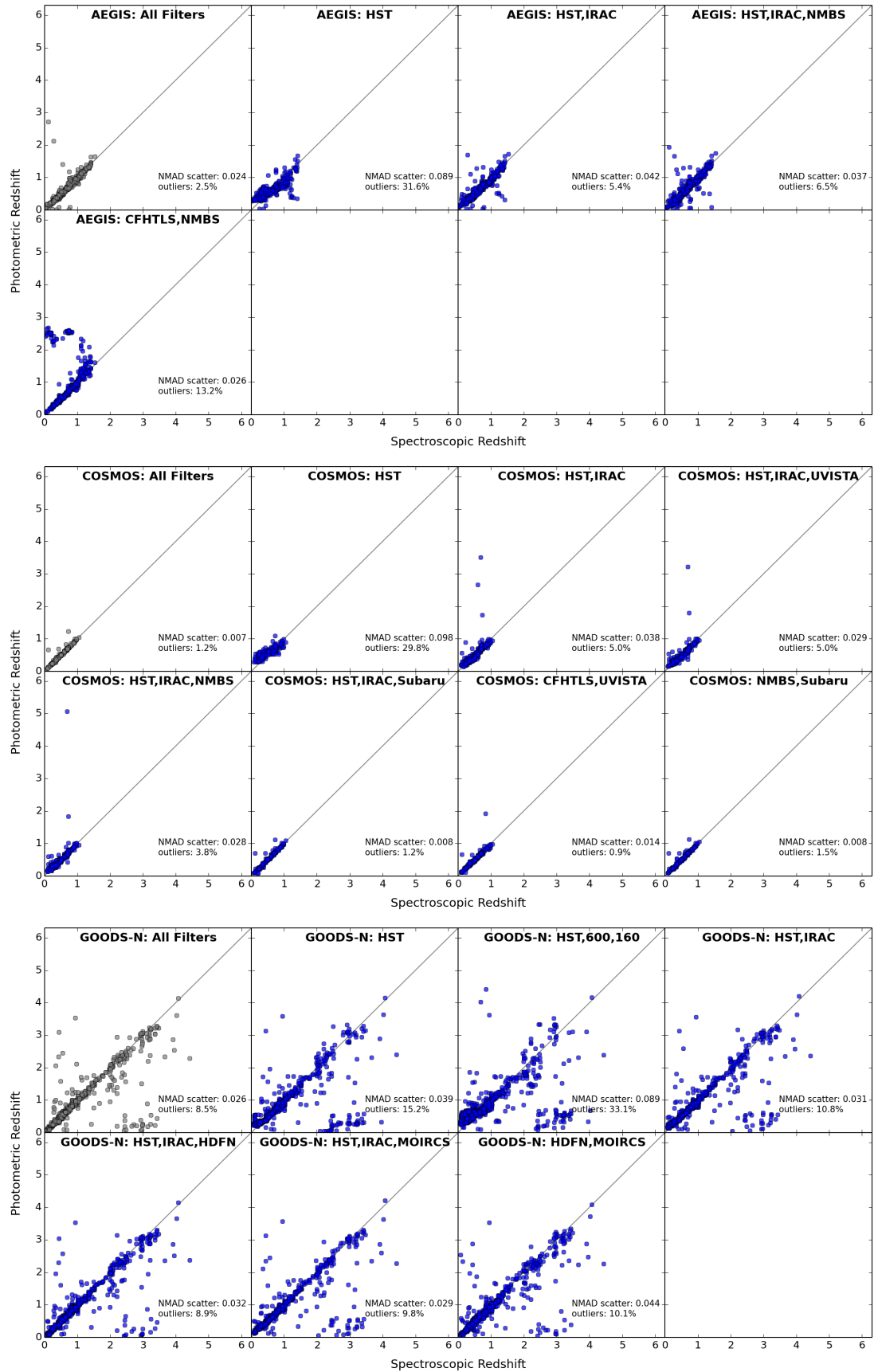

\centering
\includegraphics[width=0.85\textwidth]{Fig24a.pdf}
\includegraphics[width=0.85\textwidth]{Fig24b.pdf}
\includegraphics[width=0.85\textwidth]{Fig24c.pdf}
\caption{Photometric versus Spectroscopic redshifts in the 3D-HST/CANDELS fields for different filter combinations. Fits to the full photometric catalogs are included as gray symbols, all other tests are included as blue symbols.}
\label{fig:scatt_fields_spec}
\end{figure*}

\begin{figure*}
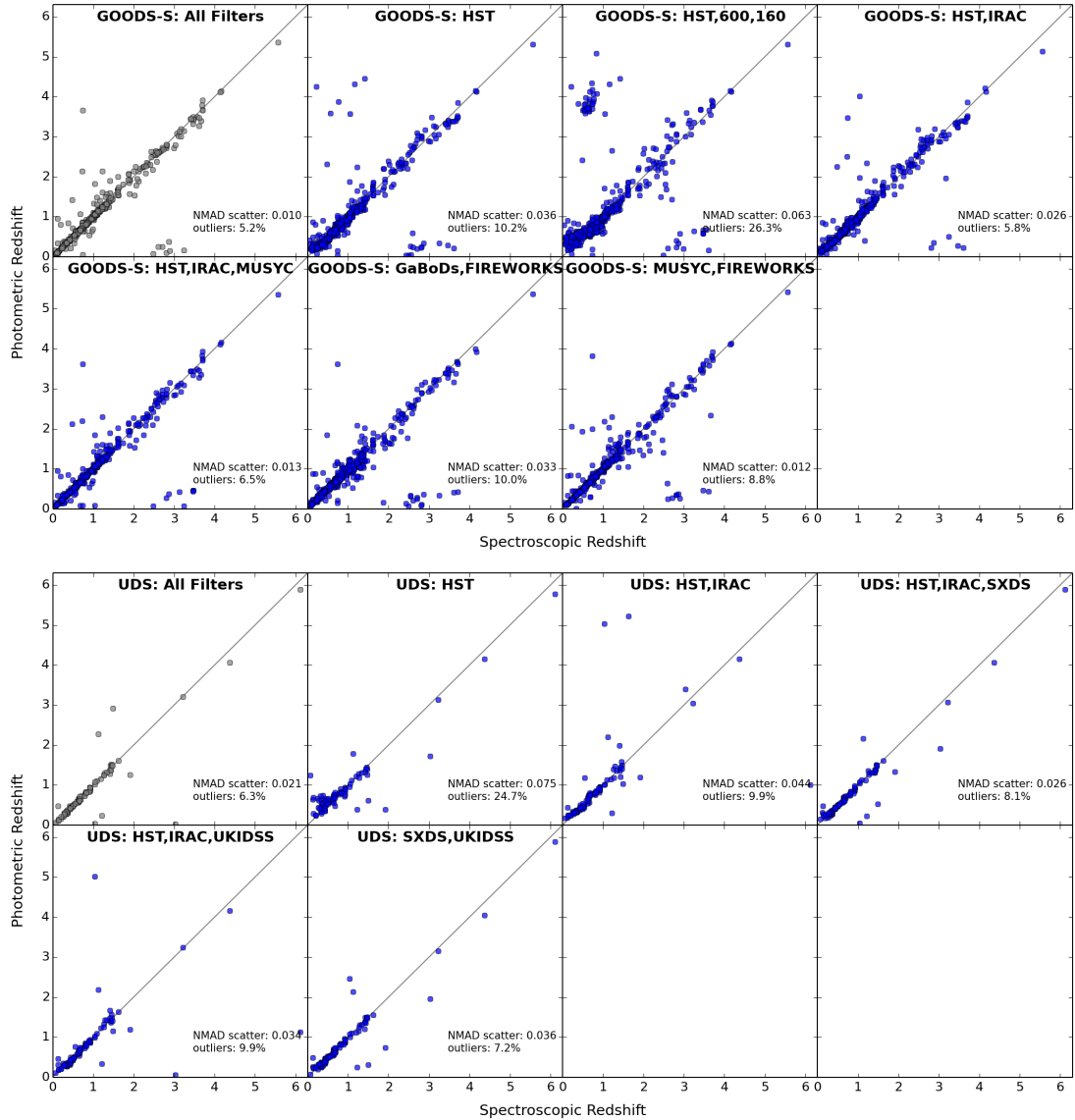

\centering
\addtocounter{figure}{-1}
\includegraphics[width=0.85\textwidth]{Fig24d.pdf}
\includegraphics[width=0.85\textwidth]{Fig24e.pdf}
\caption{(Continued) Photometric versus Spectroscopic redshifts in the 3D-HST/CANDELS fields for different filter combinations.}
\end{figure*}

\begin{figure*}
\centering
\includegraphics[width=0.85\textwidth]{Fig25a.pdf}
\includegraphics[width=0.85\textwidth]{Fig25b.pdf}
\includegraphics[width=0.85\textwidth]{Fig25c.pdf}
\caption{Photometric versus grism redshifts in the 3D-HST/CANDELS fields for different filter combinations.}
\label{fig:scatt_fields_gris}
\end{figure*}

\begin{figure*}
\centering
\addtocounter{figure}{-1}
\includegraphics[width=0.85\textwidth]{Fig25d.pdf}
\includegraphics[width=0.85\textwidth]{Fig25e.pdf}
\caption{(Continued) Photometric versus grism redshifts in the 3D-HST/CANDELS fields for different filter combinations.}
\end{figure*}

\begin{deluxetable*}{cccccc}[]\label{tbl:fields}
\tabletypesize{\scriptsize}
\tablecaption{Photometric Redshift Accuracy with Filters}
\tablehead{
& & \multicolumn{2}{c}{Spectroscopic Redshift Comparison} & \multicolumn{2}{c}{Grism Redshift Comparison} \\
 \colhead{Field} & \colhead{Subsets} &\colhead{$\sigma_{NMAD}$} & \colhead{Outlier \%} &\colhead{$\sigma_{NMAD}$} & \colhead{Outlier \%}}
\startdata
AEGIS & all & $0.024\pm0.001$ & $2.5\%\pm0.6$  & $0.027\pm0.001$ & $5.0\%\pm0.5$ \\ 
 & HST & $0.089\pm0.006$ & $31.6\%\pm2.3$  & $0.090\pm0.003$ & $31.7\%\pm1.0$ \\ 
 & HST,IRAC & $0.042\pm0.002$ & $5.4\%\pm0.8$  & $0.050\pm0.002$ & $15.6\%\pm0.8$ \\ 
 & HST,IRAC,NMBS & $0.037\pm0.002$ & $6.5\%\pm0.9$  & $0.042\pm0.001$ & $12.9\%\pm0.7$ \\ 
 & CFHTLS,NMBS & $0.026\pm0.001$ & $13.2\%\pm1.3$  & $0.050\pm0.002$ & $21.3\%\pm1.0$ \\ 
  \\
COSMOS & all & $0.007\pm0.001$ & $1.2\%\pm0.6$  & $0.012\pm0.000$ & $2.1\%\pm0.3$ \\ 
 & HST & $0.098\pm0.005$ & $29.8\%\pm3.0$  & $0.103\pm0.003$ & $35.5\%\pm1.2$ \\ 
 & HST,IRAC & $0.038\pm0.002$ & $5.0\%\pm1.1$  & $0.058\pm0.002$ & $19.3\%\pm0.9$ \\ 
 & HST,IRAC,UVISTA & $0.029\pm0.003$ & $5.0\%\pm1.2$  & $0.028\pm0.001$ & $8.4\%\pm0.7$ \\ 
 & HST,IRAC,NMBS & $0.028\pm0.003$ & $3.8\%\pm1.0$  & $0.032\pm0.001$ & $10.4\%\pm0.8$ \\ 
 & HST,IRAC,Subaru & $0.008\pm0.001$ & $1.2\%\pm0.6$  & $0.018\pm0.001$ & $3.7\%\pm0.5$ \\ 
 & CFHTLS,UVISTA & $0.014\pm0.001$ & $0.9\%\pm0.5$  & $0.026\pm0.001$ & $4.2\%\pm0.5$ \\ 
 & NMBS,Subaru & $0.008\pm0.001$ & $1.5\%\pm0.7$  & $0.017\pm0.001$ & $4.2\%\pm0.5$ \\ 
 \\
GOODS-N & all & $0.026\pm0.001$ & $8.5\%\pm0.9$  & $0.024\pm0.001$ & $4.4\%\pm0.4$ \\ 
 & HST & $0.039\pm0.002$ & $15.2\%\pm1.2$  & $0.036\pm0.001$ & $8.8\%\pm0.6$ \\ 
 & HST,600,160 & $0.089\pm0.005$ & $33.1\%\pm1.5$  & $0.053\pm0.002$ & $16.4\%\pm0.8$ \\ 
 & HST,IRAC & $0.031\pm0.001$ & $10.8\%\pm1.0$  & $0.028\pm0.001$ & $5.9\%\pm0.5$ \\ 
 & HST,IRAC,HDFN & $0.032\pm0.001$ & $8.9\%\pm0.9$  & $0.029\pm0.001$ & $5.1\%\pm0.5$ \\ 
 & HST,IRAC,MOIRCS & $0.029\pm0.001$ & $9.8\%\pm1.0$  & $0.025\pm0.001$ & $5.4\%\pm0.5$ \\ 
 & HDFN,MOIRCS & $0.044\pm0.002$ & $10.1\%\pm1.0$  & $0.050\pm0.001$ & $11.7\%\pm0.7$ \\ 
 \\
 GOODS-S & all & $0.010\pm0.001$ & $5.2\%\pm0.9$  & $0.014\pm0.000$ & $3.1\%\pm0.4$ \\ 
 & HST & $0.036\pm0.002$ & $10.2\%\pm1.2$  & $0.035\pm0.001$ & $10.9\%\pm0.7$ \\ 
 & HST,600,160 & $0.063\pm0.005$ & $26.3\%\pm1.7$  & $0.051\pm0.002$ & $21.6\%\pm0.9$ \\ 
 & HST,IRAC & $0.026\pm0.002$ & $5.8\%\pm0.9$  & $0.029\pm0.001$ & $9.0\%\pm0.6$ \\ 
 & HST,IRAC,MUSYC & $0.013\pm0.001$ & $6.5\%\pm1.0$  & $0.018\pm0.001$ & $5.6\%\pm0.5$ \\ 
 & GaBoDs,FIREWORKS & $0.033\pm0.002$ & $10.0\%\pm1.1$  & $0.056\pm0.002$ & $20.6\%\pm0.9$ \\ 
 & MUSYC,FIREWORKS & $0.012\pm0.001$ & $8.8\%\pm1.1$  & $0.025\pm0.001$ & $15.0\%\pm0.7$ \\ 
 \\
UDS & all & $0.021\pm0.002$ & $6.3\%\pm2.3$  & $0.025\pm0.001$ & $4.0\%\pm0.4$ \\ 
 & HST & $0.075\pm0.009$ & $24.7\%\pm4.5$  & $0.071\pm0.002$ & $20.1\%\pm0.9$ \\ 
 & HST,IRAC & $0.044\pm0.007$ & $9.9\%\pm2.8$  & $0.051\pm0.002$ & $13.9\%\pm0.8$ \\ 
 & HST,IRAC,SXDS & $0.026\pm0.002$ & $8.1\%\pm2.6$  & $0.029\pm0.001$ & $5.2\%\pm0.5$ \\ 
 & HST,IRAC,UKIDSS & $0.034\pm0.006$ & $9.9\%\pm2.9$  & $0.035\pm0.001$ & $10.8\%\pm0.7$ \\ 
 & SXDS,UKIDSS & $0.036\pm0.003$ & $7.2\%\pm2.5$  & $0.034\pm0.001$ & $5.8\%\pm0.5$ \\ 
\label{tbl:photoz_filters}
\end{deluxetable*}


\begin{thebibliography}{90}
\expandafter\ifx\csname natexlab\endcsname\relax\def\natexlab#1{#1}\fi

\bibitem[{{Abdalla} {et~al.}(2011){Abdalla}, {Banerji}, {Lahav}, \&
  {Rashkov}}]{abdalla:11}
{Abdalla}, F.~B., {Banerji}, M., {Lahav}, O., \& {Rashkov}, V. 2011, \mnras,
  417, 1891

\bibitem[{{Akiyama} {et~al.}(2015){Akiyama}, {Ueda}, {Watson}, {Furusawa},
  {Takata}, {Simpson}, {Morokuma}, {Yamada}, {Ohta}, {Iwamuro}, {Yabe},
  {Tamura}, {Moritani}, {Takato}, {Kimura}, {Maihara}, {Dalton}, {Lewis},
  {Lee}, {Curtis Lake}, {Macaulay}, {Clarke}, {Silverman}, {Croom}, {Ouchi},
  {Hanami}, {D{\'{\i}}az Tello}, {Yoshikawa}, {Fujishiro}, \&
  {Sekiguchi}}]{akiyama:15}
{Akiyama}, M., {et~al.} 2015, \pasj

\bibitem[{{Ascaso} {et~al.}(2015){Ascaso}, {Mei}, \&
  {Ben{\'{\i}}tez}}]{ascaso:15}
{Ascaso}, B., {Mei}, S., \& {Ben{\'{\i}}tez}, N. 2015, ArXiv e-prints

\bibitem[{{Ashby} {et~al.}(2013){Ashby}, {Willner}, {Fazio}, {Huang}, {Arendt},
  {Barmby}, {Barro}, {Bell}, {Bouwens}, {Cattaneo}, {Croton}, {Dav{\'e}},
  {Dunlop}, {Egami}, {Faber}, {Finlator}, {Grogin}, {Guhathakurta},
  {Hernquist}, {Hora}, {Illingworth}, {Kashlinsky}, {Koekemoer}, {Koo},
  {Labb{\'e}}, {Li}, {Lin}, {Moseley}, {Nandra}, {Newman}, {Noeske}, {Ouchi},
  {Peth}, {Rigopoulou}, {Robertson}, {Sarajedini}, {Simard}, {Smith}, {Wang},
  {Wechsler}, {Weiner}, {Wilson}, {Wuyts}, {Yamada}, \& {Yan}}]{ashby:13}
{Ashby}, M.~L.~N., {et~al.} 2013, \apj, 769, 80

\bibitem[{{Astropy Collaboration} {et~al.}(2013){Astropy Collaboration},
  {Robitaille}, {Tollerud}, {Greenfield}, {Droettboom}, {Bray}, {Aldcroft},
  {Davis}, {Ginsburg}, {Price-Whelan}, {Kerzendorf}, {Conley}, {Crighton},
  {Barbary}, {Muna}, {Ferguson}, {Grollier}, {Parikh}, {Nair}, {Unther},
  {Deil}, {Woillez}, {Conseil}, {Kramer}, {Turner}, {Singer}, {Fox}, {Weaver},
  {Zabalza}, {Edwards}, {Azalee Bostroem}, {Burke}, {Casey}, {Crawford},
  {Dencheva}, {Ely}, {Jenness}, {Labrie}, {Lim}, {Pierfederici}, {Pontzen},
  {Ptak}, {Refsdal}, {Servillat}, \& {Streicher}}]{astropy}
{Astropy Collaboration} {et~al.} 2013, \aap, 558, A33

\bibitem[{{Banerji} {et~al.}(2015){Banerji}, {Jouvel}, {Lin}, {McMahon},
  {Lahav}, {Castander}, {Abdalla}, {Bertin}, {Bosman}, {Carnero}, {Kind}, {da
  Costa}, {Gerdes}, {Gschwend}, {Lima}, {Maia}, {Merson}, {Miller}, {Ogando},
  {Pellegrini}, {Reed}, {Saglia}, {S{\'a}nchez}, {Allam}, {Annis}, {Bernstein},
  {Bernstein}, {Bernstein}, {Capozzi}, {Childress}, {Cunha}, {Davis}, {DePoy},
  {Desai}, {Diehl}, {Doel}, {Findlay}, {Finley}, {Flaugher}, {Frieman},
  {Gaztanaga}, {Glazebrook}, {Gonz{\'a}lez-Fern{\'a}ndez}, {Gonzalez-Solares},
  {Honscheid}, {Irwin}, {Jarvis}, {Kim}, {Koposov}, {Kuehn}, {Kupcu-Yoldas},
  {Lagattuta}, {Lewis}, {Lidman}, {Makler}, {Marriner}, {Marshall}, {Miquel},
  {Mohr}, {Neilsen}, {Peoples}, {Sako}, {Sanchez}, {Scarpine}, {Schindler},
  {Schubnell}, {Sevilla}, {Sharp}, {Soares-Santos}, {Swanson}, {Tarle},
  {Thaler}, {Tucker}, {Uddin}, {Wechsler}, {Wester}, {Yuan}, \&
  {Zuntz}}]{banerji:15}
{Banerji}, M., {et~al.} 2015, \mnras, 446, 2523

\bibitem[{{Barger} {et~al.}(2008){Barger}, {Cowie}, \& {Wang}}]{barger:08}
{Barger}, A.~J., {Cowie}, L.~L., \& {Wang}, W.-H. 2008, \apj, 689, 687

\bibitem[{{Barmby} {et~al.}(2008){Barmby}, {Huang}, {Ashby}, {Eisenhardt},
  {Fazio}, {Willner}, \& {Wright}}]{barmby:08}
{Barmby}, P., {Huang}, J.-S., {Ashby}, M.~L.~N., {Eisenhardt}, P.~R.~M.,
  {Fazio}, G.~G., {Willner}, S.~P., \& {Wright}, E.~L. 2008, \apjs, 177, 431

\bibitem[{{Benjamin} {et~al.}(2010){Benjamin}, {van Waerbeke}, {M{\'e}nard}, \&
  {Kilbinger}}]{benjamin:10}
{Benjamin}, J., {van Waerbeke}, L., {M{\'e}nard}, B., \& {Kilbinger}, M. 2010,
  \mnras, 408, 1168

\bibitem[{{Bezanson} {et~al.}(2015){Bezanson}, {Franx}, \& {van
  Dokkum}}]{bezanson:15}
{Bezanson}, R., {Franx}, M., \& {van Dokkum}, P.~G. 2015, \apj, 799, 148

\bibitem[{{Bezanson} {et~al.}(2013){Bezanson}, {van Dokkum}, {van de Sande},
  {Franx}, {Leja}, \& {Kriek}}]{bezanson:13b}
{Bezanson}, R., {van Dokkum}, P.~G., {van de Sande}, J., {Franx}, M., {Leja},
  J., \& {Kriek}, M. 2013, \apjl, 779, L21

\bibitem[{{Bielby} {et~al.}(2012){Bielby}, {Hudelot}, {McCracken}, {Ilbert},
  {Daddi}, {Le F{\`e}vre}, {Gonzalez-Perez}, {Kneib}, {Marmo}, {Mellier},
  {Salvato}, {Sanders}, \& {Willott}}]{bielby:12}
{Bielby}, R., {et~al.} 2012, \aap, 545, A23

\bibitem[{{Bower} {et~al.}(2006){Bower}, {Benson}, {Malbon}, {Helly}, {Frenk},
  {Baugh}, {Cole}, \& {Lacey}}]{bower:06}
{Bower}, R.~G., {Benson}, A.~J., {Malbon}, R., {Helly}, J.~C., {Frenk}, C.~S.,
  {Baugh}, C.~M., {Cole}, S., \& {Lacey}, C.~G. 2006, \mnras, 370, 645

\bibitem[{{Brammer} {et~al.}(2008){Brammer}, {van Dokkum}, \& {Coppi}}]{eazy}
{Brammer}, G.~B., {van Dokkum}, P.~G., \& {Coppi}, P. 2008, \apj, 686, 1503

\bibitem[{{Brammer} {et~al.}(2011){Brammer}, {Whitaker}, {van Dokkum},
  {Marchesini}, {Franx}, {Kriek}, {Labb{\'e}}, {Lee}, {Muzzin}, {Quadri},
  {Rudnick}, \& {Williams}}]{brammer:11}
{Brammer}, G.~B., {et~al.} 2011, \apj, 739, 24

\bibitem[{{Brammer} {et~al.}(2012){Brammer}, {van Dokkum}, {Franx},
  {Fumagalli}, {Patel}, {Rix}, {Skelton}, {Kriek}, {Nelson}, {Schmidt},
  {Bezanson}, {da Cunha}, {Erb}, {Fan}, {F{\"o}rster Schreiber}, {Illingworth},
  {Labb{\'e}}, {Leja}, {Lundgren}, {Magee}, {Marchesini}, {McCarthy},
  {Momcheva}, {Muzzin}, {Quadri}, {Steidel}, {Tal}, {Wake}, {Whitaker}, \&
  {Williams}}]{3dhst}
---. 2012, \apjs, 200, 13

\bibitem[{{Bruzual} \& {Charlot}(2003)}]{bc:03}
{Bruzual}, G., \& {Charlot}, S. 2003, \mnras, 344, 1000

\bibitem[{{Capak} {et~al.}(2004){Capak}, {Cowie}, {Hu}, {Barger}, {Dickinson},
  {Fernandez}, {Giavalisco}, {Komiyama}, {Kretchmer}, {McNally}, {Miyazaki},
  {Okamura}, \& {Stern}}]{capak:04}
{Capak}, P., {et~al.} 2004, \aj, 127, 180

\bibitem[{{Cardamone} {et~al.}(2010){Cardamone}, {van Dokkum}, {Urry},
  {Taniguchi}, {Gawiser}, {Brammer}, {Taylor}, {Damen}, {Treister}, {Cobb},
  {Bond}, {Schawinski}, {Lira}, {Murayama}, {Saito}, \&
  {Sumikawa}}]{cardamone:10}
{Cardamone}, C.~N., {et~al.} 2010, \apjs, 189, 270

\bibitem[{{Chabrier}(2003)}]{chabrier:03}
{Chabrier}, G. 2003, \pasp, 115, 763

\bibitem[{{Chen} {et~al.}(2003){Chen}, {Marzke}, {McCarthy}, {Martini},
  {Carlberg}, {Persson}, {Bunker}, {Bridge}, \& {Abraham}}]{chen:03}
{Chen}, H.-W., {et~al.} 2003, \apj, 586, 745

\bibitem[{{Cohen}(2001)}]{cohen:01}
{Cohen}, J.~G. 2001, \aj, 121, 2895

\bibitem[{{Cohen} {et~al.}(2000){Cohen}, {Hogg}, {Blandford}, {Cowie}, {Hu},
  {Songaila}, {Shopbell}, \& {Richberg}}]{cohen:00}
{Cohen}, J.~G., {Hogg}, D.~W., {Blandford}, R., {Cowie}, L.~L., {Hu}, E.,
  {Songaila}, A., {Shopbell}, P., \& {Richberg}, K. 2000, \apj, 538, 29

\bibitem[{{Cole} {et~al.}(2000){Cole}, {Lacey}, {Baugh}, \& {Frenk}}]{cole:00}
{Cole}, S., {Lacey}, C.~G., {Baugh}, C.~M., \& {Frenk}, C.~S. 2000, \mnras,
  319, 168

\bibitem[{{Cooper} {et~al.}(2012){Cooper}, {Newman}, {Davis}, {Finkbeiner}, \&
  {Gerke}}]{cooper:12}
{Cooper}, M.~C., {Newman}, J.~A., {Davis}, M., {Finkbeiner}, D.~P., \& {Gerke},
  B.~F. 2012, {spec2d: DEEP2 DEIMOS Spectral Pipeline}, astrophysics Source
  Code Library

\bibitem[{{Cowie} {et~al.}(2004){Cowie}, {Barger}, {Hu}, {Capak}, \&
  {Songaila}}]{cowie:04}
{Cowie}, L.~L., {Barger}, A.~J., {Hu}, E.~M., {Capak}, P., \& {Songaila}, A.
  2004, \aj, 127, 3137

\bibitem[{{Dahlen} {et~al.}(2013){Dahlen}, {Mobasher}, {Faber}, {Ferguson},
  {Barro}, {Finkelstein}, {Finlator}, {Fontana}, {Gruetzbauch}, {Johnson},
  {Pforr}, {Salvato}, {Wiklind}, {Wuyts}, {Acquaviva}, {Dickinson}, {Guo},
  {Huang}, {Huang}, {Newman}, {Bell}, {Conselice}, {Galametz}, {Gawiser},
  {Giavalisco}, {Grogin}, {Hathi}, {Kocevski}, {Koekemoer}, {Koo}, {Lee},
  {McGrath}, {Papovich}, {Peth}, {Ryan}, {Somerville}, {Weiner}, \&
  {Wilson}}]{dahlen:13}
{Dahlen}, T., {et~al.} 2013, \apj, 775, 93

\bibitem[{{Dawson} {et~al.}(2001){Dawson}, {Stern}, {Bunker}, {Spinrad}, \&
  {Dey}}]{dawson:01}
{Dawson}, S., {Stern}, D., {Bunker}, A.~J., {Spinrad}, H., \& {Dey}, A. 2001,
  \aj, 122, 598

\bibitem[{{Dickinson} {et~al.}(2003){Dickinson}, {Papovich}, {Ferguson}, \&
  {Budav{\'a}ri}}]{dickinson:03}
{Dickinson}, M., {Papovich}, C., {Ferguson}, H.~C., \& {Budav{\'a}ri}, T. 2003,
  \apj, 587, 25

\bibitem[{{Diehl} \& {For Dark Energy Survey Collaboration}(2012)}]{diehl:12}
{Diehl}, T., \& {For Dark Energy Survey Collaboration}. 2012, Physics Procedia,
  37, 1332

\bibitem[{{Dobos} {et~al.}(2012){Dobos}, {Csabai}, {Yip}, {Budav{\'a}ri},
  {Wild}, \& {Szalay}}]{dobos:12}
{Dobos}, L., {Csabai}, I., {Yip}, C.-W., {Budav{\'a}ri}, T., {Wild}, V., \&
  {Szalay}, A.~S. 2012, \mnras, 420, 1217

\bibitem[{{Erben} {et~al.}(2005){Erben}, {Schirmer}, {Dietrich}, {Cordes},
  {Haberzettl}, {Hetterscheidt}, {Hildebrandt}, {Schmithuesen}, {Schneider},
  {Simon}, {Deul}, {Hook}, {Kaiser}, {Radovich}, {Benoist}, {Nonino}, {Olsen},
  {Prandoni}, {Wichmann}, {Zaggia}, {Bomans}, {Dettmar}, \&
  {Miralles}}]{erben:05}
{Erben}, T., {et~al.} 2005, Astronomische Nachrichten, 326, 432

\bibitem[{{Erben} {et~al.}(2009){Erben}, {Hildebrandt}, {Lerchster}, {Hudelot},
  {Benjamin}, {van Waerbeke}, {Schrabback}, {Brimioulle}, {Cordes}, {Dietrich},
  {Holhjem}, {Schirmer}, \& {Schneider}}]{erben:09}
---. 2009, \aap, 493, 1197

\bibitem[{{Fioc} \& {Rocca-Volmerange}(1997)}]{fioc:97}
{Fioc}, M., \& {Rocca-Volmerange}, B. 1997, \aap, 326, 950

\bibitem[{{Flaugher}(2005)}]{flaugher:05}
{Flaugher}, B. 2005, International Journal of Modern Physics A, 20, 3121

\bibitem[{{Fumagalli} {et~al.}(2014){Fumagalli}, {Labb{\'e}}, {Patel}, {Franx},
  {van Dokkum}, {Brammer}, {da Cunha}, {F{\"o}rster Schreiber}, {Kriek},
  {Quadri}, {Rix}, {Wake}, {Whitaker}, {Lundgren}, {Marchesini}, {Maseda},
  {Momcheva}, {Nelson}, {Pacifici}, \& {Skelton}}]{fumagalli:14}
{Fumagalli}, M., {et~al.} 2014, \apj, 796, 35

\bibitem[{{Furusawa} {et~al.}(2008){Furusawa}, {Kosugi}, {Akiyama}, {Takata},
  {Sekiguchi}, {Tanaka}, {Iwata}, {Kajisawa}, {Yasuda}, {Doi}, {Ouchi},
  {Simpson}, {Shimasaku}, {Yamada}, {Furusawa}, {Morokuma}, {Ishida}, {Aoki},
  {Fuse}, {Imanishi}, {Iye}, {Karoji}, {Kobayashi}, {Kodama}, {Komiyama},
  {Maeda}, {Miyazaki}, {Mizumoto}, {Nakata}, {Noumaru}, {Ogasawara}, {Okamura},
  {Saito}, {Sasaki}, {Ueda}, \& {Yoshida}}]{furusawa:08}
{Furusawa}, H., {et~al.} 2008, \apjs, 176, 1

\bibitem[{{Geach} {et~al.}(2007){Geach}, {Simpson}, {Rawlings}, {Read}, \&
  {Watson}}]{geach:07}
{Geach}, J.~E., {Simpson}, C., {Rawlings}, S., {Read}, A.~M., \& {Watson}, M.
  2007, \mnras, 381, 1369

\bibitem[{{Giavalisco} {et~al.}(2004){Giavalisco}, {Ferguson}, {Koekemoer},
  {Dickinson}, {Alexander}, {Bauer}, {Bergeron}, {Biagetti}, {Brandt},
  {Casertano}, {Cesarsky}, {Chatzichristou}, {Conselice}, {Cristiani}, {Da
  Costa}, {Dahlen}, {de Mello}, {Eisenhardt}, {Erben}, {Fall}, {Fassnacht},
  {Fosbury}, {Fruchter}, {Gardner}, {Grogin}, {Hook}, {Hornschemeier}, {Idzi},
  {Jogee}, {Kretchmer}, {Laidler}, {Lee}, {Livio}, {Lucas}, {Madau},
  {Mobasher}, {Moustakas}, {Nonino}, {Padovani}, {Papovich}, {Park},
  {Ravindranath}, {Renzini}, {Richardson}, {Riess}, {Rosati}, {Schirmer},
  {Schreier}, {Somerville}, {Spinrad}, {Stern}, {Stiavelli}, {Strolger},
  {Urry}, {Vandame}, {Williams}, \& {Wolf}}]{giavalisco:04}
{Giavalisco}, M., {et~al.} 2004, \apjl, 600, L93

\bibitem[{{Grogin} {et~al.}(2011){Grogin}, {Kocevski}, {Faber}, {Ferguson},
  {Koekemoer}, {Riess}, {Acquaviva}, {Alexander}, {Almaini}, {Ashby}, {Barden},
  {Bell}, {Bournaud}, {Brown}, {Caputi}, {Casertano}, {Cassata}, {Castellano},
  {Challis}, {Chary}, {Cheung}, {Cirasuolo}, {Conselice}, {Roshan Cooray},
  {Croton}, {Daddi}, {Dahlen}, {Dav{\'e}}, {de Mello}, {Dekel}, {Dickinson},
  {Dolch}, {Donley}, {Dunlop}, {Dutton}, {Elbaz}, {Fazio}, {Filippenko},
  {Finkelstein}, {Fontana}, {Gardner}, {Garnavich}, {Gawiser}, {Giavalisco},
  {Grazian}, {Guo}, {Hathi}, {H{\"a}ussler}, {Hopkins}, {Huang}, {Huang},
  {Jha}, {Kartaltepe}, {Kirshner}, {Koo}, {Lai}, {Lee}, {Li}, {Lotz}, {Lucas},
  {Madau}, {McCarthy}, {McGrath}, {McIntosh}, {McLure}, {Mobasher},
  {Moustakas}, {Mozena}, {Nandra}, {Newman}, {Niemi}, {Noeske}, {Papovich},
  {Pentericci}, {Pope}, {Primack}, {Rajan}, {Ravindranath}, {Reddy}, {Renzini},
  {Rix}, {Robaina}, {Rodney}, {Rosario}, {Rosati}, {Salimbeni}, {Scarlata},
  {Siana}, {Simard}, {Smidt}, {Somerville}, {Spinrad}, {Straughn}, {Strolger},
  {Telford}, {Teplitz}, {Trump}, {van der Wel}, {Villforth}, {Wechsler},
  {Weiner}, {Wiklind}, {Wild}, {Wilson}, {Wuyts}, {Yan}, \& {Yun}}]{candels}
{Grogin}, N.~A., {et~al.} 2011, \apjs, 197, 35

\bibitem[{{Hildebrandt} {et~al.}(2009){Hildebrandt}, {Pielorz}, {Erben}, {van
  Waerbeke}, {Simon}, \& {Capak}}]{hildebrandt:09}
{Hildebrandt}, H., {Pielorz}, J., {Erben}, T., {van Waerbeke}, L., {Simon}, P.,
  \& {Capak}, P. 2009, \aap, 498, 725

\bibitem[{{Hildebrandt} {et~al.}(2008){Hildebrandt}, {Wolf}, \&
  {Ben{\'{\i}}tez}}]{hildebrandt:08}
{Hildebrandt}, H., {Wolf}, C., \& {Ben{\'{\i}}tez}, N. 2008, \aap, 480, 703

\bibitem[{{Hildebrandt} {et~al.}(2006){Hildebrandt}, {Erben}, {Dietrich},
  {Cordes}, {Haberzettl}, {Hetterscheidt}, {Schirmer}, {Schmithuesen},
  {Schneider}, {Simon}, \& {Trachternach}}]{hildebrandt:06}
{Hildebrandt}, H., {et~al.} 2006, \aap, 452, 1121

\bibitem[{{Hildebrandt} {et~al.}(2010){Hildebrandt}, {Arnouts}, {Capak},
  {Moustakas}, {Wolf}, {Abdalla}, {Assef}, {Banerji}, {Ben{\'{\i}}tez},
  {Brammer}, {Budav{\'a}ri}, {Carliles}, {Coe}, {Dahlen}, {Feldmann}, {Gerdes},
  {Gillis}, {Ilbert}, {Kotulla}, {Lahav}, {Li}, {Miralles}, {Purger},
  {Schmidt}, \& {Singal}}]{hildebrandt:10}
---. 2010, \aap, 523, A31

\bibitem[{{Hogg} {et~al.}(1998){Hogg}, {Cohen}, {Blandford}, {Gwyn},
  {Hartwick}, {Mobasher}, {Mazzei}, {Sawicki}, {Lin}, {Yee}, {Connolly},
  {Brunner}, {Csabai}, {Dickinson}, {Subbarao}, {Szalay}, {Fern{\'a}ndez-Soto},
  {Lanzetta}, \& {Yahil}}]{hogg:98}
{Hogg}, D.~W., {et~al.} 1998, \aj, 115, 1418

\bibitem[{{Hsieh} {et~al.}(2012){Hsieh}, {Wang}, {Hsieh}, {Lin}, {Yan}, {Lim},
  \& {Ho}}]{hsieh:12}
{Hsieh}, B.-C., {Wang}, W.-H., {Hsieh}, C.-C., {Lin}, L., {Yan}, H., {Lim}, J.,
  \& {Ho}, P.~T.~P. 2012, \apjs, 203, 23

\bibitem[{{Ivezic} {et~al.}(2008){Ivezic}, {Tyson}, {Abel}, {Acosta},
  {Allsman}, {AlSayyad}, {Anderson}, {Andrew}, {Angel}, {Angeli}, {Ansari},
  {Antilogus}, {Arndt}, {Astier}, {Aubourg}, {Axelrod}, {Bard}, {Barr},
  {Barrau}, {Bartlett}, {Bauman}, {Beaumont}, {Becker}, {Becla}, {Beldica},
  {Bellavia}, {Blanc}, {Blandford}, {Bloom}, {Bogart}, {Borne}, {Bosch},
  {Boutigny}, {Brandt}, {Brown}, {Bullock}, {Burchat}, {Burke}, {Cagnoli},
  {Calabrese}, {Chandrasekharan}, {Chesley}, {Cheu}, {Chiang}, {Claver},
  {Connolly}, {Cook}, {Cooray}, {Covey}, {Cribbs}, {Cui}, {Cutri}, {Daubard},
  {Daues}, {Delgado}, {Digel}, {Doherty}, {Dubois}, {Dubois-Felsmann},
  {Durech}, {Eracleous}, {Ferguson}, {Frank}, {Freemon}, {Gangler}, {Gawiser},
  {Geary}, {Gee}, {Geha}, {Gibson}, {Gilmore}, {Glanzman}, {Goodenow},
  {Gressler}, {Gris}, {Guyonnet}, {Hascall}, {Haupt}, {Hernandez}, {Hogan},
  {Huang}, {Huffer}, {Innes}, {Jacoby}, {Jain}, {Jee}, {Jernigan},
  {Jevremovic}, {Johns}, {Jones}, {Juramy-Gilles}, {Juric}, {Kahn}, {Kalirai},
  {Kallivayalil}, {Kalmbach}, {Kantor}, {Kasliwal}, {Kessler}, {Kirkby},
  {Knox}, {Kotov}, {Krabbendam}, {Krughoff}, {Kubanek}, {Kuczewski},
  {Kulkarni}, {Lambert}, {Le Guillou}, {Levine}, {Liang}, {Lim}, {Lintott},
  {Lupton}, {Mahabal}, {Marshall}, {Marshall}, {May}, {McKercher}, {Migliore},
  {Miller}, {Mills}, {Monet}, {Moniez}, {Neill}, {Nief}, {Nomerotski},
  {Nordby}, {O'Connor}, {Oliver}, {Olivier}, {Olsen}, {Ortiz}, {Owen}, {Pain},
  {Peterson}, {Petry}, {Pierfederici}, {Pietrowicz}, {Pike}, {Pinto}, {Plante},
  {Plate}, {Price}, {Prouza}, {Radeka}, {Rajagopal}, {Rasmussen}, {Regnault},
  {Ridgway}, {Ritz}, {Rosing}, {Roucelle}, {Rumore}, {Russo}, {Saha},
  {Sassolas}, {Schalk}, {Schindler}, {Schneider}, {Schumacher}, {Sebag},
  {Sembroski}, {Seppala}, {Shipsey}, {Silvestri}, {Smith}, {Smith}, {Strauss},
  {Stubbs}, {Sweeney}, {Szalay}, {Takacs}, {Thaler}, {Van Berg}, {Vanden Berk},
  {Vetter}, {Virieux}, {Xin}, {Walkowicz}, {Walter}, {Wang}, {Warner},
  {Willman}, {Wittman}, {Wolff}, {Wood-Vasey}, {Yoachim}, {Zhan}, \& {for the
  LSST Collaboration}}]{ivezic:08}
{Ivezic}, Z., {et~al.} 2008, ArXiv e-prints

\bibitem[{{Kajisawa} {et~al.}(2010){Kajisawa}, {Ichikawa}, {Yamada},
  {Uchimoto}, {Yoshikawa}, {Akiyama}, \& {Onodera}}]{kajisawa:10}
{Kajisawa}, M., {Ichikawa}, T., {Yamada}, T., {Uchimoto}, Y.~K., {Yoshikawa},
  T., {Akiyama}, M., \& {Onodera}, M. 2010, \apj, 723, 129

\bibitem[{{Kajisawa} {et~al.}(2011){Kajisawa}, {Ichikawa}, {Tanaka}, {Yamada},
  {Akiyama}, {Suzuki}, {Tokoku}, {Katsuno Uchimoto}, {Konishi}, {Yoshikawa},
  {Nishimura}, {Omata}, {Ouchi}, {Iwata}, {Hamana}, \& {Onodera}}]{kajisawa:11}
{Kajisawa}, M., {et~al.} 2011, \pasj, 63, 379

\bibitem[{{Koekemoer} {et~al.}(2011){Koekemoer}, {Faber}, {Ferguson}, {Grogin},
  {Kocevski}, {Koo}, {Lai}, {Lotz}, {Lucas}, {McGrath}, {Ogaz}, {Rajan},
  {Riess}, {Rodney}, {Strolger}, {Casertano}, {Castellano}, {Dahlen},
  {Dickinson}, {Dolch}, {Fontana}, {Giavalisco}, {Grazian}, {Guo}, {Hathi},
  {Huang}, {van der Wel}, {Yan}, {Acquaviva}, {Alexander}, {Almaini}, {Ashby},
  {Barden}, {Bell}, {Bournaud}, {Brown}, {Caputi}, {Cassata}, {Challis},
  {Chary}, {Cheung}, {Cirasuolo}, {Conselice}, {Roshan Cooray}, {Croton},
  {Daddi}, {Dav{\'e}}, {de Mello}, {de Ravel}, {Dekel}, {Donley}, {Dunlop},
  {Dutton}, {Elbaz}, {Fazio}, {Filippenko}, {Finkelstein}, {Frazer}, {Gardner},
  {Garnavich}, {Gawiser}, {Gruetzbauch}, {Hartley}, {H{\"a}ussler},
  {Herrington}, {Hopkins}, {Huang}, {Jha}, {Johnson}, {Kartaltepe},
  {Khostovan}, {Kirshner}, {Lani}, {Lee}, {Li}, {Madau}, {McCarthy},
  {McIntosh}, {McLure}, {McPartland}, {Mobasher}, {Moreira}, {Mortlock},
  {Moustakas}, {Mozena}, {Nandra}, {Newman}, {Nielsen}, {Niemi}, {Noeske},
  {Papovich}, {Pentericci}, {Pope}, {Primack}, {Ravindranath}, {Reddy},
  {Renzini}, {Rix}, {Robaina}, {Rosario}, {Rosati}, {Salimbeni}, {Scarlata},
  {Siana}, {Simard}, {Smidt}, {Snyder}, {Somerville}, {Spinrad}, {Straughn},
  {Telford}, {Teplitz}, {Trump}, {Vargas}, {Villforth}, {Wagner}, {Wandro},
  {Wechsler}, {Weiner}, {Wiklind}, {Wild}, {Wilson}, {Wuyts}, \&
  {Yun}}]{candelsb}
{Koekemoer}, A.~M., {et~al.} 2011, \apjs, 197, 36

\bibitem[{{Kriek} {et~al.}(2009){Kriek}, {van Dokkum}, {Labb{\'e}}, {Franx},
  {Illingworth}, {Marchesini}, \& {Quadri}}]{kriek:09}
{Kriek}, M., {van Dokkum}, P.~G., {Labb{\'e}}, I., {Franx}, M., {Illingworth},
  G.~D., {Marchesini}, D., \& {Quadri}, R.~F. 2009, \apj, 700, 221

\bibitem[{{Lilly} {et~al.}(2007){Lilly}, {Le F{\`e}vre}, {Renzini}, {Zamorani},
  {Scodeggio}, {Contini}, {Carollo}, {Hasinger}, {Kneib}, {Iovino}, {Le Brun},
  {Maier}, {Mainieri}, {Mignoli}, {Silverman}, {Tasca}, {Bolzonella},
  {Bongiorno}, {Bottini}, {Capak}, {Caputi}, {Cimatti}, {Cucciati}, {Daddi},
  {Feldmann}, {Franzetti}, {Garilli}, {Guzzo}, {Ilbert}, {Kampczyk}, {Kovac},
  {Lamareille}, {Leauthaud}, {Borgne}, {McCracken}, {Marinoni}, {Pello},
  {Ricciardelli}, {Scarlata}, {Vergani}, {Sanders}, {Schinnerer}, {Scoville},
  {Taniguchi}, {Arnouts}, {Aussel}, {Bardelli}, {Brusa}, {Cappi}, {Ciliegi},
  {Finoguenov}, {Foucaud}, {Franceschini}, {Halliday}, {Impey}, {Knobel},
  {Koekemoer}, {Kurk}, {Maccagni}, {Maddox}, {Marano}, {Marconi}, {Meneux},
  {Mobasher}, {Moreau}, {Peacock}, {Porciani}, {Pozzetti}, {Scaramella},
  {Schiminovich}, {Shopbell}, {Smail}, {Thompson}, {Tresse}, {Vettolani},
  {Zanichelli}, \& {Zucca}}]{lilly:07}
{Lilly}, S.~J., {et~al.} 2007, \apjs, 172, 70

\bibitem[{Marchesini {et~al.}(2009)Marchesini, van Dokkum,
  F{\"o}rster~Schreiber, Franx, Labbe, \& Wuyts}]{marchesini:09}
Marchesini, D., van Dokkum, P.~G., F{\"o}rster~Schreiber, N.~M., Franx, M.,
  Labbe, I., \& Wuyts, S. 2009, \apj, 701, 1765

\bibitem[{{Marchesini} {et~al.}(2010){Marchesini}, {Whitaker}, {Brammer}, {van
  Dokkum}, {Labb{\'e}}, {Muzzin}, {Quadri}, {Kriek}, {Lee}, {Rudnick}, {Franx},
  {Illingworth}, \& {Wake}}]{marchesini:10}
{Marchesini}, D., {et~al.} 2010, \apj, 725, 1277

\bibitem[{{Marchesini} {et~al.}(2014){Marchesini}, {Muzzin}, {Stefanon},
  {Franx}, {Brammer}, {Marsan}, {Vulcani}, {Fynbo}, {Milvang-Jensen}, {Dunlop},
  \& {Buitrago}}]{marchesini:14}
---. 2014, \apj, 794, 65

\bibitem[{{McCracken} {et~al.}(2012){McCracken}, {Milvang-Jensen}, {Dunlop},
  {Franx}, {Fynbo}, {Le F{\`e}vre}, {Holt}, {Caputi}, {Goranova}, {Buitrago},
  {Emerson}, {Freudling}, {Hudelot}, {L{\'o}pez-Sanjuan}, {Magnard}, {Mellier},
  {M{\o}ller}, {Nilsson}, {Sutherland}, {Tasca}, \& {Zabl}}]{mccracken:12}
{McCracken}, H.~J., {et~al.} 2012, \aap, 544, A156

\bibitem[{{McCracken} {et~al.}(2015){McCracken}, {Wolk}, {Colombi},
  {Kilbinger}, {Ilbert}, {Peirani}, {Coupon}, {Dunlop}, {Milvang-Jensen},
  {Caputi}, {Aussel}, {B{\'e}thermin}, \& {Le F{\`e}vre}}]{mccracken:15}
---. 2015, \mnras, 449, 901

\bibitem[{{McMahon} {et~al.}(2013){McMahon}, {Banerji}, {Gonzalez}, {Koposov},
  {Bejar}, {Lodieu}, {Rebolo}, \& {VHS Collaboration}}]{mcmahon:13}
{McMahon}, R.~G., {Banerji}, M., {Gonzalez}, E., {Koposov}, S.~E., {Bejar},
  V.~J., {Lodieu}, N., {Rebolo}, R., \& {VHS Collaboration}. 2013, The
  Messenger, 154, 35

\bibitem[{{Momcheva} {et~al.}(2015){Momcheva}, {Brammer}, {van Dokkum},
  {Skelton}, {Whitaker}, {Nelson}, {Fumagalli}, {Maseda}, {Leja}, {Franx},
  {Rix}, {Bezanson}, {Da Cunha}, {Dickey}, {F{\"o}rster Schreiber},
  {Illingworth}, {Kriek}, {Labb{\'e}}, {Ulf Lange}, {Lundgren}, {Magee},
  {Marchesini}, {Oesch}, {Pacifici}, {Patel}, {Price}, {Tal}, {Wake}, {van der
  Wel}, \& {Wuyts}}]{momcheva3dhst}
{Momcheva}, I.~G., {et~al.} 2015, arXiv:1510.02106

\bibitem[{{Muzzin} {et~al.}(2013){Muzzin}, {Marchesini}, {Stefanon}, {Franx},
  {McCracken}, {Milvang-Jensen}, {Dunlop}, {Fynbo}, {Brammer}, {Labb{\'e}}, \&
  {van Dokkum}}]{muzzin:13}
{Muzzin}, A., {et~al.} 2013, \apj, 777, 18

\bibitem[{{Myers} {et~al.}(2009){Myers}, {White}, \& {Ball}}]{myers:09}
{Myers}, A.~D., {White}, M., \& {Ball}, N.~M. 2009, \mnras, 399, 2279

\bibitem[{{Newman}(2008)}]{newman:08}
{Newman}, J.~A. 2008, \apj, 684, 88

\bibitem[{{Newman} {et~al.}(2013){Newman}, {Cooper}, {Davis}, {Faber}, {Coil},
  {Guhathakurta}, {Koo}, {Phillips}, {Conroy}, {Dutton}, {Finkbeiner}, {Gerke},
  {Rosario}, {Weiner}, {Willmer}, {Yan}, {Harker}, {Kassin}, {Konidaris},
  {Lai}, {Madgwick}, {Noeske}, {Wirth}, {Connolly}, {Kaiser}, {Kirby},
  {Lemaux}, {Lin}, {Lotz}, {Luppino}, {Marinoni}, {Matthews}, {Metevier}, \&
  {Schiavon}}]{newman:13}
{Newman}, J.~A., {et~al.} 2013, \apjs, 208, 5

\bibitem[{{Nonino} {et~al.}(2009){Nonino}, {Dickinson}, {Rosati}, {Grazian},
  {Reddy}, {Cristiani}, {Giavalisco}, {Kuntschner}, {Vanzella}, {Daddi},
  {Fosbury}, \& {Cesarsky}}]{nonino:09}
{Nonino}, M., {et~al.} 2009, \apjs, 183, 244

\bibitem[{{Ono} {et~al.}(2010){Ono}, {Ouchi}, {Shimasaku}, {Akiyama}, {Dunlop},
  {Farrah}, {Lee}, {McLure}, {Okamura}, \& {Yoshida}}]{ono:10}
{Ono}, Y., {et~al.} 2010, \mnras, 402, 1580

\bibitem[{{Ouchi} {et~al.}(2008){Ouchi}, {Shimasaku}, {Akiyama}, {Simpson},
  {Saito}, {Ueda}, {Furusawa}, {Sekiguchi}, {Yamada}, {Kodama}, {Kashikawa},
  {Okamura}, {Iye}, {Takata}, {Yoshida}, \& {Yoshida}}]{ouchi:08}
{Ouchi}, M., {et~al.} 2008, \apjs, 176, 301

\bibitem[{{Papovich} {et~al.}(2010){Papovich}, {Momcheva}, {Willmer},
  {Finkelstein}, {Finkelstein}, {Tran}, {Brodwin}, {Dunlop}, {Farrah}, {Khan},
  {Lotz}, {McCarthy}, {McLure}, {Rieke}, {Rudnick}, {Sivanandam}, {Pacaud}, \&
  {Pierre}}]{papovich:10}
{Papovich}, C., {et~al.} 2010, \apj, 716, 1503

\bibitem[{{Quadri} \& {Williams}(2010)}]{quadri:10}
{Quadri}, R.~F., \& {Williams}, R.~J. 2010, \apj, 725, 794

\bibitem[{{Quadri} {et~al.}(2008){Quadri}, {Williams}, {Lee}, {Franx}, {van
  Dokkum}, \& {Brammer}}]{quadri:08}
{Quadri}, R.~F., {Williams}, R.~J., {Lee}, K.-S., {Franx}, M., {van Dokkum},
  P., \& {Brammer}, G.~B. 2008, \apjl, 685, L1

\bibitem[{{Reddy} {et~al.}(2006){Reddy}, {Steidel}, {Erb}, {Shapley}, \&
  {Pettini}}]{reddy:06}
{Reddy}, N.~A., {Steidel}, C.~C., {Erb}, D.~K., {Shapley}, A.~E., \& {Pettini},
  M. 2006, \apj, 653, 1004

\bibitem[{{Retzlaff} {et~al.}(2010){Retzlaff}, {Rosati}, {Dickinson},
  {Vandame}, {Rit{\'e}}, {Nonino}, {Cesarsky}, \& {GOODS Team}}]{retzlaff:10}
{Retzlaff}, J., {Rosati}, P., {Dickinson}, M., {Vandame}, B., {Rit{\'e}}, C.,
  {Nonino}, M., {Cesarsky}, C., \& {GOODS Team}. 2010, \aap, 511, A50

\bibitem[{{Sanders} {et~al.}(2007){Sanders}, {Salvato}, {Aussel}, {Ilbert},
  {Scoville}, {Surace}, {Frayer}, {Sheth}, {Helou}, {Brooke}, {Bhattacharya},
  {Yan}, {Kartaltepe}, {Barnes}, {Blain}, {Calzetti}, {Capak}, {Carilli},
  {Carollo}, {Comastri}, {Daddi}, {Ellis}, {Elvis}, {Fall}, {Franceschini},
  {Giavalisco}, {Hasinger}, {Impey}, {Koekemoer}, {Le F{\`e}vre}, {Lilly},
  {Liu}, {McCracken}, {Mobasher}, {Renzini}, {Rich}, {Schinnerer}, {Shopbell},
  {Taniguchi}, {Thompson}, {Urry}, \& {Williams}}]{sanders:07}
{Sanders}, D.~B., {et~al.} 2007, \apjs, 172, 86

\bibitem[{{Simpson} {et~al.}(2006){Simpson}, {Almaini}, {Cirasuolo}, {Dunlop},
  {Foucaud}, {Hirst}, {Ivison}, {Page}, {Rawlings}, {Sekiguchi}, {Smail}, \&
  {Watson}}]{simpson:06}
{Simpson}, C., {et~al.} 2006, \mnras, 373, L21

\bibitem[{{Simpson} {et~al.}(2012){Simpson}, {Rawlings}, {Ivison}, {Akiyama},
  {Almaini}, {Bradshaw}, {Chapman}, {Chuter}, {Croom}, {Dunlop}, {Foucaud}, \&
  {Hartley}}]{simpson:12}
---. 2012, \mnras, 421, 3060

\bibitem[{{Skelton} {et~al.}(2014){Skelton}, {Whitaker}, {Momcheva}, {Brammer},
  {van Dokkum}, {Labb{\'e}}, {Franx}, {van der Wel}, {Bezanson}, {Da Cunha},
  {Fumagalli}, {F{\"o}rster Schreiber}, {Kriek}, {Leja}, {Lundgren}, {Magee},
  {Marchesini}, {Maseda}, {Nelson}, {Oesch}, {Pacifici}, {Patel}, {Price},
  {Rix}, {Tal}, {Wake}, \& {Wuyts}}]{skelton3dhst}
{Skelton}, R.~E., {et~al.} 2014, \apjs, 214, 24

\bibitem[{{Smail} {et~al.}(2008){Smail}, {Sharp}, {Swinbank}, {Akiyama},
  {Ueda}, {Foucaud}, {Almaini}, \& {Croom}}]{smail:08}
{Smail}, I., {Sharp}, R., {Swinbank}, A.~M., {Akiyama}, M., {Ueda}, Y.,
  {Foucaud}, S., {Almaini}, O., \& {Croom}, S. 2008, \mnras, 389, 407

\bibitem[{{So{\l}tan} \& {Chodorowski}(2015)}]{soltan:15}
{So{\l}tan}, A.~M., \& {Chodorowski}, M.~J. 2015, \mnras, 453, 1013

\bibitem[{{Springel} {et~al.}(2005){Springel}, {White}, {Jenkins}, {Frenk},
  {Yoshida}, {Gao}, {Navarro}, {Thacker}, {Croton}, {Helly}, {Peacock}, {Cole},
  {Thomas}, {Couchman}, {Evrard}, {Colberg}, \& {Pearce}}]{springel:05}
{Springel}, V., {et~al.} 2005, \nat, 435, 629

\bibitem[{{Steidel} {et~al.}(2003){Steidel}, {Adelberger}, {Shapley},
  {Pettini}, {Dickinson}, \& {Giavalisco}}]{steidel:03}
{Steidel}, C.~C., {Adelberger}, K.~L., {Shapley}, A.~E., {Pettini}, M.,
  {Dickinson}, M., \& {Giavalisco}, M. 2003, \apj, 592, 728

\bibitem[{{Taniguchi} {et~al.}(2007){Taniguchi}, {Scoville}, {Murayama},
  {Sanders}, {Mobasher}, {Aussel}, {Capak}, {Ajiki}, {Miyazaki}, {Komiyama},
  {Shioya}, {Nagao}, {Sasaki}, {Koda}, {Carilli}, {Giavalisco}, {Guzzo},
  {Hasinger}, {Impey}, {LeFevre}, {Lilly}, {Renzini}, {Rich}, {Schinnerer},
  {Shopbell}, {Kaifu}, {Karoji}, {Arimoto}, {Okamura}, \&
  {Ohta}}]{taniguchi:07}
{Taniguchi}, Y., {et~al.} 2007, \apjs, 172, 9

\bibitem[{{Treu} {et~al.}(2005){Treu}, {Ellis}, {Liao}, \& {van
  Dokkum}}]{treu:05}
{Treu}, T., {Ellis}, R.~S., {Liao}, T.~X., \& {van Dokkum}, P.~G. 2005, \apjl,
  622, L5

\bibitem[{{van Breukelen} {et~al.}(2007){van Breukelen}, {Cotter}, {Rawlings},
  {Readhead}, {Bonfield}, {Clewley}, {Ivison}, {Jarvis}, {Simpson}, \&
  {Watson}}]{breukelen:07}
{van Breukelen}, C., {et~al.} 2007, \mnras, 382, 971

\bibitem[{{van de Sande} {et~al.}(2013){van de Sande}, {Kriek}, {Franx}, {van
  Dokkum}, {Bezanson}, {Bouwens}, {Quadri}, {Rix}, \& {Skelton}}]{sande:13}
{van de Sande}, J., {et~al.} 2013, \apj, 771, 85

\bibitem[{{Wake} {et~al.}(2011){Wake}, {Whitaker}, {Labb{\'e}}, {van Dokkum},
  {Franx}, {Quadri}, {Brammer}, {Kriek}, {Lundgren}, {Marchesini}, \&
  {Muzzin}}]{wake:11}
{Wake}, D.~A., {et~al.} 2011, \apj, 728, 46

\bibitem[{{Whitaker} {et~al.}(2012){Whitaker}, {Kriek}, {van Dokkum},
  {Bezanson}, {Brammer}, {Franx}, \& {Labb{\'e}}}]{whitaker:12a}
{Whitaker}, K.~E., {Kriek}, M., {van Dokkum}, P.~G., {Bezanson}, R., {Brammer},
  G., {Franx}, M., \& {Labb{\'e}}, I. 2012, \apj, 745, 179

\bibitem[{Whitaker {et~al.}(2011)Whitaker, Labbe, van Dokkum, Brammer, Kriek,
  Marchesini, Quadri, Franx, Muzzin, Williams, Bezanson, Illingworth, Lee,
  Lundgren, Nelson, Rudnick, Tal, \& Wake}]{whitaker:11}
Whitaker, K.~E., {et~al.} 2011, \apj, 735, 86

\bibitem[{{Wirth} {et~al.}(2004){Wirth}, {Willmer}, {Amico}, {Chaffee},
  {Goodrich}, {Kwok}, {Lyke}, {Mader}, {Tran}, {Barger}, {Cowie}, {Capak},
  {Coil}, {Cooper}, {Conrad}, {Davis}, {Faber}, {Hu}, {Koo}, {Le Mignant},
  {Newman}, \& {Songaila}}]{tkrs}
{Wirth}, G.~D., {et~al.} 2004, \aj, 127, 3121

\bibitem[{{Wuyts} {et~al.}(2008){Wuyts}, {Labb{\'e}}, {Schreiber}, {Franx},
  {Rudnick}, {Brammer}, \& {van Dokkum}}]{wuyts:08}
{Wuyts}, S., {Labb{\'e}}, I., {Schreiber}, N.~M.~F., {Franx}, M., {Rudnick},
  G., {Brammer}, G.~B., \& {van Dokkum}, P.~G. 2008, \apj, 682, 985

\bibitem[{{Yamada} {et~al.}(2005){Yamada}, {Kodama}, {Akiyama}, {Furusawa},
  {Iwata}, {Kajisawa}, {Iye}, {Ouchi}, {Sekiguchi}, {Shimasaku}, {Simpson},
  {Tanaka}, \& {Yoshida}}]{yamada:05}
{Yamada}, T., {et~al.} 2005, \apj, 634, 861

\bibitem[{{Yoshikawa} {et~al.}(2010){Yoshikawa}, {Akiyama}, {Kajisawa},
  {Alexander}, {Ohta}, {Suzuki}, {Tokoku}, {Uchimoto}, {Konishi}, {Yamada},
  {Tanaka}, {Omata}, {Nishimura}, {Koekemoer}, {Brandt}, \&
  {Ichikawa}}]{yoshikawa:10}
{Yoshikawa}, T., {et~al.} 2010, \apj, 718, 112

\end{thebibliography}
\end{document}